\documentclass[11pt,preprint]{aastex}
 
\received{}
\revised{}
\accepted{}
\ccc{}
\cpright{}{}
\newcommand{\bb}{\beta}
\newcommand{\dd}{\delta}
\newcommand{\DD}{\Delta}

\newcommand{\Ga}{\Gamma}

\newcommand{\sig}{\sigma}

\newcommand{\Th}{\Theta}
\newcommand{\lra}{\longrightarrow}

\newcommand{\gp}{$\gamma p \lra e^- e^+p$\/}
\newcommand{\gggg}{$\gamma \gamma \lra e^- e^+$\/}
\begin{document}
 
\title{PRODUCTION OF THE HIGH ENERGY-MOMENTUM SPECTRA OF
               QUASARS 3C 279 AND 3C 273 USING
                  THE PENROSE MECHANISM}
\author{Reva Kay Williams} 

\affil{Department of Astronomy, University of Florida,
Gainesville, FL 32611} 

\email{revak@astro.ufl.edu}

\begin{abstract}
Theoretical and numerical (Monte Carlo) {\it N-particle} computer 
model simulations
show that Penrose Compton scattering \allowbreak
 (PCS) near the event horizon
and Penrose pair production (PPP) at or near the photon
orbit, in
the ergosphere  of a supermassive ($M=10^8 M_\odot$) rotating black
hole,  can generate
the necessary energy-momentum spectra to explain the origin
of the mysterious fluxes of ultrarelativistic
 electrons, inferred
from observations to emerge
from the cores of Quasars 3C 279 and 3C 273, and
other active galactic nuclei (AGNs).
Particles from an accretion disk surrounding the black hole fall into
the ergosphere and scatter off particles that are in trapped or
bound unstable orbits.
The Penrose mechanism allows rotational energy
of a Kerr black hole, and energy-momentum produced
by its strong gravitational field, to be extracted by
scattered particles escaping from the
ergosphere to infinity (i.e., large distances from the black hole).
The results of
these model calculations show that the Penrose mechanism is capable of
producing the observed high energy particles
($\sim $~GeV) emitted by quasars and other AGNs.  This mechanism can
extract hard X-ray/$\gamma$-ray
photons from PCS of initially infalling
low energy UV/soft X-ray photons by target orbiting electrons in the
ergosphere.  The PPP (\gggg) allows  the escape of
relativistic $e^- e^+$ pairs---produced by infalling low energy
photons interacting with  highly blueshifted target
photons at the photon
orbit.
These $e^- e^+$ pairs emerge with maximum Lorentz factor $\sim 10^4$,
which are consistent with relativistic beaming models used to explain
the high energy spectra of so-called blazars, such as 3C~279.
Moreover, and importantly, the emission of scattered particles by this
mechanism naturally produces relativistic jets
collimated about the polar axis, and in most cases one-sided or
asymmetrical,
agreeing with observations of AGNs.  In these fully relativistic
calculations, the energy-momentum four vectors (or four-momenta)
of the scattered particles
are obtained.
Based on the consistency of the four-momenta of the scattered
Penrose particles with observations of quasars and other
objects we observe to be powered by black holes, i.e., the high
energies and luminosities, asymmetrical and collimated
jets, it appears that the geodesic treatment of many individual
particles is sufficient to describe the ``flow'' of the particles,
at least very
close to the event horizon. 
\end{abstract}

%
\section{INTRODUCTION}
\label{sec:intro}

Blazars [optically violently variable (OVV) quasars
and BL Lac objects] are probably ``normal'' 
active galactic nuclei (AGNs) 
with relativistic jets pointed somewhat in the 
direction of the observer,
in which Doppler-boosted radiation plays a large part
in their observed spectra.  
Many of the popular models of blazars involve 
relativistic beaming
(Blandford, McKee, \& 
Rees 1977;
Cohen et~al. 1977; Bassani \& Dean 1981; Unwin et~al. 1989; 
Dermer, Schlickeiser, \&  Mastichiadis 1992; 
Maraschi, Ghisellini, \& Celotti 1992; 
Dermer \& Schlickeiser 1993; Zdziarski \& Krolik 1993;
Sikora, Begelman, \& Rees 1994). 
These models can explain (1)
how the relativistic expansion  of the jets avoids electron-positron
($e^-e^+$) pair production that attenuates $\gamma$-rays in compact 
sources (Bassani \& Dean 1981; Dermer et~al. 1992) and (2) how   
the expansion avoids the so-called Compton catastrophe that raises
radio photons to X-ray photons by self-Compton scattering (i.e.,
the expansion provides the conditions for us to 
observe large fluxes of $\gamma$-rays and radio photons).
In addition, these models appear to give an explanation for 
the observed one-sided jets,
while also explaining why matter appears to be
ejected with speeds greater than the speed of light
(superluminal motion). 
Such models are no doubt important.  
However, these models are less successful in 
explaining the origin of the ultrarelativistic electrons 
emerging from the   
nucleus with Lorentz factors up to $\gamma\sim 10^4$. Such high energy 
particles it appears are needed in relativistic beaming models to 
initiate
the expansion.  That is, one must first
find a mechanism to create copious ultrarelativistic 
electrons with maximum Lorentz 
factors $\gamma_{\rm max} \sim 5-7\times 10^3$ (Sikora, Begelman, \&
Rees 1994) in order to make 
these superluminal expansion models
operate efficiently to reproduce the observed spectra.  
Here in this paper, I present such a mechanism.

In this paper, I present a
4-dimensional, fully relativistic, theoretical and numerical
(Monte Carlo) model calculation of Penrose (1969)
scattering processes in 
the ergosphere of a supermassive rotating black hole, which allows
the extraction of  orbital energy and momentum from particles  
in bound trapped orbits near the event horizon. This energy
and momentum escape from the black hole to large distances in
the form of X-rays, $\gamma$-rays, and $e^-e^+$
pairs.  This black hole model 
for the energy source of AGNs (1) naturally 
produces one-sided jets---a purely relativistic effect related
to the extreme inertial frame 
dragging in the ergosphere (Williams 2002a, 2002b, 2001);  
(2) shows that AGNs may or may not
be strong $\gamma$-ray emitters, depending on the form of the accretion
disk and the jet axis orientation relative to the observers line of
sight; (3) $e^-e^+$ pairs emerge from the nucleus with Lorentz 
factors up to 
$\sim 10^5$, enabling relativistic beaming models to yield
apparent energies in the TeV range;
and (4) produces jets of escaping energetic 
vortical orbiting particles collimated about the rotation
axis of the black hole
(de Felice \& Carlotto 1997; Williams 2002b, 2001)---this also is a
purely relativistic effect caused by the frame dragging.
This frame dragging (sometimes referred to as the Lense-Thirring 
effect) is 
 due to the 
gravitational force associated with a rotating mass (Thirring \&
Lense 1918).

In general, past models have attempted to explain both 
items (1) and (2) above
solely by relativistic beaming; and 
because of this,
inconsistencies have arisen: observations 
seem to suggest that these
features (the one-sided jets and the sometimes strong $\gamma$-ray
emission) may not just be optical illusions, as in the case of
relativistic beaming models,
but instead may also be intrinsic 
to the source (Barthel 1987).  
I have found this to be true in the
theoretical and numerical analysis of the Penrose mechanism 
presented in 
this paper. 
In addition, relativistic
beaming models can not explain one-sided jets when the observed
radiation from the jet does not 
appear to be near the observer's line of sight, i.e.,
when the source exhibits little superluminal expansion
(Jones 1987).
Moreover, findings by Dennett-Thorpe et al. (1997),
after studying the asymmetry of jets, lobe length, and spectral 
index in a sample of quasars, with well defined jets, led them  
to conclude that the simplest models of relativistic beaming fail to   
explain the ``present observations''; and that only 
if the intrinsic spectrum is curved (i.e., with the spectral index 
increasing with frequency), such that the approaching lobe is seen 
at a significant lower frequency, will asymmetry of the right sense
be produced.     
This supports my findings 
that some asymmetry in the jet/counter jet may be intrinsic to 
the energy source, and that, relativistic beaming 
may just serve as an important enhancement 
mechanism (Williams 2002a, 2002b).    
Further,  observations of 
so-called microquasars in our Galaxy show some asymmetry
in the jet and counter jet distributions (Mirabel \& Rodriguez 1998);
and not  all of the observed 
asymmetry can be explained by relativistic beaming, suggesting 
that some asymmetry is intrinsic (Hjellming \& Rupen 1995; Mirabel
\& Rodiguez 1996).

As related to item (3) above,
some attempts have been made to generate 
the ultrarelativistic particles in 
standard accretion disk models using electrodynamics  
(Blandford 1976, Blandford \& Znajek 1977; Macdonald
\& Thorne 1982; Phinney 1982; Blandford \& Payne 1982;
Lovelace 1976, Lovelace \& Scott 1982; Lovelace \& Burns 1982;
Punsly \& Coroniti 1989, 1990a, 1990b; Punsly 1991).
In such models, it is proposed that,
the dipolar-like  magnetic field
lines---due to the accretion disk---thread through the highly
conductive disk plasma, inducing an electric field while
torquing the rotating black hole or the accretion disk,
creating an electromagnetic
field which acts like a dynamo, resulting in the generation
of two oppositely-directed jets of Poynting fluxes 
[$\sim c/4\pi({\bf E}\times\bf B)$] of
electromagnetic energy.
These past models are complex and have problems as pointed out
by Punsly and Coroniti (1989, 1990a,  1990b).  The 
model (Punsly 1991)
in which the disk magnetic field lines thread the ergospheric disk,
extracting rotational energy by means of magnetohydrodynamics (MHD),
although complex, appears to be potentially powerful; 
however, 
it in itself, does not appear to
produce the large Lorentz factors and luminosities needed to 
universally explain
observations.
Moreover, such models,  as of yet, are not easy 
to qualify (Eilek 1980, 1991), even until today. 
The electrodynamics of these models probably best describe 
collimation and 
acceleration of relativistic particles that 
emerge from  the nucleus 
by some other
process, such as the Penrose process described here; 
and thus, the electrodynamics could possibly sustain 
the jets of relativistic particles out to their 
observed distances, while
producing the beaming effect. 

Now, it has been generally believed in the past that
inflow from the accretion disk into the black hole is a 
MHD (or hydrodynamic) process; and one
would think that postulating particles in geodesic  orbits 
is inconsistent with 
participating in the inflow.
However, these present model calculations are not saying that the 
Penrose scattering particles do not 
participate in the inflow---yielding inconsistency; 
what they are saying is that the   
geodesic characteristics of the particles appear to be the best way 
to define the inflow and outflow of plasma 
close to the event horizon. This flow, of course, can no 
doubt be treated as a fluid: whose velocity pattern is controlled 
by the density, pressure, temperature, magnetic field, and 
gravitational field---however, MHD may not be the 
best treatment for this already extremely complex problem,
i.e., where gravity is dominant.

Moreover, based on the consistency of the energy-momentum
four-vectors (or four-momenta) of the 
scattered
Penrose particles with the observations of quasars and other
objects we observe to be powered by black holes (i.e., high 
energies and luminosities, asymmetrical and collimated
jets), it appears that the geodesic treatment of many individual
particles is sufficient to describe the flow, at least very
close to the event horizon.  At most, MHD models (whenever
the major problems associated with them are worked out, such as,
how to convert from electromagnetic energy to particle energy,
i.e., from a Poynting flux to a relativistic particle flux)
will duplicate what I have already found from these Penrose 
processes (see for example the 
discussing in Appendix A), and will in no doubt describe    
the flow of the jets away from the black hole, at a distance
where electromagnetic collimation due to accretion disk
properties may become important. 
However, the single-particle approach is essential close to 
the black 
hole even in a more realistic MHD treatment, especially when 
the fluid is in a supersonic regime (de Felice \& Carlotto 1997).
Gravity apparently causes the bulk of  fluid elements to 
behave like the bulk motion of individual particles moving
along geodesics. 

Further, considering the Blandford-Znajek (1977) model 
referenced above, although recently, once again, has risen
in popularity, this model which attempts to extract
rotational energy from a rotating black hole, 
as related to the Penrose mechanism, is not 
tenable, as pointed out by Punsly and Coroniti (1989, 
1990a,  1990b), according to the ``no-hair'' theorem 
(Carter 1973; Misner, Thorne, \& Wheeler 1973) and the 
``vacuum infinity'' horizon (Regge \& Wheeler 1957; Punsly 1989).
Specifically, the Blandford-Znajek 
model appears to be untenable for at least two reasons:
(1) it relies on energy and angular momentum being transported
out of the event horizon by an electric current; however,
the assumption that the
electric current can freely flow into and out of the horizon is
not valid;
(2)  the horizon is causally decoupled 
from the accretion disk plasma, including its magnetic field, 
 thus making
this model an ineffective way of extracting  rotational 
energy (Punsly 1991, 1996).   
In other words, the magnetic field lines of the 
surrounding accretion disk cannot anchor to the 
``vacuum infinity,'' and in general, charge neutral horizon, 
before being expelled or redshifted away (Bi\v{c}\'{a}k 2000;
Bi\v{c}\'{a}k \& Ledvinka 2000),
thereby being unable to
torque the black hole, generating an electric field, leading
to a Poynting flux.      That is, just as the Kerr metric 
(\S~2)
asymptotically goes to a Minkowski (flat spacetime) metric
at infinity, such that an observer at infinity is not affected
by the black hole, so is the vacuum infinity horizon to the 
magnetic field and neutral plasma of the accretion disk, 
suggesting that Blandford-Znajek type models may be 
important at an ``effective'' faraway distance from the event 
horizon.
This presents a problem for MHD models
of AGNs 
that rely on the Blandford-Znajek process to
extract electromagnetic energy from the rotating black hole
(e.g.,  Blandford \& Levinson 1994; Hirotani \&
Okamoto 1998; Li 2002). 

Still other attempts have been made to create the ultrarelativistic
electrons by shocks in the jets themselves, 
associated with either dense clouds
accelerated by the flow or with an unsteady velocity field in the jet
(Blandford \& McKee 1977; Jones \& Tobin 1977; Marscher 1978).
Such shocks could possibly explain the local acceleration along the
jet at a  distance from the nucleus, perhaps even out to 
the lobes. 
However, these models are not self-contained; they must assume 
that the initial
blast, which initiates the shocks, originated within the nucleus of
the compact source as a result of an explosion or instability.

In this paper, I present an analysis of the Penrose mechanism 
by which the central 
black hole source can eject a ``blast'' of ultrarelativistic 
particles, escaping into the polar regions, mainly
$e^-e^+$ pairs and some lower energy hard X-ray and $\gamma$-ray photons. 
Subsequently, the Penrose produced pairs
can possibly interact with a
surrounding electromagnetic 
field, through synchrotron and curvature radiation
processes, while undergoing
Comptonization, thereby, adding to the initial Penrose
 relativistic expansion,
and thus, giving rise to shock waves and beaming effects.  

The Penrose mechanism, or black hole model for AGNs, presented here, 
invokes the ultrarelativistic $e^-e^+$ pair production as a direct 
result of the high temperatures 
in the accretion disk.  Shapiro, Lightman, and Eardley (1976) and
other authors (Eilek 1980; Eilek and Kafatos 1983) 
calculate a model for a two-temperature disk in which the protons in
the inner 
region
of the disk attain a temperature of 
$\sim 2\times 10^{12}$~K due to the 
absence of any efficient cooling mechanism (see \S~4.1 for details).
Such high temperatures ($k_{\rm B}T\sim 200$~MeV) can result
in copious pion production (Eilek 1980; Mahadevan, Narayan, \&
Krolik 1997), in which neutral pion 
($\pi^0$) decays, produce  
abundant $\gamma$-rays with energies up to 
$\sim 100$~MeV, narrowly peaked
around $\approx 75$~MeV, assuming that the initial protons have a
thermal distribution.  Such decay 
products are candidates for the Penrose
mechanism.
Williams (1995, hereafter Paper~1) has shown that 
Penrose pair production
(PPP), in the ergosphere of a supermassive ($10^8 M_\odot$) 
rotating Kerr
black hole (KBH), at the photon orbit, can be 
significant, wherein, the  $\gamma$-rays, resulting from
$\pi^0$ decays,  are used as seed photons  
to populate the photon orbit.  Moreover, such accretion
disks with thermally unstable inner regions (Piran 1978),
which could vary between unstable configurations 
(Narayan 1997), would be conducive to a blast of 
ultrarelativistic particles.

Eilek and Kafatos (1983) used the structure of the 
two-temperature disk in
a canonical Kerr metric (Thorne 1974) to calculate 
the high-energy spectrum of such disks. 
They predict $\gamma$-ray luminosity, 
due to the $\pi^0$ decays, as high as 10\% of 
the bolometric luminosity
in sub-Eddington models.  In these present calculations, 
I assume such $\gamma$-ray  
luminosity [calculated to be 
$\sim 10^{43}~{\rm erg\,s^{-1}}$ at $\sim 75-100$~MeV,
yielding a $\gamma$-ray rate of $\sim 7\times 10^{46}~{\rm s^{-1}}$
(Eilek \& Kafatos 1983)]
to populated the photon orbit in the highest energy 
PPP (\gggg) processes.
These infalling $\gamma$-rays are assumed to  populate the photon orbit 
by a method similar to that  
assumed by Eilek and Kafatos (1983), 
and earlier by Leiter and Kafatos 
(1978), to supply the seed photons for PPP (\gp) 
in the Coulomb field of orbiting protons
near the event horizon, specifically 
between the radii of marginally bound 
$r_{\rm mb}$ and
marginally stable $r_{\rm ms}$ orbits. 
Such seed photons, proposed by these authors, after being 
blueshifted in energy by a factor of 
$\sim 30$ (see below), due to rotation of the 
KBH and the curvature of spacetime,
will achieve energies as high as $\sim $~GeV. However, it is 
shown in Paper~1 that the PPP (\gp)
in the Coulomb field of a proton
as initially proposed by Leiter and Kafatos (1978), 
proved to be unfruitful
(see discussion in \S~2). 
Nevertheless, it is also shown in Paper~1, that, 
when these 
$\gamma$-rays (produced by the $\pi^0$ decays) are assumed to populate
the photon orbit, upon  being blueshifted in energy 
by a factor of $\sim 52$ (see below), the 
PPP (\gggg) can occur in which large 
fluxes of ultrarelativistic $e^- e^+$ pairs of 
energy $\sim $~GeV
are allowed to
 escape from the nucleus
into the 
polar regions.  An aim of this paper is to show that such 
ultrarelativistic PPP electrons (with $\gamma\sim 10^4$) can
undergo inverse Comptonization  with surrounding lower
energy photons, say from the accretion disk, 
boosting the energy of these photons up to $\sim $~GeV,
or apparently higher due to beaming effects,
and the resulting $\gamma$-ray luminosities are consistent
with the observed spectra of quasars 3C 279 and 3C 273.

Note, the blueshift factor $e^{-\nu}$, mentioned above,
can be found in the transformation
Eq. (2.8d) of Paper~1, and 
is given by the Kerr metric coefficient Eq. (2.10a), also of
Paper~1.  It is a well
known parameter in the theory of Kerr metric black holes.
Its inverse is sometimes referred to as the ''redshift
factor'' (Thorne et al. 1986; 
see also Eq. (2.7d) of Paper~1).  This blueshift factor 
gives the blueshift in the energy of a particle,
at a specific radius, as measured by a local frame
 observer. 
Such blueshifted energies of the particles (due to the
fact that any particle falling into the ergosphere must
rotate in the direction of the KBH) 
can be extracted by these Penrose scattering processes (as
will be described in this paper).
The calculated values of the blueshift factors at the radii 
$r_{\rm ms}$, 
$r_{\rm mb}$, and $r_{\rm ph}$, the photon orbit,
are approximately 10.7, 32, 
and 52, respectively, for $a=0.998 M$ 
(where $a$ is the angular momentum
per unit mass parameter of the KBH, with
$G=c=1$), 

In general, the observed 
spectra of quasars show intense
nonthermal continuum radiation with highly excited emission lines.
The broad-band photon spectrum shows  a characteristic mixture
of thermal and nonthermal
emission, from radio to $\gamma$-ray frequencies.  
The
nonthermal continuum spectrum appears to be associated with the nucleus
and the astrophysical jets, and  
the thermal emission appears to be associated with the 
accretion disk.
Features in 
the radio, X-ray, and $\gamma$-ray  spectra, and the astrophysical  
jets associated with radio loud quasars 
seem to distinguish  two populations: radio loud and 
radio quiet quasars. For example, 
in radio loud quasars, the infrared 
baseline power law that always drops in the millimeter band 
(the mm-break) drops only $\sim$ 2 decades. The great majority of 
quasars, i.e., the radio quiet quasars, have a much stronger mm-break 
$\sim 5$ to 6 decades (Elvis 1988).  The size of the mm-break does not 
seem to affect 
the infrared (IR) to UV continuum shapes.  However, there is
clearly a correlation between radio loudness and the X-ray 
emission slope: a weak mm-break implies a strong X-ray turn up.
Moreover,
observations (Fichtel 1992, 1993) suggest that the 
high energy $\gamma$-ray spectra (${>\atop\sim}\, 30$~MeV) detected 
in radio loud quasars further distinguish the two populations, since
only radio loud quasars were detected in this energy range. 
In this paper, I show that the 
density and thermal instabilities occurring in the inner
region of the accretion disk [which increase the
temperature, giving rise to the so-called
two-temperature, bistable
disk model, in which the disk can in principle exist in two phases: 
a thin disk and ion torus (Lightman \& Eardley 1974; Shapiro, 
Lightman, \& Eardley 1976)] could be closely related to there being
radio loud and radio
quiet quasars.  
Disk models of this sort are also
referred to as thin disk/ion corona models, or 
more recently, the ``hot'' phase of 
these type of disks have been called advection dominated 
accretion flows (ADAFs): further details are given in 
\S~4.  In ADAFs, the energy released by viscous 
dissipation is stored in the gas as entropy and advected
into the black hole with a small fraction being radiated away
(Narayan \&Yi 1994, 1995; Narayan, Kato, Honma 1997; 
Narayan et al. 1998).   Note that, I will 
sometimes interchangeably
refer to the ion corona as the  ion torus or ADAF.
   
Energies as high as $\sim 4-10$~GeV have been 
observed in the spectral observations of quasars 3C 279, 3C 273,
PKS 0208$-$512, and  PKS 0528$+$134
(Hartman et~al. 1992; von Montigny et~al. 1993; 
Hermsen et~al. 1993; Hunter et~al. 
1993). 
Even higher energies $\sim$~TeV have been observed in the 
BL Lac $\gamma$-ray source Markarian 421 (Punch et~al. 1992;
Zdziarski \& Krolik 1993). 
Figure 1 shows  schematic energy spectra of 3C 279 and 3C 273, 
based on a number of observations.  Details of Figure 1 will be discussed 
in later sections of this paper.

Overall,
in this paper, I present results of theoretical 
model calculations using the Monte Carlo numerical 
method to treat Penrose scattering processes in the ergosphere of a 
KBH: the cases considered here are
inverse Compton scattering 
and $\gamma$-ray---$\gamma$-ray pair production (\gggg). These processes are 
assumed to  occur in material falling into the ergospheric region of 
a rotating black hole from a surrounding accretion disk that extends
inside the ergosphere to approximately 
the marginally stable radius (Eilek 1980).  
Resulting from these scattering processes, high energy particles are
emitted with calculated luminosities that are  in agreement with 
the general observed spectra of AGNs; and in this model,
relativistic beaming can be  used as an enhancement mechanism, giving 
rise to the objects we call blazars.
The theoretical background for these computations are presented in
Paper~1 (I refer the reader to Paper~1 for a thorough 
description of the model).  Here, I concentrate
on the astrophysical results, using these Penrose processes to 
explain the observed energy spectra of 3C 279 and 3C 273: 
these quasars were chosen because of the availability of the data on 
them, while assuming that they can serve to represent in general all 
AGNs.  In \S~2,  a summary is given of the general formalism of the
Penrose mechanism considered in Paper~1.  In \S~3, relevant 
observations of 3C~279, 3C~273, and other AGNs are presented.
In \S~4, I summarize the overall results of the 
Penrose Compton scattering (PCS) and 
PPP processes considered in Paper~1.
This summary includes an astrophysical discussion
of the PCS and PPP (\gggg) processes,
as correlated  with popular theoretical accretion 
disk models for AGNs, in 
particular the classical
thin disk and the thin disk/ion corona. 
In \S~5, I compare the 
model results with 
observations of 3C 279 and 3C 273.  
Specifically, I compare
the observed luminosity spectra with the calculated luminosity
spectra due to the Penrose processes.
Finally, in \S~6, I present conclusions
and future investigations.

\section{THE PENROSE PROCESS} 
\label{sec:penrose}

A mechanism suggested by Roger Penrose (1969) permits rotational 
energy to be extracted from a KBH.  The $\it classical$ Penrose 
process  utilizes the existence of retrograde particle orbits 
(with respect to the rotation of the KBH) in the ergosphere. 
In general, it is
possible for a particle, say $p_1$, that has fallen inwardly from
infinity into the ergosphere to scatter off another particle, say
$p_2$, initially in a bound direct orbit. 
If the orbit
of the scattering particle $p_2$ changes into a retrograde orbit
(of negative  energy), then the scattered particle $p_1$ can escape
to infinity with more
mass-energy than the sum of the initial energies of $p_1$ and $p_2$.
Since the  orbit of the
initially bound particle $p_2$ is dependent on the angular momentum
of the KBH and the curvature of spacetime defined by the
mass $M$ of the black hole, then, consequently, when $p_2$ gives up
energy to the escaping particle $p_1$ and falls through the event
horizon, the KBH loses energy in the form of rotational energy.
A possible class of 
Penrose processes occur when particles 
already inside the ergosphere (say particles from an 
accretion disk) undergo local relativistic scatterings
(Piran, Shaham, \& Katz 1975; Paper~1). 
If one of the scattering particles is initially in a bound 
orbit, as in the general description above,
then it is possible for the 
other initially unbound scattered particle to escape to infinity 
with rotational energy from the KBH. This process 
allows scattered ergospheric particles to (1) escape to infinity 
with more mass-energy than they would have had 
if the scattering occurred outside the ergosphere; and (2) escape to 
infinity with mass-energy initially trapped by the KBH, mass-energy 
that more than likely had no other way of escaping, save these Penrose
processes.  This is because 
for a particle initially trapped by the KBH, an escaping
orbit is said to occur only when some physical process near the 
hole injects the  particle into such an orbit 
(Bardeen, Press, \& Teukolsky 1972).

Since the inner edge of an accretion disk around a 
KBH with $a{>\atop\sim}0.95 M$ 
can 
extend inward to the radius of  marginal stable orbits, 
$r_{\rm ms}\simeq 1.2M$,
which is inside the ergosphere,
then a large enough 
fraction of the disk luminosity 
can possibly reach the radii of 
importance for the occurrences of Penrose 
scattering processes. 
That is, the fraction of the disk luminosity intersecting at the
the scattering radii will be larger than for the case in which
the accretion disk does not extend inside the ergosphere, such as 
in the cases
of nonrotating ($a$=0) and slow rotating black holes.

An expression for the fraction of the disk 
luminosity 
that  falls onto the scattering radii of the black hole 
can be found.
We assume that the disk can be represented by 
cylindrical surfaces of various
radii
$r$, concentric about a spherical shaped region representing 
near the event horizon of
the black hole.  An element of area representing the disk
at a given radius $r=r_d$, centered on the black hole, as measured by
an observer at infinity,  can be expressed
as
\begin{equation}
da_d=r_d d\Phi d{\rm Z}, 
\label{eq:1}
\end{equation}
where $r_d$ is the inner radius of the disk.
Similarly, an element of area at the scattering 
radius $r_s$ on the surface
of the sphere representing the region near 
the black hole is given by
\begin{equation}
da_s=r_s^2\sin\Th d\Phi d\Th. 
\label{eq:2}
\end{equation}
We assume also  that the disk, which extends inside the ergosphere,
radiates into the volume element of space 
$\sim (r_d^2-r_s^2)d\Phi d{\rm Z}$,
for sufficiently small $d{\rm Z}$ and $r_s\sim r_d$,  
such that the flux of the radiation
from the disk remains nearly constant inward to the scattering radii.
The above assumptions allow an expression for the photon
luminosity intersecting the scattering radii:
\begin{equation}
L_s\approx{r_s^2\over r_d\,h}(\cos\theta_1 -\cos\theta_2)L_\gamma,
\label{eq:3}
\end{equation}
where $L_\gamma$ is the disk photon luminosity  at the cylindric 
surface of radius $r_d$ and height $h/2$ above and below 
the equatorial plane.
Equation~(3) gives the fraction of the disk luminosity impinging
on a solid angle with polar bandwidth $\DD\theta= 
\theta_2-\theta_1$ subtended
the black hole at the scattering 
radius $r_s$, where $\theta_1$ and $\theta_2$ are polar angles above
and below the equatorial plane.  From equation~(3), we see that 
$\theta_1$ and $\theta_2$ can be used as 
free parameters in these model
calculations, however, within limits, as explained below.

Note, equation~(3), only gives 
a fraction (or probability), and, therefore, the ``strong
bending''  of photon trajectories need not be considered here, 
since this fraction
is only used to show what is needed to fit the model
calculated luminosity, due to the Penrose mechanism, with 
observations (see \S~5).  
That is, the calculated Penrose luminosity assumes  
that every particle in $L_\gamma$ scatters; 
the fraction given by equation~(3) 
 in part accounts for this unrealistic assumption, since
in a realistic situation we know that only a fraction of the 
particles will scatter.  

Now, the luminosity of photons from the disk used to
participate in the Penrose scattering processes,
defined by $L_\gamma$ in equation~(3), is assumed
to be a power law distribution based on spectra observations
of 3C 279 and 3C 273 (as described in \S~5.1).   In the 
trajectories of the infalling
photons, that actually participate in the individual scattering 
events, the so-called bending of the light or the warping of 
spacetime by the rotating black hole has been taken into account, 
since they are assumed to fall radially along null geodesics in the  
Kerr metric, with the conserved azimuthal angular momentum 
as measure by an observer
at infinity 
$L=P_\Phi$ set equal to zero. 
This assumption has also been used by other authors (Piran \&
Shaham 1977a; Leiter \& Kafatos 1978; Kafatos \& Leiter 1979) for  
maximum energy gain in a Penrose scattering process.  
In a more realistic situation, $L$ will not just 
be zero, i.e.,  the incident angle of the photon relative to the motion
of the target particle will generally be 
anywhere from $0^\circ$ to $90^\circ$.  
(Note, L=0 means that the incident
angle is $90^\circ$.) 

If we normalize equation~(3) to unity at the inner radius of the 
accretion disk, by setting
$r_s=r_d$ and  
the luminosity $L_s= L_\gamma$, we get the condition 
that $(r_d/h)(\cos\theta_1-
\cos\theta_2)\approx 1$. This puts a constraint 
on the values of $\theta_1$ and
$\theta_2$.  For example,   
for the two-temperature 
accretion disk  used in these calculations, of say height 
$h\simeq 0.779M$, 
normalized to unity at
$r_d\simeq 1.3M$
(see \S~4.1), it is required that
$\theta_1{>\atop\sim}73^\circ$, $\theta_2{<\atop\sim}108^\circ$, implying 
that $\DD\theta$ must always be ${<\atop\sim}35^\circ$.  
If these requirements are satisfied, then equation~(3)
gives that,
from
$\sim 85$\% at $r_s\equiv r_{\rm ms}\simeq 1.2M$ to $\sim 68$\% at 
$r_s\equiv r_{\rm ph}\simeq 1.074M\/$
of the photons generated in the accretion disk at $r_d=1.3M$
can fall onto the 
black hole at these scattering radii ($r_s$), between
$\DD\theta\approx 35^\circ$.
These radii are where the PCS 
and the
PPP (\gggg) take place.  
So, we see that, if the accretion disk 
extends inside the ergosphere ($\simeq 2M$ for $a=0.998M$), 
a large enough
fraction of the disk luminosity reaches the 
scattering radii to make
the Penrose mechanism a feasible process in the 
contribution to the observed
spectra of AGNs, contrary to what is estimated to 
occur in reference to 
the accretion disk being far away from the ergospheric 
region, where, in this case, only $\sim 1-10$\% 
 (for $r_d\gg r_s$) of the disk luminosity
would reach the scattering radii (Eilek \& Kafatos 1983).

However, as we shall see in this investigation, the ratio
found in equation~(3) need not be larger that $L_s/L_\gamma\simeq 4$\%
in order for  the Penrose mechanism described here 
to be an effective energy source
for AGNs.

Overall, this Penrose mechanism could very well be the important 
source of 
the relativistic particles 
emerging
from the nucleus of quasars and other AGNs. 
The physical background needed 
for these Penrose processes to occur is described below. 

The Kerr metric (Kerr 1963), a stationary, axially symmetric, 
asymptotically flat spacetime solution to the vacuum Einstein field 
equations, describes the geometry outside of a rotating massive body. 
This metric, conveniently written in the Boyer-Lindquist coordinates 
(Boyer \& Lindquist 1967),  defines the spacetime separation 
between events near a rotating black hole 
as measured by an observer at infinity---referred to as the 
Boyer-Lindquist coordinate frame (BLF).  The general orbits of 
particles (including photons) near the KBH are best described in the
BLF. The BLF admits three constants of motion as measured by an 
observer at infinity (Carter 1968). They are the total energy $E$, 
the angular momentum parallel to the symmetry axis $L$, and the 
``$Q$ value'' 
given by
\begin{equation}
Q=P_\Th^2 + \cos^2\Th \biggl[a^2 \biggl(\mu_o^2-E^2 \biggr)   
+{L^2 \over \sin^2 \Th} \biggr], 
\label{eq:4} 
\end{equation}
where $\mu_o$ is the rest mass energy of the particle, which is a 
trivial 
fourth constant of motion. 
The value of $Q$ is zero for particles 
whose motions are confined to the equatorial plane.  The nonzero 
values of $Q$ belong to particles which are moving in the 
$\Th$-direction (i.e., polar direction) and/or are not confined 
to the equatorial plane.

In general, the parameter $a$ of equation~(4)
can have values $0 \le (a/M) \le 1$, 
values which allow for the existence of an event horizon. When $a=0$,
the Kerr metric reduces to the form of the Schwarzschild metric for a 
nonrotating massive body.  For definiteness, a canonical KBH is used, 
with its limiting value ${a/M}=0.998$, as defined by Thorne (1974) 
in his investigation of an accretion disk around a KBH. 
It is found that, if $a/M$ is 
initially very close to $1$ ($0.999\leq {a/M}\leq 1$), a small 
amount of accretion (${\DD M/M}\leq 0.05$) through a disk quickly 
spins the hole down to a limiting state  ${a/M}\simeq 0.998$.  
Conversely, if $a/M$ is initially below this limiting value, 
accretion spins the hole up toward it.

Upon approaching a KBH from infinity, a limit or a region is reached 
where the angular momentum of the KBH 
causes inertial frames to be dragged around in the direction that the 
black hole rotates. This means that, it is impossible for an observer 
inside this so-called stationary limit to remain at rest relative to 
distant observers.  This limit is given analytically by 
\begin{equation}
r=r_o=M+\Bigl(M^2-a^2\cos^2\Th \Bigr)^{1/2}. 
\label{eq:5} 
\end{equation}
The region between this stationary limit and the event 
horizon [located at $r=r_+=M+\Bigl(M^2-a^2\Bigr)^{1/2}$] is
called the ergosphere. 
This means that, inside the ergosphere, the Kerr metric in the BLF 
does not allow an observer to be stationary (in the sense of the 
observer being at 
rest with $r,~\Th,~\Phi={\rm constant}$) because of the dragging of 
inertial frames; this makes physical processes difficult to 
describe in the 
BLF. In order to examine physical processes inside the ergosphere, 
Bardeen, Press, \& Teukolsky (1972)
devised a frame of reference called the local nonrotating frame 
(LNRF). Observers in this frame rotate with the KBH in such away 
that the frame dragging effect of the rotating black hole is canceled 
as much as possible.  In the LNRF, special relativity applies since 
locally spacetime has Lorentz or flat spacetime geometry. 
That is, the LNRF is used as a convenient inertial frame to describe 
physical processes inside the ergosphere---then one must 
transform back to the BLF to find out what is measured by an 
observer at infinity (see Paper~1 for details).

Now, of the three Penrose scattering processes presented in Paper~1: 
PCS, PPP (\gggg), and PPP (\gp), 
the most rewarding ones were the 
PCS and the PPP (\gggg).  These rewarding processes are summarized in 
details in \S~4 of this present paper. 
Unfortunately, the PPP (\gp), as mentioned earlier,
 as suggested by some authors (Leiter \&
Kafatos 1978; Kafatos \&
Leiter 1979; Eilek \& Kafatos 1983) did not allow any of the 
scattered $e^-e^+$ pairs to escape, mainly because of the large
acquired inward radial momenta of the infalling blueshifted 
incident photons at the scattering radius, 
too large for the produced pairs to 
be scattered outward by the protons. This large inward radial 
momentum arises because of the assumption that the angular 
momentum $L$ is zero for the infalling incident photons. 
[Note, $L=0$ was assumed only for maximum energy
gain in a Penrose scattering process.  This
assumption is not made for the infalling $\gamma$-rays, 
mentioned in \S~1,
that are assumed to populate the photon 
orbit in the PPP  (\gggg).
These $\gamma$-rays will indeed acquire large $L$ as well large $E$.]  
I refer the reader 
to Paper~1 for details concerning the PPP (\gp).  

\section{OBSERVATIONS} 
\label{sec:observations}
\subsection{General} 
\label{sec:general}

Since the discovery of quasars in 1963, numerous observations 
have been 
made of them and other AGNs.  
Below I 
present some of the general observed features that 
are relevant to the present model calculations.
Quasars are observed to be the most distant objects,
 with redshifts 
 up to $z\sim 5.5$ (Schmidth 1992; Stern et al. 2000).  It is 
generally accepted by the 
astrophysical community
that these objects and other AGNs are powered by 
massive black holes at their cores.  
Even before 1994, the year in which 
NASA's Hubble Space Telescope
began confirming the existence of black holes in the 
center of numerous galaxies,   
a large amount of circumstantial evidence for the presence of massive
black holes ($M{>\atop\sim}10^6 M_\odot $) in the nuclei of bright 
nearby galaxies ($d\leq 20$ Mpc) had accumulated (Filippenko 1988).
Further,
the overall nuclear activities in galaxies 
seem to decrease as observational 
distances approach nearby galaxies.    
This should be of no surprise if one believes that many
nearby galaxies are
evolved quasars that have outlived their quasar phases;
i.e., their nuclear activities have decreased as the available 
matter they 
``feed'' off has depleted.

Some observed spectral properties of AGNs are the following:
At energies greater than $2$ keV, the X-ray spectra of AGNs show a
remarkable uniformity with almost all objects, well fixed by simple
power laws in the $2-100$ keV band with narrow dispersion in spectral
index $\alpha\sim 0.7\pm 0.15$, as related to the flux density $f_\nu$ of
the photons
having a power law $k\nu^{-\alpha}$ distribution.
This universal 
X-ray power law also exists in lower mass galactic black hole 
candidates, which suggests that this feature may be due to processes 
inherent to the central power source.  Moreover, it is generally
believed that
X-ray emission in radio selected QSOs (quasi-stellar objects) is
likely to have two components: one thermal related to the optical
luminosity ($\alpha\sim 0.5$),
and one nonthermal related to the radio luminosity ($\alpha\sim 0.8$;
Isobe et~al. 1988).  The thermal component could very well be
related to the inner region of the accretion disk, and the nonthermal
to Penrose processes, as described in \S~5.2 of this paper.

AGNs emit the major fraction of their power in the $\gamma$-ray
region of the electromagnetic spectrum (Bassani  et~al. 1985).
In this region, 
the 
observed spectral slopes of quasars 
range from  $1.7\,{<\atop\sim}\,\alpha\,{<\atop\sim}\, 2.6$ (Hartman 
et~al. 1992; Hunter et~al. 1993; Bertsch et~al. 1993; 
von Montigny et~al. 1993; Hermsen et~al. 1993), as in the photon
emission
power law  $N_E=kE^{-\alpha}$.  
The steep spectra up to  
$\alpha\simeq 2.6$ may be an indication 
either that photons
with a broad range of spectral index are emitted from quasars or that 
the photon spectrum is a function of the angle from the jet axis
(Hunter et~al. 1993), or perhaps a combination of both.  
Since a large fraction of emitted power is contained at $\gamma$-ray
energies above $\sim 1$~MeV, this suggests that the production of 
the $\gamma$-ray photons is intimately related to the central
power source.  Also, recall that, energies as high as 
$\sim$ TeV have been observed (Punch et~al. 1992;
Zdziarski \& Krolik 1993).

Observations show AGNs to be variable on time scales ranging 
from less than an hour (Hartman et~al. 1992) 
up to years. In general, this variability appears to
be related to accretion disk instabilities.  Blazars 
are the most variable AGNs; and they are among the most luminous 
known objects.  From variability 
studies we know that much of their luminosity is generated inside 
volumes at most $\sim 0.008$~pc across (Jones 1988). 
Because they are linear
polarized at several percent, the prevailing view is that the 
emission is synchrotron. These objects make up only
$\sim 2\%$ of all known AGNs.  As mentioned earlier,
relativistic beaming is believed to 
be important in these object, wherein 
we observe an apparent increase in source luminosity and energy,
and an apparent decrease in size.

When comparing a number of different AGNs observations
(Hermsen et~al. 1992; Fichtel 1992; Hunter et~al. 1993), there 
appears to be five universal regions in the 
luminosity 
spectrum where energy is definitely emitted: 
at
$\nu\sim 10^{9-12}$~Hz (radio/mm), $\nu\sim 10^{13}-2\times 
10^{15}$~Hz 
(infrared/optical/UV), $\nu\sim 10^{17-20}$~Hz (X-ray),
$\nu\sim 10^{21-22}$~Hz (soft $\gamma$-ray), and $\nu\sim 10^{22-24}$
($\gamma$-ray); compare Figure~1. The exact 
location of emission, in these regions 
on the observed luminosity spectrum, probably depends on the 
characteristics of the accretion disk and to
what extent the source is superluminous.  The observed
$\sim $~TeV emission in Mrk 412 and a few other blazars
(Stecker, de Jager, \& Salamon 1996) appear, from these present
model calculations, to be extreme cases of conditions
that can occur to effect the observed spectrum (namely, Penrose
processes, relativistic beaming, and an ADAF)---as will be discussed 
in \S~6.

\subsection{3C 279}
\label{sec:3c279} 

Quasar 3C 279 (Fig.~1$a$) is an OVV quasar, which means, in general, 
that observations 
show intense optical variability, 
and a highly polarized optical continuum; 3C 279 is also radio loud. 
It has a redshift $z=0.538$, implying a 
distance $d\simeq {cz/H_0\simeq }2.2\times 10^3$~Mpc
($H_0=75~{\rm km\,s^{-1}\,Mpc^{-1}}$).  
This quasar was the first 
object in which superluminal motion was detected 
(Whitney et~al. 1971).  
A change of $20\% $ in X-ray intensity ($2-20$~keV) on a time 
scale of less than an hour has been observed.  In general, 
the X-ray luminosity
is observed to vary around
$L_{\rm x}\sim 10^{46}~ {\rm erg\,s^{-1}}$, with flux
$f_\nu$ spectral index $\alpha\sim 0.5$ (i.e.,  photon $N_E$ or $N_\nu$ 
spectral index $\alpha\sim 1.5$).
In $\gamma$-ray intensity ($>100$~MeV), 3C~279 has 
been observed to be time
 variable on 
the order of a day (Fichtel 1993).
The energy output per logarithmic energy interval observed by the
Energetic Gamma Ray Experiment Telescope (EGRET) aboard the  
Compton Gamma Ray Observatory (GRO) in the
high-energy $\gamma$-ray band above $100$~MeV
is larger than in any other frequency band (Hartman et~al. 1992).
The spectrum of the detected emission is well represented by a
power law ($kE^{-\alpha}$) of photon spectral index $\alpha=2.02\pm 0.07$.
If the emission from 3C~279 is isotropic, its $\gamma$-ray luminosity
between $100$~MeV and $10$~GeV is about $L_\gamma\simeq 1.6\times10^
{51}$~photons s$^{-1}$, or $\simeq 1.1\times 10^{48}~{\rm erg\,
s^{-1}}$,  for $H_0=75 ~{\rm km\, s^{-1}Mpc^{-1}}$.  
However, it may very well be that  this emission is beamed; 
then for 1 steradian, the true luminosity would
be $L_\gamma\sim 9\times 10^{46}~{\rm erg\,s^{-1}}$.  For 
comparison, our own Galaxy
emits $\sim 5\times 10^{38}~ {\rm erg\,s^{-1}}$.  
The overall characteristics of the observed spectrum in the 
$\gamma$-ray regime suggest the presence of an 
extraordinary source of relativistic particles with an energy spectrum 
at least as hard as that observed for charged cosmic rays near Earth.
There is no way of knowing from the  GRO data whether this 
spectrum extends 
across the many additional decades to the extreme energy required by
the extragalactic part of the cosmic ray spectrum.
Many of the 20 AGNs detected by EGRET appear to have flat spectra 
and might be expected to be detected at TeV energies (Weekes 1993).

\subsection{3C 273} 
\label{sec:3c273}

Quasar 3C 273 (Fig.~1$b$) is radio loud; unlike 3C~279, 
it is not an OVV quasar.
It has a redshift $z=0.158$, implying a 
distance $d\simeq 6.3 \times 10^2$~Mpc ($H_0=75~{\rm km\,s^{-1}\,
Mpc^{-1}}$).  
At X-ray ($2-10$ keV), with $H_0=50~{\rm km\,s^{-1}\,Mpc^{-1}}$, 
3C 273 has luminosity of 
$\sim 10^{46}~{\rm erg\,s^{-1}}$ (Bassani et~al. 1985). 
In the $2-60$~keV range, the 
data is well represented by a power law spectrum having a photon 
index of $\alpha=1.4\pm 0.02$.  At X-ray ($20-200$~keV), 
$L_{\rm x}\sim 5\times 10^{46}~{\rm erg\,s^{-1}}$, a very hard spectrum
($\alpha=1.2$), with no apparent break or cut-off up to around
200~keV.  The soft X-ray is well known for large factor flux variations
on timescales $\sim$ hours to days.  At hard X-ray energies, 
observations show the 
luminosity increasing by a factor of 3 on a timescale of 2 years.
At $\gamma$-ray ($50-800$~MeV), $L_\gamma\sim 4\times 
10^{46}~{\rm erg\,s^{-1}}$.  
The GRO observations done by EGRET (1991) show a 
small variation in the flux density from that of the COS-B 
observations (1976-1980) in the energy range $\sim 50-800$~MeV
(compare Fig. 1$b$).
The COMPTEL (1991) aboard the GRO 
measured a photon spectral index $\alpha\simeq 2.3$ for 
energies between $\sim 10-30$~MeV (Hermsen et~al. 1993).
EGRET (1991) observations of 3C 273 (von Montigny et~al. 1993)
measured $\alpha \simeq 2.4$ for energies between $\sim 100~{\rm MeV}
- 3$~GeV.  
%

\section{SCENARIO TO EXPLAIN THE ENERGY SOURCE}
\label{sec:energy}

\subsection{ADAFs and Two-Temperature Bistable Accretion
Disk Models}
\label{sec:disk}

It is assumed that the accretion disk surrounding the KBH 
is the two-temperature, bistable disk model 
[also referred to as the  thin disk/ion corona, 
and recently, the corona phase called an advection dominated 
accretion flow
(ADAF)]. Such an accretion 
disk can possibly exist in either phase, 
thin disk or ion torus,
or both (thin disk/ion corona)---while oscillating between the 
two (Chen \& Taam 1996; Narayan 1997; see also Esin, McClintock, \& 
Narayan 1997 and
references therein).
The two-temperature disk model was first
proposed by Shapiro, Lightman and Eardley (1976) to account 
for the {\it high} and {\it low} states of the Galactic 
X-ray source Cygnus X-1.
In the standard thin disk model (Shakura \& Sunyaev 1973; 
Novikov \& Thorne 1973), it is assumed that the gas at each 
radius is in a nearly 
Keplerian orbit.  Slow radial infall occurs as viscosity
[including magnetic (Eardley \& Lightman 1975; 
Balbus \& Hawley 1991)] 
transfers angular momentum outward; that is, the gas must give up 
some of its angular momentum before it can cross the event horizon.  
The binding energy released by accretion may then be retained 
by the gas of the disk in the form of heat or radiated away.   
An accretion flow which is able to radiate away most of its 
binding energy forms a thin accretion disk; the alternative 
form is a geometrical thick disk or an ion corona---sometimes 
referred to as a hot torus (Rees  et~al. 1982). 
The possible formation of a thick disk was first realized through 
Lightman's (1974a, 1974b) investigation of 
the instability of a thin, time-dependent
 accretion disk around a black hole. His calculations show that as 
time progresses, density and thermal instabilities develop in the 
thin disk,
causing the material to clump into high-density zones and low-density 
zones---that become optically thin, bringing about extremely high 
temperatures, which cause the vertical height $h$ of the inner region 
to rise quickly.  However, without a satisfactory relativistic
time-dependent 
model to treat $h>r$, we can only speculate about properties 
of the disk as $h $ becomes relatively large.
Among 
these speculations are the following:  (1)~The rapidly rising internal 
temperatures could cause the temperature
gradient to become superadiabatic, and thus, drive convective 
(Lightman 1974a; Narayan \& Yi 1994) or advective (Abramowicz et al. 1988;
Abramowicz et al. 1996) energy 
transport at a rate necessary to counterbalance the energy 
released, and an ion corona/hot torus (or ADAF) 
forms in the inner radii. 
(2)~The rings probably destroy themselves by explosion or supersonic 
turbulence 
(Thorne \& Price 1975), causing the density in the optically thin 
regions to once again rise, making cooling processes more efficient;
and after this, the disk will most likely
settle back down to its thin disk structure, subsequently repeating 
the instability cycle.  Note, the above mentioned density and thermal
instabilities are commonly
referred to as the ``Lightman instabilities.''
Such stability and instability phases of the accretion disk, proposed 
 in the items above, can generally be explained in the context of the 
accretion rate $\dot{M}$ versus surface density $\Sigma$ plane
(Abramowicz et al. 1988; Abramowicz et al. 1995).

Thin disks are primarily thermal emitters, with spectra peaking in 
the UV;  however, the inner region of a thin disk is capable of 
producing both 
an UV thermal component and a power law continuum, if surrounded by a 
corona (hence, the thin disk/ion corona model), since the corona 
can be heated to temperatures $>3\times 10^9$~K by the Lightman 
instabilities. 
A detailed model of the thin disk (without the Lightman
instabilities) surrounding a KBH 
gives a 
surface temperature in the range of 
$\sim 4\times 10^6$~K ($0.4$~keV) to $4\times 10^7$~K ($3.5$~keV) 
for the inner region
(Novikov \& Thorne 1973). 
But once the instabilities set in, the ionized corona can have 
electron and proton temperatures as high as 
$\sim 10^9$~K and $\sim 10^{12}$~K, respectively (Shapiro, Lightman,
\& Eardley 1976).
Because of the high 
electron and ion (or proton) temperatures attainable in the corona, 
the basic 
output radiation is mostly nonthermal $\gamma$-rays.  The ion corona is 
a poor radiator; and if the binding energy released is small, 
funnels can form 
about the polar axes, which could enable collimation and the 
formation of jets (Rees et~al. 1982).  In addition, it has been
found independently by Williams (1991, Paper~1), and de Felice
and Calvani (1972); de Felice 
and Curir (1992), 
that the KBH naturally gives rise to vortical escaping 
orbits, which are collimated about the polar axis; for example, 
the Penrose 
scattered particles discussed here, as we shall see in \S~4, 
exhibit such
collimation. From these findings,
and a subsequent detailed analysis (de Felice \& Carlotto 1997),
one can safely say that the KBH is most likely responsible for 
the initial
collimation of the jets.

Before proceeding, I want to pause briefly to explain the origin
of this collimation.  
The physical mechanism responsible for the collimation 
is the inertial frame dragging, sometimes
referred to as the 
Lense-Thirring effect: due to the 
gravitational force associated with a rotating mass---it is 
a purely general relativistic effect.  This component of the 
gravitational force is sometimes called the ''gravitomagnetic''
force (Thorne, Price, \& Macdonald  1986); it is the
the gravitational 
analog of a magnetic field, which distorts spacetime and 
acts on the space momentum vector of a test particle in its field.
The Lense-Thirring effect or the frame dragging is an intrinsic 
part of the Penrose mechanism.  Inside the ergosphere, where  
Penrose processes operate, as matter is made 
to rotate in the direction of the KBH, it acquires  
large angular momentum and energy 
from the black hole.  The particles inside the  
ergosphere that have  
nonequatorially confined orbits will have in addition nonzero  
polar coordinate momenta (Wilkins 1972; Paper~1; Williams 
2002a, 2001).  If these highly blueshifted  particles 
were allowed to escape, they would escape along vortical trajectories,
as point out by de Felice \& Carlotto (1997), i.e., 
due to the inertial spacetime frame dragging of their
orbits.  These authors define
such trajectories as gravitational spacetime
``geometry-induced collimation.'' Generally, in these Penrose 
processes, the incident particles, after scattering in the
ergosphere, take with them 
the energy and momentum of the orbiting targets. This,
along with the gravitomagnetic force, enables  the 
escaping Penrose scattered particles to  leave the KBH  in the form of 
vortically orbiting relativistic particles, collimated about 
the polar axis (Williams 2002a, 2001).  
In summary, the collimation is the 
characteristic structure of geodesic orbits escaping from the 
ergosphere or region close to the event horizon of a KBH: this is due
to the Lense-Thirring effect, and,
hence, is seen here in the four-momenta of these Penrose scattered
escaping particles.

Now continuing, on Table~1, model parameters are given for the 
inner region of a thin disk/ion corona  
two-temperature, bistable, fully relativistic accretion flow 
surrounding a KBH
($M=10^8M_\odot,~a/M=0.998$). 
These model parameters
have been calculated from several sources (Novikov \& Thorne 1973;
Eardley \& Lightman 1975; Eilek 1980; Eilek \& Kafatos 1983).  Note,
Eardley \&  Lightman (1975) provide the conditions for instability
to set in, i.e., in  going from a thin disk to ADAF (see Appendix C of 
that paper). 
The parameters are defined as follows: $\alpha$ is the viscosity parameter
($0<\alpha\leq 1$); $y\equiv(4kT_e/m_ec^2){\rm max}(\tau_{_{\rm es}},
\tau_{_{\rm es}}^2)$ 
is the dimensionless parameter that characterizes
Comptonization, commonly called the Kompaneets parameter; 
$\dot M$ is 
the sub-Eddington accretion rate;
$T$ and $T_s$ are the thermal and surface temperatures, 
respectively; $T_e$ and $T_i$ are the electron and ion temperatures, 
respectively; 
$r_{\rm inner}$ and $r_{\rm instability}$ give approximate 
values for the extent of
the inner region and the hot optically thin 
 region of instability, respectively; $L_{\rm bol}$ is
the bolometric luminosity; $L_\gamma$ is the luminosity
of the $\pi^0$-induced $\gamma$-rays in the ADAF; 
and $h$ is the height
of the disk.  
The radius at which these values are calculated is $r=1.3M
\approx r_{\rm ms}$ [model  implied
inner edge of the dynamically stable thin disk: with a soft
X-ray inner region, i.e, where the calculated relativistic correction
function   ${\cal Q}(r) \lra 0$ (Novikov \& Thorne 1973; 
Page \& Thorne 1974)\footnote{The behavior of ${\cal Q}(r)$ in the inner 
region between the limits $1.2\la r/M\la 1.28$ needs further 
investigation: for it is not clear why ${\cal Q}(r) \lra 0$ at the
end points and is $<0$ within the limits, irrespective of $M$.}].
Note, unsaturated Comptonization gives $y\sim 1$ and is appropriate
whenever there is a copious source of soft photons in the inner hot
region (Shapiro \& Lightman 1976), whereas 
saturated
Comptonization characterized by $y\gg 1$ is appropriate
in the inner region of the cool disk model (Shakura \& Sunyaev
1973).  The input parameters are $\alpha$, $y$, $\dot M$, $M$, $r$,
and $\bb$.
All other values are calculated from the references cited
above.
The extent of the inner and optically thin regions
depend on $a$, $\alpha$, $M$, and $\dot M$. 
Specifically, the 
regions $r_{\rm inner}$ and $r_{\rm instability}$  
are calculated
from Eardley and Lightman (1975).
Moreover, notice  the following: 
(1) Increasing $\alpha$, the  viscosity
parameter, as in model cases
3 and~6, decreases $T$, $T_e$, $T_i$, and  increases $T_s$,
$L_\gamma$, and 
the extents of the
regions $r_{\rm inner}$ and $r_{\rm instability}$.  (2) The
bolometric luminosity 
$L_{\rm bol}$ increases with increasing
$\bb$, $M$, or $\dot M$---unstable accretion flows could 
cause $\bb$ to vary.  
(3) In all cases, the height of the disk $h$ 
after instability sets 
in is less than $r_s$, where $r_s$ is the assumed radius at which 
the scattering 
takes place.

The numerical disk model of Table~1
gives values inward to the photon orbit $r_{\rm ph}$
($\simeq 1.074M$) for the region of instability.
However, the gravitational physics of the rotating KBH will set the
minimum radii for the location of the disk particles.   
Normally, for a steady-state accretion disk
surrounding a canonical KBH, the plasma moves in slowly in 
Keplerian fashion, until it reaches $r_{\rm ms}$.   The inner
region of the disk ends here and material particles quickly 
free-falls into
the black hole; hence, the density of matter builds to a maximum 
at ${r_{\rm ms}}$ and then falls quickly to zero inside this 
radius (Novikov \& Thorne 1973; Leiter \& Kafatos 1978); 
$r_{\rm ms}$ is sometimes
referred to as the last stable orbit. Now, if Lightman
instabilities develop
in the disk, large amounts of plasma may attain temporary
Keplerian orbits within the region between $r_{\rm ms}$ and 
$r_{\rm mb}$ for material particles, and inward to $r_{\rm ph}$ for
massless particles.
Such disk particles could very well populate the orbits
of the target particles participating in the Penrose scattering
processes (described in details in Paper~1, and summarized
in the following sections).  On Table~1, I have set the 
minimum inward radius for the region of instability 
equal to $r_{\rm mb}$;
however, trapped photons will be  able to 
exist beyond this radius,  to $r_{\rm ph}$, without falling 
quickly into the event horizon.  

\subsection{Penrose Scattering Processes}
\label{sec:penrose2}

Now summarizing the astrophysical results of Paper~1,  
Monte Carlo computer simulations of Compton scattering 
and $e^- e^+$ pair production (\gggg) in the ergosphere of a KBH are 
presented: Specifically, I calculate 
the energy-momentum four-vectors of 
the resulting Penrose scattered particles.
Particles compatible with a thin disk/ion corona
accretion surrounding the black hole are assumed to
infall along radial geodesics in  the equatorial plane inside 
the ergosphere and scatter off tangential target
particles that are in bound orbits.  These orbits consist of
equatorially confined ($Q=0$; see eq.~[4]) and nonequatorially 
confined ($Q>0$) orbits.
The equations that govern the orbital
trajectory of a particle
about a KBH are solved (Paper~1) to determine analytic expressions
for the conserved energy $E$ and azimuthal angular momentum $L$ 
of material and
massless particles that have orbits not confined to the equatorial
plane.  
The escape conditions
to determine whether or not a particle escapes from the potential
well of the KBH are applied to the scattered particles.  
As stated previously, the Penrose mechanism in general 
allows rotational energy of the 
KBH to be extracted by scattered particles escaping from the
ergosphere to large distances from the black hole (or to infinity).
The results of these model calculations, as we shall see 
in more details below, show that the Penrose mechanism is capable of 
producing 
the observed high energy particles emitted by
quasars and other AGNs. 
This mechanism, as
applied in the model of Paper~1, can extract
hard X-ray/$\gamma$-ray photons from
Penrose Compton scatterings of initially low energy
UV/soft X-ray photons
by target orbiting electrons inside the ergosphere (discussed further
in \S~4.2.1).
These model calculations
also allow  relativistic $e^- e^+$ pairs to escape, 
after being produced by
infalling low energy photons interacting with highly
blueshifted energetic 
target photons (in bound 
nonequatorially confined 
orbits) at the photon orbit: This PPP (\gggg)
process 
may be the
origin of the ultrarelativistic electrons 
inferred from
observations to emerge from the cores of AGNs (discussed further 
in \S~4.2.2).

\subsubsection{Penrose Compton Scattering}
\label{sec:pcs}

In the PCS processes investigated, photons are assumed to be 
emitted from material that has fallen inward from 
infinity and scatter off tangentially equatorially 
(or nonequatorially) confined 
orbiting electron rings of completely ionized plasma, revolving near 
the event horizon, between the marginally stable ($r_{\rm ms}
\simeq 1.2M$) and the 
marginally bound  ($r_{\rm mb}\simeq 1.089M$) circular orbits.
The bound unstable orbits
are chosen for the target particles 
because such orbits can be made, if 
perturbed slightly, to spiral inward or escape outward,
thus increasing the efficiency of the Penrose mechanism.  
For simplicity, the conditions that the infalling photon 
encounters in its free fall through the ergosphere before 
arriving at the designated scattering radius are ignored in 
these calculations.  Otherwise, conditions such as the particle 
density in the ergosphere and radiative transfer effects would 
have to be incorporated into the calculations along the null 
geodesic of the photon.  It is assumed that these conditions 
cause little qualitative change in the results.   This assumption
has been to some extent validated in the Penrose analysis by 
Piran and Shaham (1977b), and was used also by these authors
in their analysis (Piran \& Shaham 1977a, 1977b).

Now, some of 
the photons, after being scattered by the electrons,  escape to 
infinity. An observer 
at infinity, observes a low-energy (in most cases 
$< \mu_e=0.511$~MeV) photon being 
scattered by a direct orbiting electron, after which the photon comes 
out with a higher energy (inverse Compton scattering). Subsequently, 
the target electron may recoil to another direct orbit of lesser energy
(defined as a $\it quasi$-Penrose process), 
or the electron may be put on a retrograde orbit of negative energy
(a classical Penrose process); 
in both cases, the target electron gives up energy as measured by an 
observer at infinity. However, to a 
particular local~frame observer this is just a normal Compton scattering 
process in which the photon loses energy to the electron, since the photon 
arrives at the local~frame with initial energy higher than $\sim \mu_e$.  

Putting the PCS results (discussed in details in Paper~1)
into an astrophysical context, it is concluded that
this process may play an
important role in the upgrading of UV photons $\sim 5$~eV,
say from a thin disk/ion corona accretion, to X-ray photons
$\sim 15-218$~keV; and in upgrading
soft X-ray/hard X-ray
photons $\sim 0.511-150$~keV, also from a
thin disk/ion corona accretion,
to hard X-ray/$\gamma$-ray photons
$\sim 53$~keV$-$$12$~MeV: contributing to the
high energy observed spectra of AGNs.
In addition, because many of the escaping $\gamma$-rays have
negative radial momenta (out from the  
$\Th =90^\circ$ plane) 
and nonzero polar and azimuthal coordinate angular momenta,
 these $\gamma$-rays
could possible aid,
along with magnetic fields and relativistic electrons, in the
formation of the observed astrophysical jets commonly seen in AGNs
(see Williams 2002a).
Such vortical-like escaping orbits which are naturally collimated 
about the axis of symmetry (the rotation axis) due to the 
spin of the KBH have 
been described by de Felice 
and Curir 
(1992) and de Felice and Carlotto 
(1997).   Their description gives support to the Penrose 
theoretical
model calculations 
presented in this present paper.  
A detailed discussion of these vortical orbits produced 
in Penrose
processes is given by Williams (2002b; see also 2001).

Features of the PCS photons 
are shown in the energy-momentum spectra of Figures~2
and~3.
The radial and azimuthal momentum components are plotted 
in Figures~2$a$ 
and~2$b$
for a typical PCS of 2000 incoming photons.
Depicted in Figure~3 are polar momentum components
for two PCS cases.
Compare Figures~3$a$ and 3$b$;
notice the asymmetry in the polar directions: the one-sided
distribution favors the positive $\Th$-direction. This feature
could be extremely important in jet formation (Paper~1).
Such asymmetry
in the distribution of the scattered particles is a  general
relativistic effect, due to the gravitomagnetic (GM) field, 
which causes the extreme dragging
of inertial frames in the ergosphere of a rotating black hole,
and exerts a force on the momentum vector of a particle in its 
field
(Williams 2002a, 1999), as mentioned earlier.  
Williams (2002a) presents a detailed
investigation of the gravitomagnetic field and 
Penrose processes, showing that the GM field
alters incident and outgoing angles of 
the Penrose scattered particles, appearing to   
break the reflection symmetry of the Kerr metric above and below 
the equatorial plane.  This asymmetry, due to the GM field, has 
recently been confirmed by an independent analysis 
(Bini et al. 2002).  

The corresponding   
polar angles of the escaping particles, above and below the 
equatorial plane ($\Th=90^\circ$), 
as measured by an observer at infinity, 
for the cases
in Figure~3, are displayed in Figure~4.  These polar angles
 are derived 
from equation~(4) by letting $P_\Th\lra 0$ (see also
Piran \& Shaham 1977a); thus, in general,
\begin{equation}
\dd=\arccos{\left[ {-T+\sqrt{T^2-4SU}\over 2S} \right] }^{1/2},
\label{eq:6}
\end{equation}
where
\begin{eqnarray*}
S&\equiv &a^2(\mu_o^2-E^2); \\
T&\equiv &Q +a^2(E^2-\mu_o^2)+L^2;\\
U&\equiv &-L^2
\end{eqnarray*}
($\dd$ is measured relative to the equatorial plane).

Figures~2$a$, 3 and~4 show that the KBH 
scatters most of the photons  
into  the polar direction,  at an angle
$\Th_{\rm ph}^\prime$ ($= 90^\circ\mp\dd_{\rm ph}$, for 
escaping above
``$-$'' or below ``$+$'' the equatorial plane, where 
$\dd=\dd_{\rm ph}$ is given by eq.~[6]),  
and naturally produces 
one-sided jets of particles; see Williams (2002a) for details.
Moreover, the particles that are scattered with 
nonzero azimuthal  
and nonzero polar  coordinate 
momentum components (compare Figs.~2$b$ and~3$b$) 
give rise to vortical 
orbits concentric about the polar axis, with 
corresponding polar
angles of escape (Fig.~4$b$) indicating 
a coil-like collimation (Williams 2002b, 2001).  
Note that,  photons escaping with 
the highest energy, after being scattered by the nonequatorially
confined orbiting target electrons, 
have polar angles of escape approximately equal 
to that of the maximum latitudinal angle---relative to
the equatorial plane---of the target electrons 
($\dd\sim 30^\circ$; compare Fig.~4$b$, and see 
\S~4.4 for further explanation of the latitudinal angle).  
Similarly, the highest energy PCS photons by equatorially
confined target electrons have $\dd_{\rm ph}$ 
values near the equatorial plane
(compare Fig.~4$a$). 

The calculated luminosity spectra resulting from the above
described PCS processes
are presented in \S~5, where they are compared with observations.


\subsubsection{Gamma Ray-Gamma Ray Penrose Pair 
Production (PPP)}
\label{sec:ppp}

This PPP (\gggg) consists of  collisions inside the ergosphere, 
between radially infalling photons and bound circularly 
orbiting photons, 
at the radius of the  photon  orbit (Bardeen et al. 
1972): 
\begin{equation}
r_{\rm ph}=2M \lbrace 1+ \cos {[(2/3) \arccos{(-a/M)}]}
\rbrace,
\label{eq:7}
\end{equation}
where  $r_{\rm ph}$ represents an
unstable circular orbit.  This orbit is
the innermost boundary of circular orbits for particles
i.e., for massless particles 
(the marginally bound
orbit $r_{\rm mb}$ 
is the 
innermost
orbit for material particles). 
The initial energies used in these calculations, as measured by an 
observer at infinity, for the 
infalling photon and the orbiting photon, $E_{\gamma 1}$ and $E_{\gamma 2}$, 
respectively, are in the following ranges: 
$3.5 {\rm~keV}\leq E_{\gamma 1}\leq 1$~MeV 
and $3.4~{\rm MeV}\,{<\atop\sim}\, E_{\gamma 2}\,
{<\atop\sim}\, 3.89$~GeV [defined by 
eqs.~(A20) and~(A21) of Paper~1 for given $Q$ values].  These 
ranges are consistent with those produced by thin disk/ion corona models. 
The high energy range for $E_{\gamma 2}$ is chosen 
based on the expected blueshift in the energy of a photon at the photon 
orbit  (recall from \S~\ref{sec:intro} that the blueshift factor 
at the photon 
orbit for a KBH with $a=0.998M$ 
is $\simeq 52$).  The $\gamma$-rays produced in Eilek's hot 
accretion disk model (Eilek 1980; Eilek \& Kafatos 1983), 
with energies up to $\sim 100 $~MeV,
can very well 
be seed photons (now blueshifted and bound at the photon 
orbit) for these 
PPP (\gggg)
processes (Eilek 1991). In addition, some of the PCS 
photons, described in the last section, can also be 
seed particles for these 
processes (Paper~1).  The target photons having orbital
energies, $E_{\gamma 2}$, within the range given above, have
corresponding $Q$ values within the range of $0.099\,M m_e
\leq \sqrt{Q_{\gamma 2}}\leq 113 \,Mm_e$, 
and their maximum 
latitudinal orbiting
angles (relative to the equatorial plane)
are ${\approx}\,0.5^\circ$,
as given by 
equation~(6),
irrespective of their individual
values of $E_{\gamma 2}$, $L_{\gamma 2}$, and $Q_{\gamma 2}$.

The results of these PPP (\gggg) processes (Paper~1), 
and the calculated 
luminosity presented in \S~5,
show that this can be an
important way, if not the dominant way, to extract the extremely
relativistic
$e^- e^+$ pairs that are inferred from observations to be
emerging from the cores of AGNs.

Note that, these PPP (\gggg) processes are classified 
as quasi-Penrose
processes (Paper~1). That is, the target 
particle is not specifically put
on a negative energy orbit as in the 
classical Penrose extraction of rotational energy, 
 yet rotational energy is still extracted, however, indirectly,
 from the highly energetic photons, that have been blueshifted 
 by the KBH, at (or near) the photon 
 orbit;
see Paper~1 for further details.

In general, a possible picture of what takes place
in these PPP (\gggg)
processes to an observer at
infinity (i.e., the BLF) is the following:
low energy unbound infalling photons interact with high energy
bound photons at the photon orbit  
producing pairs of relativistic electrons  
with energies as high as $\sim 4$~GeV or 
higher---depending on the type of accretion flow; 
about $50 \%$ of these produced $e^-e^+$ pairs 
escape from the potential well of the KBH. 
Typical radial and azimuthal energy-momentum
spectra of the $e^-e^+$ pairs are shown in  Figure~5.  
These copiously produced
relativistic escaping electrons can participate in astrophysical
processes intrinsic to observations of AGNs, such as the production
of the synchrotron radiation and the formation of the jets. 
Notice in Figure 5$a$ that only the positive radially directed $e^-e^+$
pairs escape; this is to be expected since all 
negatively directed material particles at radii less than 
$r_{\rm mb}$ 
must fall directly into the KBH.  Also, it is
important to note here that if these PPP (\gggg) processes had not
occurred, photons in
the photon orbit would,  most likely, not have had any
other way of escaping, and the energy released in
these processes would have been forever trapped by the KBH.
The reason for this is given by Bardeen et al. (1972). 
It is pointed out that,  
in order for a photon to escape from the circular photon orbit,
since nothing can come out of the hole, some physical process
near the hole is necessary to inject a particle into an escaping
trajectory, i.e., from a plunging orbit into an escaping one.

Depicted in Figure~6 are polar coordinate momentum 
components for various PPP (\gggg) cases; the corresponding angles of
escape, of the $e^-e^+$ pairs, above and below the equatorial plane, as
given by equation~(6), are shown in Figure~7.  
Notice in Figures~6 and~7, the increase in the asymmetry of the 
$e^-e^+$ pair 
distributions as the escaping energies 
increase; this is consistent with the effects caused by the GM
field (Williams 2002a).  Moreover, such one-sidedness 
could be an important factor in 
explaining the asymmetry in the observed jets of AGNs.
Thus, 
as in the PCS case,
 we find that the KBH in these PPP (\gggg) processes
naturally produces uneven fluxes of relativistic particles above
and below the equatorial plane.  
In addition, also as in the PCS, we find that
the $e^-e^+$ pairs escape 
along 
vortical 
orbits concentric about the polar axis, and highest energy
pairs escape with polar helical angle $\dd=\dd_\mp\sim 0.5^\circ$ 
(eq.~[6]); this suggests strong 
coil-like jet collimation (Williams 2002b, 2001).  
Note that,  this angle of escape for the highest energy 
$e^-e^+$ pairs 
is approximately equal 
to that of the maximum latitudinal angle---relative to
the equatorial plane---of the target nonequatorially confined 
orbiting photons
(compare Figure~7; and see \S~4.4 for further explanation of 
the latitudinal angle).
 
The above findings are consistent with observations of AGNs.
In the following sections, I shall discuss further this PPP (\gggg)
and its probable role in  astrophysical black hole sources. 

\subsection{Penrose Processes and the Classical Thin
Disk}
\label{sec:thin_disk}

In the PCS and PPP (\gggg) processes considered in this
paper, 
the lowest energies ($0.511$~keV$-3.5$~keV) used
for the infalling photons are consistent with the surface temperature
($T_s\simeq 4 \times 10^6$~K $-$ $4\times 10^7$~K, at
$r\approx 1.3M$, for $M=10^8M_\odot$) of the classical thin disk
accretion model about a KBH (Novikov \& Thorne 1973).
Now, the scenario for a $10^8M_\odot$ KBH, surrounded by a classical
thin disk (i.e., before the Lightman instabilities set in)
is the following.  About 80\% of the initially infalling
soft X-rays ($\sim 3.5$~keV), that have undergone PCS by equatorially
confined orbiting target electrons (at $\sim r_{\rm mb}$, 
$E_e\sim 0.54$~MeV),  can escape with
boosted energies as high as hard X-rays ($\sim 0.26$~MeV), while
the others either fall into the KBH,  or become  bound (i.e.,
with a turning point) at the photon
orbit---acquiring blueshifted energies  as high as $\sim 14$~MeV.
The polar coordinate momentum components $(P_{\rm ph}^\prime)_\Th$ 
of the scattered photons reveal that indeed such PCS photons
can acquire the necessary $Q$ values to populate the photon
orbit by the criteria discussed in Paper~1 [i.e., the 
$Q$ value must match that of the photon 
orbit whose
energy is consistent with the expected blueshifted energy, within
reason, as defined by equation~(2.26) of Paper~1; see also discussion in 
Williams (2002a)].
These blueshifted
photons, now assuming to have populated the photon orbit, can undergo
PPP (\gggg) processes with  infalling soft X-rays
[for maximum absolute efficiency (Paper~1)], producing $e^-e^+$
pairs with energies as high as
$\sim 13$~MeV.
About half of the produced pairs escape along vortical helical-like
orbits collimated about the polar axis.  The helical angle of
escape $\dd_{\rm esc}$, defined relative to the equatorial
plane (see equation~[6] and Fig.~7$a$), 
is $\sim 5^\circ$ down to $\sim 0.5^\circ$
for the highest energy particles, implying strong coil-like
collimation (Williams 2001, 2002b).  
Thus, I conclude, that, due to PCS and PPP (\gggg) processes, in a
classical thin disk, the highest particle energies attainable for
the PCS photons are $\sim 260$~keV (hard X-rays); 
and for
the relativistic PPP (\gggg) electrons, the highest energies attainable
 are $\sim 13$~MeV.  Yet, without these Penrose processes, as can be
seen above from the surface temperature $T_s$, the highest energy
radiated by a classical thin disk surrounding a supermassive KBH is
$\sim 3.5$~keV, and, of course, no pair production (compare Table~1).

\subsection{Penrose Processes and the Thin Disk/Ion
Corona}
\label{sec:corona}

Since the disk can exist between two phases 
[thin disk/ion corona (or hot torus)]
the occurrence of the Lightman instabilities can act
to enhance the Penrose process, thereby contributing to 
observed
variabilities and particle energies up to $\sim $~GeV. We shall
see this in the following 
scenario for a $10^8M_\odot$ KBH surrounded
by a thin disk/ion corona.  In this scenario, X-rays
($\sim 0.03-0.15~{\rm MeV}$), after undergoing PCS by
equatorially confined orbiting target 
electrons, can escape with boosted
energies ranging from $\sim 0.7$~MeV
up to $\sim 2$~MeV.  Just
as in the classical thin disk, about $80\%$ of the PCS photons 
can escape,
while the others either fall into the KBH,  or become  bound at the
photon orbit upon acquiring blueshifted energies, however, in these
cases, the energies acquired range from 
$\sim 36$~MeV to as high as $\sim 104$~MeV.
The polar coordinate momentum components of 
such PCS photons reveal
that indeed they can acquire the necessary
$Q$ values to populate the photon orbit in the above
blueshifted energy range (Paper~1). 
Subsequently, the blueshifted photons, now assuming to have
populated the photon orbit, can undergo PPP (\gggg) processes with
infalling disk soft X-rays, producing $e^-e^+$ pairs that can escape
with energies as high
as $\sim 100$~MeV.

Now we assume that the photon orbit can be populated in
addition by $\gamma$-rays from the $\pi^0$ decays 
($\pi^0\lra\gamma\gamma$), 
occurring in Eilek's (1980) 
ion corona/hot torus  model
(discussed in \S\S~1, 4.2, and Paper~1). 
These $\gamma$-rays are created with energies
narrowly peaked around
$E_\gamma\simeq 75$ MeV.  Such $\gamma$-rays 
can be blueshifted to
energies $\sim 4$~GeV, as they become populated 
or bound at the photon orbit.
Subsequently, infalling soft X-rays from the disk can undergo PPP
(\gggg) with these now bound $\gamma$-rays, producing $e^-e^+$ pairs that
can escape with energies up to $\sim 4$~GeV. 
Recently, it has been shown that if the proton distribution in
Eilek's hot torus model (or ADAF) is a power law, the energy of the 
$\gamma$-rays, produced by the $\pi^0$ decays, can reach 
energies $> 100$~MeV (Mahadevan, Narayan,  \&  Krolik 1997). 
Therefore, since the blueshift is $\simeq 52$
at the photon orbit, the PPP (\gggg) can 
produce $e^-e^+$ pairs with
energies $> 5$~GeV. 
However, here in this present analysis 
only the $\gamma$-ray particles from
Eilek's model, with energies 
${<\atop\sim}100$~MeV, before becoming blueshifted, 
need to be considered.
Such infalling particles allow these 
Penrose processes to explain the $\gamma$-ray spectra 
of 3C 279 and 3C 273.
Note, assuming that Eilek's particles (or particles from ADAFs) 
will acquire
the necessary $Q$ value appears to be 
a valid assumption, based on the
fact that the PCS photons acquire the necessary $Q$ value to
populate the photon orbit. That is, one
would expect the proton-proton nuclear reaction $\pi^0$ emission
(Eilek \& Kafatos 1983; Mahadevan et al. 1997), and the subsequent
$\pi^0$ decay:
$\pi^0\lra\gamma\gamma$,
to be  Penrose processes, and thus, the $\pi^0$-induced
$\gamma$-rays are expected to acquire the appropriate
$Q$ values.  However, confirmation of this will have to
await a Penrose process analysis of these reactions in the 
ergosphere of a KBH (Williams \& Eilek 2002).

Moreover, if we assume that pairs produced 
in Eilek's hot torus model,
which have energies peaked around
$\simeq 35$ MeV, can populate the nonequatorially confined
target electron
orbits, say in the range of $\sim 6-12$ MeV,
then PCS of infalling soft X-ray/UV photons from the disk
allows most of the scattered photons to escape with boosted
energies up to
$\sim 12$~MeV: this value is consistent with one of the general
cutoffs observed in the spectra of quasars (compare Fig.~1).  
Note that, such target electrons have  
nonequatorially confined ($Q>0$) ``spherical-like'' orbits
(Wilkins 1972), and reach latitudinal  
angles in their orbits
${<\atop\sim} 30^\circ$, above and below the equatorial plane, 
as given by equation~(6).  
The relevant algebra to determine the conserved
energy $E$ and azimuthal angular momentum $L$  
for these nonequatorially confined spherical-like orbits, for 
massless and material test particles, can be
found in  Paper~1 (see also Williams 2002a).
One of these spherical-like
orbits consists of a particle repeatedly passing
through the equatorial plane while
tracing out a helical belt lying on a sphere at constant radius.
The belt width or the maximum and minimum latitudes that the particle
achieves, in general, increase with increasing $Q$ (this does not 
appear to be true for massless particles bound at the photon
orbit; see \S~4.2.2).
Now, these PCS $\gamma$-rays up to 
$\sim 12$~MeV could contribute to the
observed spectra in this energy range (as will be shown in \S~5).
The
importance of these PCS $\gamma$-rays is that they can acquire and
escape with large polar and azimuthal 
coordinate momenta (implicating collimation), 
thus supporting the jet formation process (compare Figs.~2, 3$b$
and 4$b$).  Recently, it has been shown that these nonequatorially
confined target electron orbits can be populated by prior PPP (\gggg),
up to $\sim 2.5$~MeV (Williams 2003).

These energies of the escaping Penrose scattered particles
are to be compared with maximum energies attainable for the popular
two-temperature accretion disk models (Shapiro, Lightman, \& 
Eardley 1976; Lightman \& Eardley 1974), such as thin disk/ion
corona (ADAF). In these disk models, without Penrose
processes occurring,  the maximum particle energies attainable
are $\sim 100$~MeV for a thermal distribution of protons 
(Eilek 1980), and 
up to $\sim 31$~GeV 
for a power law distribution of protons 
(Mahadevan, Narayan,  \&  Krolik 1997). 
However, when the Penrose processes 
PCS and PPP (\gggg) are included, 
particles can possibly 
escape with energies as
high as $\sim 4$~GeV up to $\sim 54$~GeV
(attempting to go beyond this value causes the computer 
code 
to breakdown, suggesting that the model is attempting to go
beyond what is physically possible).  
In addition, since the luminosity
of the escaping $\pi^0$-induced $\gamma$~rays is so low for ADAFs, 
due to the bulk of energy being advected radially into the 
hole, rather than being radiated away, Penrose 
processes appear needed in order for the high energy particles to escape
in the form of jets (i.e., with the Penrose scattered particles
being collimated about the polar axis naturally by the KBH).
Such particles, as have been found in these calculations, were 
proposed to exist in de Felice and Curir's (1992) analysis of the Kerr
geometry.  The Penrose model calculations presented here and elsewhere
(Williams 1991, Paper~1, 2002a, 1999) appear to strongly 
validate their analysis, as well as the analysis by Bini et al. (2002):
concerning the asymmetry in the spherical-like (or nonequatorially
confined) particle orbits
(Wilkins 1972), above and below the equatorial plane, due to the GM
field.  Further analysis of the first order perturbation in latitudinal
angles $\Theta_{\rm min}$, $\Theta_{\rm max}$ (Bini et al. 2002)
should reveal that the asymmetry is directly related to the first order
perturbation in the constant radius $r$ of the spherical-like orbit 
[see eq.~(47) of Williams (2002a)], and, thus, overall related to the 
hard evidence that the GM field
 breaks the expected reflection equatorial planar symmetry of the Kerr 
metric
 (Williams 1999)\footnote{Figures 1(c) and 1(d) of Williams (1999) 
are incorrect.  Figures 4(e) and 4(h), respectively, of Williams (2002a) 
are the correct figures, where in  Figure 4(h) the target electrons have
both positive and negative equal absolute value polar coordinate
angular momenta.}.

\subsection{Secondary Processes}
\label{sec:secondary}

\subsubsection{Inverse Compton Scattering and 
the PPP Electrons} 
\label{sec:inverse}

Relativistic
inverse Compton scattering of the PCS photons, or accretion disk
photons, by the PPP  (\gggg)  electrons
is an important secondary process, which I refer to as ``secondary''
Penrose Compton scattering (SPCS). 
In general, the energy of a Compton scattered photon is given 
by (Kafatos 1980)
\begin{equation}
E_{\rm Comp}\sim\gamma_e^2 h\nu,~~~~({\rm for}~
 h\nu\la m_ec^2/\gamma_e);
\label{eq:8}
\end{equation}
or
\begin{equation}
E_{\rm Comp}\sim\gamma_e m_e c^2,~~~~({\rm for}~ 
h\nu\ga m_ec^2/\gamma_e),
\label{eq:9}
\end{equation}
where $\gamma_e$ is the Lorentz factor of the electrons
(in this case, $\gamma_e=E_\mp/m_e c^2$), and $h\nu$ is
the original energy of the photon.  
It turns out that most of the ``secondary'' Penrose Compton 
scattered   (SPCS) photons satisfy the relation of
equation~(9) for relativistic Compton scattering.  For example,
for PPP (\gggg) electrons with energies $\sim 2$~MeV, all PCS photon 
or disk photons with energies as high as
$\sim 0.26$~MeV satisfy equation~(9); and for electron pairs 
with energies
$\sim 200$~MeV, all PCS photons or disk photons with energies above 
$\sim 10^{-3}$~MeV satisfy equation~(9).  

In the calculated luminosity spectra of the 
Penrose processes, presented
in \S~5, the electrons produced by the 
PPP (\gggg) are allowed to undergo 
SPCS with infalling photons from the accretion disk. The acquired
numerically (Monte Carlo) calculated energies of the 
SPCS photons are not 
much different from the maximum attainable
energies given by equation~(9). 
In the SPCS, the numerical calculations are
similar to those used for the PCS (Paper~1),
except now the PPP (\gggg) electrons are used for the targets.

\subsubsection{Relativistic Beaming and 
the PPP Electrons} 
\label{sec:beaming}

An important aspect of the above 
secondary process is beamed relativistic 
expansion or commonly called 
relativistic beaming, 
i.e., when 
high energy radiation due to 
Compton scattering is  beamed 
into the polar directions
by means of relativistic electrons emanating  from 
the central source---such as the PPP (\gggg) electrons. 
Such PPP  electrons, which are initially
collimated by the KBH, and subsequently, perhaps, accelerated by 
shocks in the gas motions 
and electromagnetic
fields, 
can possibly give 
rise to the observed astrophysical jets. 
In relativistic beaming models, the bulk 
motion of the electrons (i.e., of a given blob of electrons) 
is assumed to move with relativistic 
speed.
Superluminal motion, produced by relativistic beaming, 
 becomes
important if the jet makes a small angle up to
$\sim 15^\circ$
 with the line of sight of the observer.  Such motion, 
in so-called  superluminal sources, can display apparent jet 
velocities greater than the speed of light,
and can enhance the observed brightness, energy, and variability
 of the source. 

Some theoretical relations for relativistic beaming from
various authors, relevant to these present model calculations,
are as follows.
The relation between the scattered photon energy 
$E_{\rm Comp}$ and the original
photon energy $h\nu$ is given by (Dermer et~al. 1992)
\begin{equation}
{E_{\rm Comp}\over h\nu}\simeq {(1-\cos\theta_s)\over
[\Ga(1-\bb\cos\theta_s)]^2}\,\gamma_e^2,
\label{eq:10}
\end{equation}
where $c\bb=v_{\rm bulk}\equiv v$ is the bulk or space velocity, 
$\Ga=(1-\bb^2)^{-1/2}$ is the Lorentz factor of the blob,
and $\theta_s$ is the angle between the velocity vector of the blob and
the direction of the observer.
Equation~(10) is for scattering in the Thomson regime, i.e.,
for $\,h\nu{<\atop\sim}m_e c^2/\gamma_e\Ga\,$ 
(Sikora, Begelman, \& Rees 1994).
The energy of the scattered photon in the Klein-Nishina
regime, i.e., for $\,h\nu{>\atop\sim}m_e c^2/\gamma_e\Ga\,$ is
given by (Dermer \& Schlickeiser 1993)
\begin{equation}
{E_{\rm Comp}\over m_e c^2}={1\over \Ga(1-\bb\cos\theta_s)}\,\gamma_e.
\label{eq:11}
\end{equation} 
Notice that equations (10) and (11) reduce to equations (8)
and (9), respectively, when $\Ga=1$ and $\theta_s=90^\circ$, 
as would be expected.

The apparent velocity of the blob in the jet as measured by
an observer at infinity
is given by (Blandford, Mckee, \& Rees 1977)
\begin{equation}
v_{\rm app}={v\sin\theta_s\over{1-(v/c)\cos\theta_s}}.
\label{eq:12}
\end{equation}

The observed luminosity $L_{\rm obs}$ relative to the intrinsic
luminosity $L_{\rm intr}$ is given by (Maraschi, Ghisellini, 
\& Celotti 1992)
\begin{equation} 
L_{\rm obs}=\dd_*^{m}L_{\rm intr},
\label{eq:13}
\end{equation}
where $\dd_*=[\Ga-(\Ga^2-1)^{1/2}\cos\theta_s]^{-1}$ 
is the usual 
Doppler factor, and $m$ ($=3+\alpha$; $\alpha\equiv$ flux spectral
index) is between 3 and~4. 

For completion,
an estimation of the intrinsic size $\bar R$ of the $\gamma$-ray
emitting region is given by the variability time scale $t_{\rm var}$:
\begin{equation}
\bar R=\dd_* c t_{\rm var}.
\label{eq:14}
\end{equation} 

For example, for angle $\theta_s\sim 5^\circ$, 
and a typical model value
of $\Ga=12$, giving $v_{\rm bulk}=0.9965c$, we find that
$E_{\rm Comp}=11.46 E_{\rm intr}$, where $E_{\rm intr}$
is the 
intrinsic photon energy (eq.~[9]);
$v_{\rm app}=11.95c$;
$L_{\rm obs}=5.1\times 10^3
L_{\rm intr}$ for $m=3.5$ and $\dd_*=11.46$; note, equations~(11)
through~(13)  have been used.

Thus, one can see from the above parameters that 
relativistic beaming near the observers line of sight
is indeed an effective enhancement mechanism.  However,
the relativistic beaming models must rely on some
preliminary mechanism to create and collimate the relativistic
electrons about the 
polar axis of the AGN.  
According to the model calculations presented in this 
paper, the Penrose process suggests the perfect
preliminary mechanism.  This mechanism has been shown
(Paper~1) to produce relativistic electrons 
($\gamma_e$ up to at least $\sim 10^4$), with vortical escaping orbits
about the polar axis
(de Felice and Curir 1992; de Felice and Carlotto 1997;
Williams 2002b), that have
asymmetrical distributions above and below
the equatorial plane---leading to asymmetrical jets
(Williams 2002a).
Recall, some specific inconsistencies encountered in 
attempting to use
relativistic beaming models alone to explain the asymmetrical
jets, without attributing any to being produced intrinsically
by the energy source, are discussed in \S~1. 

%

\section{COMPARISON WITH OBSERVATIONS AND DISK MODEL 
CORRELATIONS} 

\subsection{Model Calculated Penrose Spectra}
\label{sec:spectra}

The luminosity spectra due to Penrose
processes for the specific cases of quasars 3C~279 
and 3C~273 are plotted in
Figures~8$a$ and~8$b$, along with their observed spectra
for comparison.
The outgoing (escaping) luminosity produced by the 
Penrose scattered particles 
is given by
\begin{eqnarray}
L_\nu^{\rm esc}&\approx &4\pi d^2 F_\nu^{\rm esc}~~
({\rm erg/s\,Hz})\nonumber\\
		&\approx &4\pi d^2h\nu^{\rm esc}
f_1 f_2 \cdots f_n\,(N_\nu^{\rm in}-N_\nu^{\rm cap})
\label{eq:15}
\end{eqnarray}
(the units of $ L_\nu^{\rm esc}$ are in the parenthesis),
where $d$ is the cosmological distance 
of the black hole source;
$F_\nu^{\rm esc}$ is the flux of escaping photons;
$N_\nu^{\rm in}$ and $N_\nu^{\rm cap}$ are the emittance of 
incoming and
captured photons, respectively;
$f_n$
defines the total fraction of the particles that undergoes scattering
[$n=2$ for PCS and $n=5$ for PPP (\gggg)].
The values of $f_1,\ldots, f_n$, which fit the Penrose calculated
luminosities with observations 
for the specific cases of 3C~279
and 3C~273, are given in Tables~2 through~5. As we shall see, 
the $f_n$'s
are somewhat free parameters.  Note, the labeled points on Figure~8
correspond to the case numbers on Tables~2 through~5.

Specifically, the $f_n$'s are defined as follows:
$f_1$ is the fraction of incident disk photons, intersecting the 
scattering radius $r_s$, coming from the inner radius 
$r_d$ of the accretion
disk 
of height $h$:
\begin{equation}
f_1\approx {r_s^2\over r_d\, h}(\cos\theta_1 -\cos\theta_2)
\label{eq:16}
\end{equation}
[compare equation~(3)]; 
$f_2$ is the fraction scattered, dependent on the 
cross section---the effective area of the target particle 
[Klein-Nishina or pair production (see Paper~1)],
which  implies that
the fraction of the incoming photons that actually undergoes PCS
or PPP (\gggg) is dependent on the 
incident energies and angles; 
$f_3$ is the fraction of the incident photons from the disk 
that intersects
the scattering radius $r_s$, traveling from an inner radius of the 
accretion disk $r_d$,
that are free to undergo
SPCS with the PPP 
(\gggg) electrons 
(similar to $f_1$); 
$f_4$ is the fraction of the incident photons 
from the disk that actually undergoes
SPCS with the PPP (\gggg) electrons 
(also dependent on the Klein-Nishina cross section and
similar to $f_2$); $f_5$
is the fraction of the $e^-e^+$ pairs that undergoes  SPCS 
(dependent on the expansion rate of the jet).  
So the values of  $f_1,\ldots, f_5$, that are found to
agree with observation, indirectly may tell us something about the 
particle densities in the ergosphere, at different radii, per
frequency interval, e.g., 
number density $n_\nu$ ($\rm cm^{-3}\, Hz^{-1}$), radiation density
$u_\nu$ ($\rm erg/cm^{-3}\, Hz^{-1}$), and mass density 
$\rho_\nu$ ($\rm g\,cm^
{-3}\, Hz^{-1}$). 

The incoming surface flux density of the incident photons from
the disk participating in the Penrose scattering processes, 
\begin{equation}
F_\nu^{\rm in}=h\nu N_\nu^{\rm in}, 
\label{eq:17}
\end{equation}
is determined from spectral observations of 3C~279 and 3C~273
(see Fig.~1).
A power law distribution is assumed for the incoming 
photons:
\begin{equation}
N_\nu^{\rm in}=K\nu^{-\alpha}
~~({{\rm number~ of~ photons\over cm^2~ s~ Hz}}), 
\label{eq:18}
\end{equation}
where, the photon spectral index:
\begin{equation}
\alpha=-{\log N_{\nu_2}-\log N_{\nu_1}\over \log\nu_2-\log\nu_1},
\label{eq:19}
\end{equation}
and $K$ are determined.
The values of $\alpha$ and $K$ determined from observations, and
used in these calculations for 3C~279 and
3C~273, are the following:
\begin{eqnarray*}
\alpha&\simeq &1.48,~{\rm for~3C~279;}~{\rm between}~
\nu_1=1.166~{\rm keV},
~\nu_2=2.93~{\rm keV};\\
\alpha&\simeq &1.5~{\rm for~3C~273;}~{\rm between}~
\nu_1=0.511~{\rm keV},
~\nu_2=25~{\rm keV};
\end{eqnarray*}
and
\begin{eqnarray*}
K&\simeq& 1.29\times 10^6 {\rm cm^{-2}\,s^{-1}\,Hz^{0.48}, 
~for~3C~279};\\
K&\simeq& 1.132\times 10^7 {\rm cm^{-2}\,s^{-1}\,Hz^{0.5}, 
~for~3C~273}.
\end{eqnarray*}
So, from observations, the 
initial photon distribution laws for
3C~279 and 3C~273, used in these calculations, are given by
\begin{eqnarray}
N_{\nu_{\rm in}}^{\rm in}&=&1.29\times 10^6
\nu_{\rm in}^{-1.48}~{\rm photons\,
cm^{-2}\,s^{-1}\,Hz^{-1}};\label{eq:20}\\
N_{\nu_{\rm in}}^{\rm in}&=&1.132\times 10^7
\nu_{\rm in}^{-1.5}~{\rm photons\,
cm^{-2}\,s^{-1}\,Hz^{-1}},
\label{eq:21}
\end{eqnarray}
respectively.  Equations (20) and~(21) give the emittance of 
photons per unit area per unit time per unit frequency---based 
on spectral 
observations in the soft X-ray regime.  Here we are assuming that 
emitted energy in this regime is due to the accretion disk.

The photon initial (or incoming) distribution law $N_\nu^{\rm in}$
of equation~(18), as defined by equations~(20) and~(21),
is normalized to 2000 photons, such that, the 
weighted average of each photon per incoming frequency is given by
\begin{equation}
\langle N_{\nu_{\rm in}}^{\rm in}\rangle={N_{\nu_{\rm in}}
^{\rm in}\over n_{\rm total}^
{\rm in}}
={N_{\nu_{\rm in}}^{\rm in}\over 2000},
\label{eq:22}
\end{equation}
which states that, one incoming
 photon is equivalent to $\langle N_{\nu_{\rm in}}^{\rm in}\rangle$,
where $\nu_{\rm in}$ is the incoming frequency,  and $n_{\rm total}^
{\rm in}$ is the total number of the incoming distribution of
photons having the given
monochromatic frequency $\nu_{\rm in}$. 
It is important to note that $\langle N_{\nu_{\rm in}}^
{\rm in}\rangle$ is invariant with respect to a specific 
frequency interval, i.e., a constant, meaning that 
$\langle N_{\nu_{\rm in}}^{\rm in}\rangle$ does not change in going
from one  frequency to another.   The above is a statement of the 
conservation of photon number in Compton scattering processes
(assuming that each incoming photon scatters).

Now, using this conservation law, equation~(15) can be written as
\begin{eqnarray}
L_\nu^{\rm esc}
&\approx& 4\pi d^2h\nu^{\rm esc}
f_1 f_2\cdots f_n\,\langle N_{\nu_{\rm in}}
^{\rm in}\rangle(n_\nu-n_\nu^{\rm cap})\nonumber\\
&\approx &4\pi d^2h\nu^{\rm esc}
f_1 f_2\cdots f_n\,\langle N_{\nu_{\rm in}}
^{\rm in}\rangle n_\nu^{\rm esc},
\end{eqnarray}
where $n_\nu$ is the number of the initial photons 
($n_{\rm total}^{\rm in}\equiv 2000$)
scattered into the final frequency interval 
$\DD\nu$, i.e.,  
$n_\nu$ is 
the number
of final photons emitted into a specific 
frequency interval $\DD\nu$, 
with $\nu$ being the 
average frequency of the interval;
$n_\nu^{\rm cap}$
is the corresponding number of final 
photons captured by the black hole at the 
frequency interval $\DD\nu$; and $n_\nu^{\rm esc}$ is 
the number of final photons
that escape from the black hole at 
the frequency interval $\DD\nu$.
Note that, $n_\nu=n_\nu^{\rm cap}+n_\nu^{\rm esc}$; and
$\sum_\nu  n_\nu=n_{\rm total}^{\rm in}= 2000$.  In general, to 
determine the outgoing Penrose luminosity $L_\nu^{\rm esc}$,  
several values of $\nu_{\rm in}$, 
consistent with observations, are separately
used in equation~(22);  the result
of each distribution of 2000 infalling monochromatic 
photons
is then substituted into equation~(23), along with
Penrose model calculated values of $n_\nu^{\rm esc}$ at 
$\nu^{\rm esc}$, and other appropriate model parameters. 
From this we obtain several distributions of outgoing luminosities
over specific frequency ranges.
The peak luminosity of each outgoing distribution, at the corresponding 
$\nu^{\rm esc}$ (i.e.,  the continuum)
is then plotted, 
as discussed in the following section.

\subsection{Agreement with Observations  
of 3C~279 and 3C~273}
\label{sec:observations2}

The model luminosity spectra 
resulting from equation~(23) for the Penrose processes
considered here,
specifically for quasars 3C 279 
and 3C 273, are plotted in 
Figures 8$a$ and 8$b$, respectively.  The solid 
lower
line curves  are the observed spectra in the X-ray, soft $\gamma$-ray,
and $\gamma$-ray regimes.


On Figure~8$a$,
for 3C~279, the crosses located at 
points 
2 through 5
are from PCS at $r_{\rm ms}=1.2M$, $r=1.13M$, 
$r=1.1M$, and $r_{\rm mb}=1.089M$, 
respectively, by equatorially confined target electrons.  
These radii cover the scattering region of the ergosphere one
would expect the PCS to occur.  
In order for the PCS by equatorially confined targets 
to contribute to the observe spectrum in
the X-ray regime, the points 2 through~5
must lie directly above the observed
curve---as one can see they do not.  
Thus, it appears that PCS by equatorially confined 
target electrons contributes little,
if any, to the X-ray part of the spectrum.  
However, this is true only if we assume, 
from observations, that the infalling 
photons undergoing PCS are of the same distribution we observe 
in the part of the spectrum we believe to be intrinsic 
to the accretion disk.   
Now, the  general
curve for PCS by equatorially confined targets is shown 
[the dotted line (or curve)
between points 1 and 7]. In the general case,  
lower energy photons 
(those not observed in the spectrum of 3C 279, but are consistent
with the  two-temperature, 
bistable thin disk/ion corona accretion) are used in the 
PCS
by equatorially confined target electrons at $r_{\rm mb}$.
One sees that part of the 
general PCS curve lies directly above the observed 
curve, meaning that it could contribute to the 
observed spectrum;  but in order
for this to be possible, the infalling disk photons
would have to have energies lower than what is observed
for 3C~279.  Nevertheless, to produce the general curve, the 
infalling disk photons
have energies 
between $E_{\rm ph}=0.07$~keV 
($\log{\nu/Hz}=16.23$) and $150$~keV ($\log{\nu/Hz}=20.56$)
to give final energies in the range of 
$E_{\rm ph}^\prime\simeq7.28$~keV 
($\log{\nu/Hz}
=18.25$) and $1.5$~MeV ($\log{\nu/Hz}=20.56$),
respectively.  
Table~2  gives the values of the factors 
$f_1$ and $f_2$ of
equation~(23) used in these model calculations.  
Note, the lack of observed emission in the region attributed
to the accretion disk may be due to jet beamed emission near the 
line of sight (\S~\ref{sec:beaming}), blocking from view 
emission from the disk. 

Further, on Figure $8a$, 
the general PCS 
spectrum produced by  the nonequatorially confined 
target electrons at $r_{\rm mb}$,
and infalling monochromatic 
photons with energy $E_{\rm ph}=0.03$~MeV, is represented by the 
dotted curve in the soft $\gamma$-ray regime; the  
nonequatorially confined  target electron energies range 
from $E_e\simeq 11.79$~MeV, $\sqrt{Q_e}=24.79Mm_e$ (point 13) down to the
equatorial value of $E_e\simeq 0.539 $~MeV (point~6, i.e., where the 
general curves meet, here $Q_e=0$).   
The solid squares superimposed
on the dotted curve are the PCS 
photon distributions using nonequatorially confined 
targets, specifically
for 3C 279 
(based on the observed luminosity spectrum).  
The solid squares  
superimposed on the solid line indicate the PCS 
distributions fitted to agree with observations using the $f_n$ 
factors.

The values of $f_1$ and $f_2$ (not enclosed in brackets) on Table~2
are the values used in equation~(23) to plot the solid squares 
that are superimposed
on the upper dotted curve, labeled by the numbers 8 through~13
on Figure~8$a$; and 
the values of $f_2$ enclosed in brackets are those used 
in order to make  the  PCS  distributions
agree with observations.  The factor $f_1$ 
tells us the fraction of the luminosity
of the disk arriving at the scattering radius, and the
expected angular 
height of the scattering regime  (see eqs.~[3] and~[16]). 
The factor $f_2$ (as well as $f_4$) is defined as follows:
\begin{displaymath}
f_2\equiv {\left({\hbox{{\rm number of scattered photons
}}\over\hbox{{\rm total number of infalling photons}}}
\right)};
\end{displaymath}
thus, $f_2$ tells us the  
probability of the PCS events occurring for a
distributions of infalling photons.  This probability
will depend on the total  Klein-Nishina cross section  
$(\sig_{\rm tot})_{\rm KN}$ 
for a given set of initial conditions (see Paper~1). 
Typical values of $(\sig_{\rm tot})_{\rm KN}$ for PCS by target
equatorially and nonequatorially confined electrons are $\sim
1.8\times 10^{-25}~{\rm cm^2}$ and $\sim 3.6\times 10^{-26}
~{\rm cm^2}$, respectively.  For comparison, $\sig_{\rm tot}
\sim 2\times
10^{-26}~{\rm cm^2}$ for recombination to the ground state of
hydrogen at $T\sim 10^9$~K, and  the geometric cross section
of a nucleon is $\simeq 5\times 10^{-25}~{\rm cm^2}$.

The other
model parameters on Table~2 are defined as follows:
$r$ is the scattering radius; $E_e$ is the target
electron energy; $\nu_{\rm ph}$ is the initial infalling
photon frequency; 
$\nu_{\rm peak}$  and $L_{\rm peak}$
correspond to the points (solid squares superimposed on the
dotted line) 
which give the continuum  
luminosity resulting from  several distributions of PCS
events (each distribution has 2000 scattering
events); $L_{\rm obs}$ is the observed luminosity at 
$\nu_{\rm peak}$ (the average 
frequency of the interval $\DD\nu$ where most of the PCS
photons are emitted per 2000 scattering events).  
Each distribution of 2000
infalling photons have monochromatic energies normalized
to the  power law distribution given by equation~(20)
for 3C~279, and by equation~(21) for 3C~273.   
Overall, to produce the calculated Penrose 
luminosity spectra of Figure~8, 
from 56,000 (for 3C~279) to 74,000 (for 3C~273) infalling 
photon scattering events are used.  

The solid circles on Figure~8$a$ are due to
secondary
Penrose Compton scattering  (SPCS; described in \S~4.5.1) 
of low energy  
infalling accretion disk photons
by the highly energetic PPP (\gggg)
electrons.  These
SPCS processes are assumed to take place at (or near)
the radius of the 
photon orbit ($r_{\rm ph}\simeq1.074M$).  The energy 
acquired by the scattered disk photon is approximately
equal to that of the target PPP (\gggg)
electron, and thus is approximately
given by
equation~(9).  For example, a distribution of 2000 escaping
PPP (\gggg) electron targets with  
energies peaked around $2.57$~GeV,  
after undergoing SPCS, yields a distribution of
1964 escaping scattered photons with energies peaked
around $2.44$~GeV (point 21 on Fig.~8$a$).  The general 
spectrum produced by the SPCS photons is represented by the 
dashed line (points 14 through 21). 
The peak energies of the initial PPP (\gggg) 
electron distributions (used for the targets)
range from $1.289$~MeV ($\gamma_e=2.523$)
to $2.469$~GeV ($\gamma_e=4.832\times 10^3$), and the
initial infalling photon distribution has monochromatic energy 
$0.03$~MeV.  
To produce higher energy SPCS spectra, PPP (\gggg) electron
distributions with larger Lorentz factors can be used.
Note, the infalling incident photons (at $E_{\rm ph}=0.03$~MeV),
assuming to originate in the accretion disk,  have a
luminosity power law distribution (eq.~[20]) 
consistent with the 
observed spectrum of 3C 279. 
The solid circles superimposed on dashed line are SPCS 
specifically for 3C~279.  
The solid circles superimposed on the solid line indicate the 
SPCS photon  distributions fitted to agree with observations
using the $f_n$ factors.  

Table~3 gives the values of the factors 
$f_1,\ldots, f_5$ of
equation~(23) used in these model calculations for 3C~279.
The values of the $f_n$ factors 
enclosed in brackets are those
used to make the SPCS photon distributions
agree with observations.  The factors $f_1$ and $f_3$ give the
geometric fraction of the disk luminosity intersecting 
the scattering radius for PPP (\gggg)
and subsequently for SPCS, respectively.
The factor $f_2$, defined earlier, tells us the
probability of the scattering events occurring for
the given initial conditions, governed by
the total pair production (\gggg) 
cross section $(\sig_{\rm tot})_{\gamma\gamma}$ 
(Paper~1).  
Typical values of $(\sig_{\rm tot})_{\gamma\gamma}$, 
for the PPP (\gggg), range from  $\sim 1.7\times 10^{-26}~{\rm cm^2}$ 
to $\sim 1.3\times 10^{-27}~{\rm cm^2}$, for the orbiting target
photon energies ranging from $\sim 208$~MeV to $\sim 4$~GeV, 
respectively.  
The factor $f_4$, defined like $f_2$ above, tells us the
probability of the SPCS events occurring for
the given initial conditions, governed by
the total Klein-Nishina cross
section $(\sig_{\rm tot})_{\rm KN}$.
Typical values of $(\sig_{\rm tot})_{\rm KN}$, 
for the SPCS,  range from  $\sim 2.6\times 10^{-27}~{\rm cm^2}$ 
to $\sim 2.4\times 10^{-28}~{\rm cm^2}$ for target
PPP (\gggg) electron energies ranging from 
$\sim 200$~MeV to $\sim 3.6$~GeV, 
respectively.   
For comparison, in  the  photoproduction process: 
$\gamma p\lra \pi^0 p$, $\sig_{\rm tot}$ peaks to 
$\sim 3\times 10^{-28} ~{\rm cm^2}$ at incident photon
energy $E_\gamma\sim 340$~MeV, decreases to 
$\sim 1\times 10^{-28}~{\rm cm^2}$ at 
$E_\gamma \sim  470$~MeV, and goes to zero at $E_\gamma{<\atop\sim}
150$~MeV. 
The factor $f_5$, which relates to 
the expansion rate and, thus, the core region density, gives 
the expected fraction of $e^-e^+$ pairs that undergoes SPCS, 
while the rest of the pairs are free to be further collimated and
accelerated by a surrounding electromagnetic field.

The other
model parameters on Table~3 are defined as follows:
$r$ is the scattering radius; $(E_\mp)_{\rm peak}$ is the energy
value where most of the PPP (\gggg) electrons are created;
$\nu_{\rm peak}$  and $L_{\rm peak}$
correspond to the points which give the continuum 
luminosity resulting from several distributions of SPCS
events [each distribution has about 2000 scattering
events, depending on the number of escaping target PPP (\gggg)
electrons]; 
$L_{\rm obs}$ is the observed luminosity at 
$\nu_{\rm peak}$ (the average
frequency of the interval $\DD\nu$ where most of the SPCS
photons are emitted per $\sim 2000$ scattering events).  

On Figure~8$b$, for 3C~273,
the dotted curve shows the general distribution
for PCS 
by equatorially confined electrons
(points 1 through 7)
and by nonequatorially confined electrons 
(up from point~6 
through 13).  The solid
squares for PCS, and solid circles for  SPCS by PPP (\gggg)
target electrons, 
are calculated specifically for 3C 273, based on the observed 
photon distribution.  In this case, it appears that the PCS
by equatorially confined target electrons contributes a significant 
part to the X-ray regime of the observed spectrum (indicated by
points 2 through 5 
above the observed curve).  The solid squares at 
point 2 and points
3 through 5 are for PCS by equatorially confined targets
at $r=1.099M$ 
and $r_{\rm mb}=1.089M$, respectively;
and points 8 through 13 superimposed on 
the dotted curve in the soft $\gamma$-ray regime, directly above
the observed spectrum, are for PCS by nonequatorially confined
orbiting
targets at $r_{\rm mb}$.
The solid squares
superimposed on the solid line indicate the PCS
distributions (for the corresponding labeled points lying 
directly above)
that have been fitted to agree with observations using the 
$f_n$ factors.
Table~4 gives the values of the factors $f_1$ and 
$ f_2$ of
equation~(23) used in these model calculations for
3C~273.
The values of the $f_n$ factors enclosed in brackets
are the values used
to make the PCS distributions
to agree with observations.  Note that, the parameters of
the columns on
Table~4 are defined as on Table~2.

Further, on Figure 8$b$, recall that 
the solid circles are due to
SPCS of infalling accretion disk photons
by the PPP (\gggg) electrons.  These
SPCS processes are assumed to take place at (or near) 
the radius of the
photon orbit $r_{\rm ph}$.  The general
spectrum resulting from SPCS is represented 
by the dashed line. The
target electron distributions, resulting from the PPP (\gggg),
used in these SPCS processes,
have peak energies ranging from $1.289$~MeV ($\gamma_e=2.523$)
to $2.469$~GeV ($\gamma_e= 4.832 \times 10^3$), and the
infalling incident photon distributions have monochromatic 
energy 
$E_{\rm ph}=0.03$~MeV. The solid circles superimposed on the 
dashed line indicate SPCS photon distributions specifically for
3C~273.
The solid circles superimposed on the solid lines indicate the
SPCS photon distributions that have been 
fitted to agree with observations
using the $f_n$ factors (for each of the corresponding 
labeled points
lying directly above).
Table~5 gives the values of the $f_n$ factors
($f_1,\ldots, f_5$) of
equation~(23) used in these model calculations for 3C~273.
The values of the $f_n$ factors enclosed in brackets
are the values used to make  the SPCS photon luminosity
agree with observations.   Note that, the parameters 
of the columns on Table~5 are defined as they are on Table~3.  

From Figure~8, 
one can clearly see that
luminosities due to Penrose processes in the high energies regime are
sufficient to explain the observations of these quasars.  
Importantly, the
Penrose mechanism generates 
an abundance of energy, such that it allows for the smallness in
the cross sections at high energies.  Moreover,
notice that, the generated luminosity spectra produced by
the PCS and PPP (\gggg) 
fall off in regions 
consistent with the gaps in the radiation of the general observed
spectra of AGNs.

Now returning to Figures~1$a$ and 8$a$ to discuss the total 
observed spectrum
of 3C~279: In general, the radio/mm 
emission may be due to
the interaction of a magnetic field $B$, say of the accretion disk
($B_d\sim 10^2$~gauss), with electrons of the disk or PPP (\gggg)
electrons.
For example, for electrons
with energies between $\sim 1.8-13$~MeV, the synchrotron frequency
of the emitted photons will be $\nu_{\rm syn}\sim 5.2\times 10^9-2.7
\times 10^{11}$~Hz, where $\nu_{\rm syn}\sim 4\times 10^6 \gamma_e^2 B$ 
(Burbidge, Jones, \& O'Dell 1974).  
The infrared to UV emission spectrum can be produced by
the PPP (\gggg) electrons at $B_d\sim 10^2$~gauss, 
i.e., for electrons 
having energies between $\sim 30-2000$~MeV, 
the synchrotron frequency of the emitted photons will be 
$\nu_{\rm syn}\sim 1.5 \times 10^{12}-6.4\times 10^{15}$~Hz.  
Of course Eilek's (1980) $\, e^-e^+$ pairs, with energies peaking 
$\sim 35$~MeV, could also contribute synchrotron radiation 
to the spectrum in this regime.  
Moreover,
the accretion disk radiates thermal photons in this region 
of the spectrum, with 
maximum radiation peaking 
from $\sim 2\times 10^{14}$ to $2\times 10^{15}$~Hz. 
This thermal radiation is due to viscosity in the disk as angular 
momentum is transferred outward, thereby heating the disk, producing 
copious photons. 
Particularly, in the optical band, this is most likely the 
dominant observed emission for
``normal'' quasars, but 
not for OVV quasars, 
like 3C 279.
For the case of OVV quasars, the observed optical emission 
is probably dominated
by synchrotron emission in the jet, since, in these sources,
the jet is believed to
be pointed in a direction near the observer's line of sight;
this gives rise to 
observed features,
such as intense variability
 and high polarization in the
optical emission.
The X-ray emission ($\sim 1.3-41$~keV) can be produced by both 
PCS and the inner
region of a thin disk/ion corona accretion. In general, for a 
$10^8 M_\odot$ KBH, the PCS has emission over the total X-ray
band, and so does the inner region of the accretion disk.
However, from comparing Figure~8$a$, the accretion disk
appears to be the more 
dominant mechanism in producing
the observed X-ray emission of 3C~279 (i.e., if we limit 
the photons participating in the PCS to be only those from the
observed soft X-ray spectrum we believe to be disk radiation).
On the other hand, as mentioned earlier, the X-ray emission may be
predominantly beamed PCS.
This is not the case for 3C~273 (compare Fig.~8$b$; 
more on this below).  The soft $\gamma$-ray emission can be produced 
by infalling X-ray photons $\sim 30$~keV
($\sim 3.5\times 10^8$~K) from a thin disk/ion
corona accretion; such infalling disk X-rays can first undergo
PCS (see Fig.~3$a$), producing photons that can 
populate the photon orbit 
for PPP (\gggg) (Williams 2002a, 2002b);
these pairs can
subsequently undergo SPCS with additional infalling disk photons, 
producing $\gamma$-rays
in the 
range $\sim 1-27$~MeV, peaking in luminosity at $\sim 15$~MeV
(or $\sim 3.6\times 10^{21}$~Hz).
The soft $\gamma$-ray emission can also be produced directly
by the PCS photons (see Figs.~2 and 3$b$), with
energies extending up to, say, $\sim 13$~MeV (or $\sim 3\times 
10^{21}$~Hz). 
The cutoff of the observed soft $\gamma$-ray emission is  probably due to 
the inability of prior PCS to
populate the
photon orbit---for PPP (leading to SPCS), and/or the
inability, or unavailability, of higher energy electrons 
to populate the nonequatorially confined target
particle orbits---for PCS [from ion corona nuclear reaction processes 
(\S\S~\ref{sec:ppp} 
and~\ref{sec:corona})]. 
The higher energy
$\gamma$-ray emission  [up to $\sim 4$~GeV  ($\sim 10^{24}$~Hz)]
can be produced exclusively by the PPP (\gggg) electrons (i.e., 
after some of  the PPP 
electrons have 
undergone SPCS processes). 
These processes can contribute copious photons to the 
observed $\gamma$-ray emission spectrum, and is 
probably responsible for the large luminosities 
at very  high energies (compare Figs.~1$a$ and~8$a$).
Any observable fall off in the hard $\gamma$-ray regime
is probably due to the increasingly smaller 
cross section at higher energies.  Note, it appears that the
$\gamma$-rays produced by the $\pi^0\lra\gamma\gamma$
decays (\S\S~\ref{sec:ppp} and 4.4), satisfying  conditions to 
populate the photon orbit for PPP (\gggg), 
can safely be predicted to have energies $> 1.6$~MeV, based on 
observations:
 such $\gamma$-rays
will acquire gravitational blueshifted energies $> 85$~MeV,
with expected  $Q$ values $> 6 M^2m_e^2$.

If the radiation is preferentially beamed towards us, 
we will overestimate the true luminosity and
energy, depending on the observing angle, $\theta_s$ 
(\S~4.5.2).  This does not pose a problem for this present model,
because the luminosity and energy could undergo an apparent
shift and still be 
described by these Penrose processes.  Moreover,
according to equations~(11)
through~(14),
if the observing angle is $\theta_s=15^\circ$, and the Lorentz 
factor of the
blob is $\Ga =12$, implying a Doppler boost of 
$\dd_*\simeq 2.2$, 
 then the 
energy will shift by a factor of $\simeq 2.2$ 
(increasing from an intrinsic energy of, say, $\sim 1.8$~GeV 
to an apparent energy of $\sim 4$~GeV), 
and the intrinsic luminosity shifts upward by a factor
of $\simeq 15.8$, with $m=3.5$, to its observed value.  
The apparent velocity 
would be $\simeq 6.9c$, consistent with 
observations ($v_{\rm app}\simeq 6.9c$ for 3C~279; 
Porcas 1987),  and, for completion, 
the intrinsic size of the emitting region, say in the $\gamma$-ray 
$> 100$~MeV (\S~3.2)---for a 
variability time scale of $\sim$ a day, will be
$\bar R\sim 0.002$~pc$\sim 195~r_g$, where $r_g$ ($=2MG/c^2$)
is the so-called gravitational radius.


Referring now to Figures~1$b$ and 8$b$, 
to discuss the total observed spectrum of
3C~273: The emission in the radio to 
UV region of the spectrum 
can probably be explained by the same processes
as described above for 3C 279, i.e., Penrose, 
accretion disk,
and synchrotron emission processes.  
In the X-ray regime, particularly, from $\sim 3.5 $ to~207~keV
($\sim 10^{18}$ to $5\times 10^{19}$~Hz), 
the 
emission spectrum is probably produced by PCS and the inner region 
of an ion corona accretion disk (compare Fig.~8$b$).  
Since the X-ray emission 
extends to much greater
than $\sim 3.5$~keV, which is the 
maximum energy produced by the inner
region of a thin disk, then we conclude that the ion corona and
PCS are most likely
producing the hard X-rays.  Since the accretion
can exist in either or both of two phases: thin disk and/or 
ion corona, 
and can remain in any of these 
phases for an indefinite time, depending on the cooling mechanism,
it is possible that quasar 3C 273 is spending its 
``lifetime'' largely in the ion corona/hot torus (or the ADAF) phase.  
This is contrary to 3C 279, which, as observations suggest,  
spends its lifetime oscillation between
the thin disk and ion corona phases, on a relatively small timescale
(further comparisons of the 
spectra of 3C 279 and 
3C 273 are presented in the following section).  

An explanation for the observed $\gamma$-ray emission spectrum is 
as follows (compare Fig. 8$b$).  
The $\gamma$-ray emission from $\sim 1$ to $13$~MeV
($\sim 2\times 10^{20}$ to $3\times 10^{21}$~Hz)
is most likely due to a mixture of PCS and
relativistic 
SPCS by the PPP (\gggg) electrons,
as described above  
 for the soft $\gamma$-ray emission in 3C~279.
The $\gamma$-ray emission from $\sim 13$~MeV up to
$\sim 4$~GeV is probably due predominantly to SPCS.
Some of the $\gamma$-ray emission,
at least in the regions between $\sim 35-100$~MeV, may 
be due to Eilek's
particles (\S~4.4), as indicated in Figure~1$b$.  Finally, 
relativistic beaming may serve to boost the overall 
$\gamma$-ray emission, depending on the observing angle, since 
superluminal motion has been detected in observations of 3C~273
(Porcas 1987).

\subsection{The Compared Spectra of 3C 279 and 3C 273}
\label{sec:compared_spectra}

Now we compare further the spectra of 3C~279 and 3C~273, again using 
Figures~1 and~8.
The classification of these quasars as radio loud is evident from 
their spectra in the radio/mm region. 
The shape of the spectrum of 3C~273
looks like the ``enhanced''
 spectrum of 3C~279, except for the higher 
luminosities in 3C~279 and the radio tail in  3C~273.  
The higher luminosity and the lack of a 
radio tail in 3C~279 is
probably largely due to the radiation of 3C~279 being beamed more
in the direction of the observer than the radiation of 3C~273.  
Therefore, the spectrum of 3C~279 has been Doppler (i.e.,
apparently) blueshifted to a higher 
energy interval, and the luminosity has been apparently 
increased (\S~4.5.2).  
This is consistent with radio observations which detect more
superluminal motion (or relativistic beaming near the line of sight 
of the observer)
 in 3C~279 than in 3C~273 (Porcas 1987).
Next, notice on the schematic broad band spectral diagrams
of 3C~279 and 3C~273 (Fig.~1),
the region which I have labeled as  
the ``inner accretion disk,'' 
the emission of 3C 273 is hotter, covering  a broader band of 
emission than 3C~279. This to be expected if 3C 273 is in a predominantly 
ion torus (or advection dominated) 
state, as opposed to 3C 279 (which appears to
oscillates in a highly 
variable
fashion between the thin disk and ion corona phases---for this reason
3C~279 is classified as an OVV quasar).
Moreover, 
the hotter state of the accretion disk (ion torus), 
which is heated 
by a runaway thermal instability (\S~4),  would result in
enhanced Penrose processes [PCS and PPP (\gggg)] 
and enhanced synchrotron 
radiation due to the presence of more relativistic electrons and
$\gamma$-ray photons.
This appears to be the case always in the continuum emission of 3C~273
and sometimes in the emission spectrum of 3C~279, with 3C~279 not
quite achieving the full ion torus status of 3C~273 (compare Fig.~1).  
Thus in summary, the 
differences in the spectra of 3C~279 and 3C~273 are probably
due to (1) the more beaming effect in 3C~279, and (2) 
the predominantly
ion torus phase of 3C~273. 

\subsection{Radio Loud and Radio Quiet Quasars}
\label{sec:quasars3}

The results of these Penrose processes allow one to
make a brief statement on the evolution of radio loud 
and radio quiet 
quasars. Since radio loud galaxies are usually associated with large 
elliptical galaxies with radio lobes, and assuming 
that quasars are progenitors of at least some 
galaxies (particularly nearby AGNs), it is perhaps safe 
to say that 
radio loud galaxies evolve from radio loud quasars. 
This is based on observations and the model presented in this paper,
suggesting that radio loud quasars have
 geometrically larger, more effective 
ion tori (or coronas) than their counterparts, radio quiet quasars,
which probably possess cooler thin disks with small 
coronas (if any). The hot thick disks proposed for radio loud quasars  
are expected to contain 
more relativistic electrons (${>\atop\sim}m_e c^2$), 
thus  
leading to more synchrotron radiation in the radio, and the geometric
thickness possibly creates the 
environment
for funneling and collimating the 
emitted particles into jets (as it appears to be the 
cases for 
3C 279 and 3C~273).  
So, in this scenario, radio loud quasars, which are believed to
possess ion tori, 
eventually evolve into ``bright'' elliptical galaxies; and the 
degree of radio
loudness will depend on the phase of the stabilizing mode of
the ion torus. 
Now on the other hand, radio 
quiet galaxies are usually 
associated with spiral galaxies, in which large-scale collimated
radio jets are absent (Kotanyi, Hummel, \& van Gorkom 1987).  
However, observations of 
spiral galaxies show some evidence of radio emission strictly confined
to the optical disk.  These observations are consistent with the
model presented in this 
paper. This model suggests that radio quiet quasars may have 
thin disks, as mentioned above, with geometrically smaller ion 
coronas, if any, than radio 
loud quasars. This suggestion implies  
that radio quiet quasars eventually evolve into bright
spiral galaxies, with
radio and optical emission being confined to the disk. That is, 
without the
presence of a geometrically larger ion torus to perhaps maintain the 
collimation of the jet as in radio loud quasars
(the initial collimation due to 
Penrose processes may not be able to do it alone), 
a radio quiet quasar results, which may eventually evolve into an
AGN-type  spiral
galaxy, lenticular galaxy, or maybe even elliptical
galaxy, depending on the
size of
the geometrically smaller ion torus.   

\section{CONCLUSION AND FUTURE WORK}
\label{sec:conclusion}

I have presented a model 
that can possible explain the observed features in the
continuum spectra of 3C 279 and 3C 273: using Penrose processes
[PCS and PPP (\gggg)], 
a thin disk/ion corona (or ADAF), and 
a magnetic field strength ($\sim 10^2$~gauss) 
like that typically assumed for an accretion 
disk surrounding 
a KBH.
I have demonstrated that the general
spectra of AGNs in the high energy regime
 can be explained by this model with or 
without evoking relativistic beaming.
This makes this model feasible 
to explain the general emission spectrum of any AGN,
irrespective of the observing angle, whereas, many other 
models to date must rely on relativistic beaming. 
Now, when relativistic beaming
is observed to be important, such as through superluminal motion, 
then it can be incorporated into the modeling of the source to 
reproduce its unique spectrum.   For example, relativistic beaming 
seems to be more important in 3C~279 than in 3C~273.  That is, when 
relativistic beaming is included in the modeling of these sources, and 
3C 279 is given
a smaller observing angle relative to the axis of the jet,
it increases the observed luminosity
and energy relative to the spectrum of 3C 273; this is 
consistent with  what 
we observe (compare the $\gamma$-ray luminosities of Figs.~1$a$ and~1$b$).
 
Moreover, when relativistic beaming is included and a 
power law spectrum for the protons in ADAFs is assumed, 
which can yield $\gamma$-rays due to the $\pi^0$ decays with 
energies
$\sim 2 $~GeV 
(Mahadevan et al. 1997),
this present model, using the PPP (\gggg) and such
$\gamma$-rays, can 
explain the $\sim$~TeV $\gamma$-ray observations of the 
blazar Mrk~421.  For example, when the blueshift at 
the photon orbit is taken into account for these $\pi^0$-induced
$\gamma$-rays, the PPP (\gggg) can yield
Lorentz factors as high as
$\gamma_e\sim 10^5$ for the blobs. 
If we choose $\Ga$ to be $\sim 25$ (see eqs.~[10] to~[14]),  
such blobs when viewed at an observing angle $\sim  1^\circ$ 
will emit
$\gamma$-rays with apparent energies $\sim 2$~ TeV.  

Importantly, the model presented here applies 
to all mass size black
holes, which includes
stellar mass black holes as well. 
For example, the 
observations of the classical stellar/galactic black-hole candidate
 Cygnus X-1,  the
bright Seyfert~1 galaxy
MCG---6-30-15, and  the active giant elliptical galaxy M87 can all be 
explained by these Penrose processes [see Williams (2002b) for 
qualitative descriptions].   

Details of the collimating effects of 
these Penrose processes,  and
accelerating processes due to the associated electromagnetic field, 
which can possibly extend the jets out to very large distances, 
will be investigated 
in future work by the author.  
Further, local acceleration due 
to strong shocks, similar to the models suggested by 
Blandford and Eichler
(1987), may be important also as the jets expand 
relativistically outward.   Such models should be investigated
using the distributions of these Penrose scattered particles.

In addition, I have argued in this paper that instability in the 
accretion disk, which can cause the disk to vary between thin disk
and ion corona (or hot torus) phases, plays 
a significant role in determining
the morphology of quasars, possibly creating the two general 
classes: 
radio loud and radio
quiet quasars.   It is probable that such quasars eventually evolve
into radio loud and radio quiet galaxies.  However, these suggestions
need to be investigated further.

Finally, I point out, the consistency of this present 
model with observations
of black hole sources, in general, shows its universal application:
 suggesting it to be a complete theory in the extraction 
of energy-momentum from a black hole.


\acknowledgments
First, I thank God for His thoughts and for making this research
possible. Next, 
I wish to thank Dr.~Henry Kandrup and
Dr.~Fernando de Felice for helpful comments and discussions.  
Also, I thank Dr.~Roger Penrose for his continual encouragement.
Finally, hospitality at
the University of Florida and North Carolina A \& T State University 
is gratefully acknowledged.  This work
was supported in part 
by the National Research Council Ford Foundation, 
Bennett College
Women's Leadership Institute, and a grant from the
American Astronomical Society.

\appendix
\section*{APPENDIX}
\label{sec:appendix}

As referenced in the main text,  examples of
 results of Williams (Paper~1; 2002b)
being duplicated are discussed in the
following items:
 
1. Gammie (1999) and Krolik (1999) present
black hole physics
(in varying degrees)  already
presented in Williams' (Paper~1) Penrose process analysis,
particularly, (a) the general method used 
to extract energy,
(b) the final resulting equations due to inertial frame dragging 
showing that
energy is extracted,
(c) the knowledge of the existence of particle orbits in the 
ergosphere inside the radius
of marginal stability
$r_{\rm ms}$, and (d) application of the general conserved energy 
$E$ and azimuthal momentum
$L$ parameters as they relate to the particle's orbit, inside the
ergosphere, in a Kerr metric (see
Appendix A of Paper~1),
in their effort to make the
Blandford-Znajek (1977)
type models work.  Specifically, they consider a magnetic
torque being applied
to the disk particles at $r\la r_{\rm ms}$ (Krolik 1999). 

In Gammie's (1999)
investigation, of the
MHD version of the Penrose effect, proposed by
Takahashi et al.
(1989), he defines an
``efficiency'' of magnetic accretion:
\begin{equation}
\epsilon^B\equiv{{F_M-F_E}\over F_M}\equiv{{(-1)-F_E}\over (-1)}=1+F_E>1,
\end{equation}
for $F_E>0$, implying that energy is extracted, where $F_M$,
normalized to $-1$
(in his calculations),
is defined as the conserved ``rest mass flux,'' indicating an
{\it assumed} constant
inward flux of negative energy particles into the black hole due to the
 torque of the {\it presumed} nonnegligible accretion disk
magnetic field at $\sim r_{\rm ms}$; and
$F_E$ is the conserved
mass-energy flux, which in general is positive inside the ergosphere,
i.e., at least if it is
to escape to infinity.
Now, the general expression for the efficiency
of the release of gravitational binding energy due to accretion
of a unit of rest mass $m_o$
is
\begin{equation}
\epsilon\equiv{E_{\rm rad}\over E_{\rm tot}}={\gamma m_o c^2-E
\over \gamma m_o c^2},
\end{equation}
where $E$ is the conserved orbital energy parameter as measured by
an observer at infinity, $E_{\rm rad}$ is the energy radiated to
infinity, and $E_{\rm tot}$ is the total energy of the
 infalling mass.   Note, $\epsilon\simeq 0.42 $ for a
maximum rotating ($a=M$) KBH at $r_{\rm ms}$ (neglecting energy                 released due to viscosity).
If $\epsilon$ is $>1$, this would imply
that $E_{\rm rad}>E_{\rm tot}$, indicating that rotational
energy is extracted
 from the black hole.
This would only occur if the total
infall energy is negative or for $E < 0$, thus the reason why Gammie 
set $F_M=-1$ (compare eq.~[A1]).
Not surprisingly, in Gammie's (1999) proposed simplified so-called
MHD Penrose
model, it was
found that $\epsilon^B\approx 1.04$, particularly when we see below
how he determined $F_E$.
But, the fact that rotational energy can be extracted due to the
Penrose mechanism had already been shown in details in Williams'
(1991, Paper~1) theoretical: analytic and numerical
calculations of realistic astrophysical
Penrose  processes in the ergosphere of a KBH
[see \S\S~II.C, IV.A.1.e,
IV.D, eq.~(2.31), and Table~II of Williams
(Paper~1)].  The efficiency to show whether or not
rotational energy is extracted in a ``classical Penrose
process'' (i.e.,
 resulting in negative energy
infall) as opposed to a ``quasi-Penrose process'' (Paper~1; see
also \S~\ref{sec:pcs} of this present manuscript) is given by 
(Williams 1991, Paper~1; Piran
\& Shaham 1977)
\begin{equation}
\epsilon^{\rm PS}\equiv {{E_3-(E_1+E_2)}
                 \over E_1+E_2},
\end{equation}
where the subscripts $_1$, $_2$, and $_3$ indicate
the infalling
particle, the orbiting target particle,
and the scattered escaping particle,
respectively.  For $\epsilon^{\rm PS}>0$, rotational energy is
extracted in the
scattering event by a classical Penrose process.
  It is found that $\epsilon^{\rm PS}$ acquires values                          up to $\sim 1.6$, for escaping particle distributions, without
the need to include the magnetic field of the accretion disk
(see  Paper~1
for details).  Nevertheless, it has been shown by Wagh, S. M.,
Dhurandhar, S. V.,
\& Dadhich (1985); Parthasarathy et al.
(1986) that electromagnetic fields can increase the efficiency of
fission and nonrelativistic Penrose processes.

The expression that Gammie (1999) uses to
eliminate two degrees of freedom by expressing
$F_E$ in terms of $F_L$ and by fixing $\Omega_F$, the angular
velocity of a magnetic field line (Takahashi et al. 1990),
can be derived using the initial
conditions and the final results of my Penrose analysis (Paper~1).
For example, Gammie defines $\Omega_F$ by the angular velocity of the
orbiting fluid at constant $r$ (i.e., $u^r=0$), where $u^\mu$ is
the 4-velocity.
He (a) assumes that $\Omega_F\approx\omega_{\rm ms}$, where 
$\omega_{\rm ms}$
defines the inertial frame dragging velocity at $r_{\rm ms}$; 
(b) assumes that the ``fluid,''
 being torque by the magnetic field at $r_{\rm ms}$,
resulting in a radial
inflow of negative energy flux, initially orbits at a
constant radius ($u^r=0$), just as I initially
assume that my target particles orbit at constant $r$,
with conserved energy $E$ and azimuthal
angular momentum $L$ as measured by an observer at infinity,
given analytically by Bardeen et al. (1972) and 
Williams (Paper~1, 2002b) for 
equatorially and nonequatorially confined particle orbits,
respectively.
The final expression of the energy of particles
escaping due to Penrose scattering events [see eqs.~(3.47) and
(3.89) of Paper~1], can be adjusted  to ``fit''
Gammie's problem:
That is,
\begin{eqnarray}
E^\prime&=&e^\nu \varepsilon^\prime+\omega e^\psi(p^\prime)_\Phi\\
         &=&e^\nu \varepsilon^\prime+\omega L^\prime\nonumber\\
         &=&(E-\omega L)+\omega L^\prime, \label{eq:gammie}
\end{eqnarray}
where the primes indicate final conditions. The transformation
of the energy from the local nonrotating frame to the Boyer-Lindquist           coordinate frame [eq.~(2.8d) of Paper~1], and the assumption
that energy-momentum flux be continuous across the boundary between
the inflow and the disk (Gammie 1999)
(i.e., $\varepsilon=\varepsilon^\prime$ as measured by a local 
nonrotating frame observer) have been used.
Let $E\equiv\tilde E m_o$, $L\equiv\tilde L m_o$,
with $\Omega_F\equiv\omega$,
where $m_o$ is the particle's rest mass, then 
equation~(\ref{eq:gammie}) becomes
\begin{displaymath}
E^\prime=(\tilde E-\Omega_F \tilde L) m_o+\Omega_F L^\prime.
\end{displaymath}
To express the above equation as the flux equation given by
Gammie (1999, p.~L58), we multiply both sides by per unit (length)$^2$
per unit time
to obtain the desired equation:
\begin{equation}
F_E=(\tilde E-\Omega_F \tilde L)F_M+\Omega_F F_L,
\end{equation}
where the guiding center approximation has been used
(de Felice \&  Carlotto 1997; de Felice \&
Zonotti 2000; see also Williams 2002b).
Notice that Gammie (1999, p. L58) merely states the above equation
without saying anything about its derivation (nor where it came from),
except that
``one can show'' this to be true.
He then uses this
equation to solve a system of three equations:
$\partial_{u^r}F_E(r_f,u^r_f, F_L)=0$,
$\partial_r F_E(r_f,u^r_f, F_L)=0$, $F_E(r_f,u^r_f, F_L)-
F_E(r_{\rm ms},0, F_L)=0$,  using the conservation of energy,
to be solved numerically for the three unknowns, $F_L$ and
the fast critical points ($r_f,u^r_f$).  He thus finds in some
cases that $F_L>0$ and $F_E>0$; this is no surprise since
it appears he used results derived from radially infalling
particles, colliding with the orbiting
target particles (Paper~1), only altering them to express 
results for, instead, a radially directed
magnetic field, assumed to torque the orbiting
``target'' particles, causing them to have negative, retrograde
energy orbits, for which he represents as a negative                            ``rest mass flux''
[$F_M\approx -\dot Mr/(2H)$ ] normalized to
be $-1$.  To find that $F_L>0$ and $F_E>0 $, from my results
(Paper~1),
as modified  by Gammie,
simply means that negative particle energy is falling into the
KBH, resulting in positive particle energy and angular momentum being
extracted.  However, from his results one can not tell if, or
how, the magnetic
field torques the particles, putting them on negative energy plunges.
Only a microphysical analysis, such as that done by Williams (Paper~1),
for Penrose scattering processes, will determine what effect the
magnetic field has on the orbiting test particles.
If Gammie (1999) and Krolik (1999) are proposing
that the magnetic torque, acting on the
orbiting particles, can serve the same purpose as infalling
particle collisions with the orbiting target particles
(Paper~1),   then
the task concerning these Blandford-Znajek type models would be to
show in
details this to be true, i.e., applying the microphysics involved,
including the Lorentz
force on a particle in the local frame,  to assess
 the effectiveness of
magnetic-particle interactions.  This has somewhat been addressed by
Li (2003), as discussed in item 2 below.
 
2. The relationship between the local velocity of the
particle, the conserved specific azimuthal
angular momentum ($L/m_o$) and specific energy ($E/m_o$),
presented by Li [2003: eqs.~(24), (25) and~(26)] for 
{\it fluid} particles
are just those
previously derived
by Williams for {\it single} particles 
[Paper~1: eqs.~(3.15a) and~(3.15b)] and
the particle azimuthal circular velocity presented by
Bardeen, Press, \& Teukolsky [1972: eq.~(3.10)].
Moreover, Li (2003) finds that the dynamical effects of the
magnetic field are negligible, in the region $r<r_{\rm ms}$,                    inside the ergosphere, as
previously suggested by Williams (2002b), contrary to that
proposed by Gammie (1999) and Krolik (1999).


\clearpage


\begin{deluxetable}{lccccccccccc}
\tabletypesize{\footnotesize}
\rotate
\tablecolumns{12}
\tablewidth{8.2in}
\tablecaption{Accretion Disk Parameters}
\tablehead{
\colhead{} & \colhead{$\alpha$} & \colhead{$y$} & 
\colhead{$\dot M$} & \colhead{$M$} & \colhead{$r$} & 
\colhead{$T^{\rm \,a}$} & \colhead{$T_s^{\rm \,a}$} & 
\colhead{$T_e^{\rm \,b}$} & \colhead{$T_i^{\rm \,b}$} & 
\colhead{$T_e^{\rm \,c}$}  & \colhead{$T_i^{\rm \,c}$}\\
\colhead{Model } & \colhead{$(\alpha<1)$} & \colhead{$(y\le 1)$} & 
\colhead{$({M_\odot\over{{\rm yr}}})$} & \colhead{$(M_\odot)$} & 
\colhead{$(M)$} & \colhead{$(K)$} & \colhead{$(K)$} & 
\colhead{$(K)$} & \colhead{$(K)$} & \colhead{$(K)$} & 
\colhead{$(K)$}}
\startdata
$1^{\rm \,d}$...... &$0.1  $&$0.3  $&$~0.03 $&$10^8 $
&$1.3  $&$8.81\, (5)$
   &$4.13\, (6) $&\nodata&\nodata&\nodata&\nodata\\
2........ &$0.1  $&$0.3  $&$0.1 $&$10^8 $
&$1.3  $&$8.81\, (5)$
&$1.20\, (7) $&$4.53\,(8)$&$2.46\,(12)$&$1.64\,(9)$
&$6.79\,(11)$\\
3........ &$0.1  $&$0.3  $&$~0.35 $&$10^8 $
&$1.3  $&$8.81\, (5)$
&$3.67\, (7) $&$4.19\, (8)$&$6.12\, (12)$&$1.24\, (9)$
&$2.07\, (12)$\\
4........ &$0.1  $&$0.6  $&$~0.35 $&$10^8 $
&$1.3  $&$8.81\, (5)$
   &$3.67\, (5) $&$5.93\, (8)$&$4.33\, (12)$&$1.56\, (9)$
&$1.64\, (12)$\\
5........ &$0.1  $&$1.0  $&$~0.3 $&$10^8 $
&$1.3  $&$8.81\, (5)$
&$3.20\, (7) $&$7.72\, (8)$&$3.00\, (12)$&$1.92\, (9)$
&$1.21\, (12)$\\
6........ &$1.0  $&$0.3  $&$~0.35 $&$10^8 $
&$1.3  $&$4.95\, (5)$
&$6.11\, (7) $&$3.63\, (8)$&$3.28\, (11)$&$7.44 (8)$
&$1.60\, (11)$\\
\tableline
\\[.5in] 
\cutinhead{Additional Accretion Disk Parameters}
~~~~ &r(inner)&&r(instability)&$~~L_{\rm bol}$&$
~~L_\gamma^{\rm \,b}$
&$~~L_\gamma^{\rm \,c}$
&$~~~\bb^{\rm \,e}$&$~~~h^{\rm \,a}$&$~~~h^{\rm \,b}$
&$~~~h^{\rm \,c}$\\
Model &~~$(M)$&&$(M)$
&
$({\rm erg/ s})$&$({\rm erg/ s})$&$~({\rm erg/ s})$
&&$~(M)$&$~(M)$&$~(M)$\\[.05in]
\tableline
\\[.001in]
$1^{\rm \,d}$...... &$~1.3$ to 150&&(does not exist)&
$3.6\, (44)  $&\nodata
   &\nodata&$0.2$&$0.001$&\nodata&\nodata\\
2........ &$~1.3$ to $ 470$&&$~~\,1.089 $ to 6.73&$1.2\, (45)$
   &$3.0\, (42) $&$2.9\,(43)$&$0.2$&$0.004$&$0.587$&0.309\\
3........ &$~~1.3$ to $ 1675$&&$1.089 $ to 21&$4.2\, (45)$
   &$1.7\, (43) $&$3.5\,(43)$&$0.2$&$0.013$&$0.928$&0.539\\
4........ &$~~1.3$ to $ 1675$&&$1.089 $ to 21&$4.2\, (45)$
   &$2.6\, (43) $&$2.4\,(43)$&$0.2$&$0.013$&$0.779$&0.481\\
5........ &$~~1.3$ to $ 1450$&&$~~\,1.089$ to 17.5&$3.6\, (45)$
   &$2.6\, (43) $&$1.7\,(42)$&$0.2$&$0.012$&$0.649$&0.413\\
6........ &$~~1.3$ to $ 2100$&&$1.089$ to 55&$1.3\, (46)$
   &$1.3\, (44) $&$2.0\,(45)$&$0.6$&$0.013$&$0.215$&0.150\\
\enddata
\tablenotetext{a}{Temperature (or height) of inner region before instability
sets in.} 
\tablenotetext{b}{After instability sets in ($\tau_{\rm es}<1$).} 
\tablenotetext{c}{After instability sets in ($\tau_{\rm es}>1$).} 
\tablenotetext{d}{Inner region exist; region of instability does not exist.} 
\tablenotetext{e}{$\bb$ is the efficiency factor in the bolometric
luminosity [it is generally assumed that  $0.1<\bb <1 $ (Eilek 1980)];
$L_{\rm bol}=\bb(GM/r_{\rm ms)}\dot M$.} 
\end{deluxetable}
\clearpage

\begin{table}
\begin{center}
\caption{Model Parameters for 3C 279 (PCS)}
\begin{tabular}{crrrrrrrr}
\tableline\tableline
 &$r$~~~&$E_e$~~~~&$~~\log (\nu_{\rm ph})$&
$\log (\nu_{\rm peak})$
&$\log (L_{\rm peak})$&$\log (L_{\rm obs})$
&$f_1^{\rm \,a}~~~~$
&$f_2$~~~~~~~ \\
 Case no. &$(M)$&(MeV)&(Hz)
&(Hz)&
$({\rm erg/ s})$&$({\rm erg/ s})$&&\\
\tableline
$1^{\rm \,b}$..... &$1.089  $&$~0.539^{\rm \,c} $&$16.23 $&$18.25 $
&$~~46.73$&$\ldots~~$
   &$4.09\, (-2) $&$1.0\,$~~~~~~~~~~~~\\                    
2........&$1.2~~~  $&$0.349~~ $&$17.5~  $&$18.81 $
&$~~46.32$&$\ldots~~$
   &$4.97\, (-2) $&$1.0\,$~~~~~~~~~~~~\\
3........&$1.13~\,   $&$0.461~~$&$17.5~  $&$19.05 $
&$~~46.71$&$\ldots~~$
   &$4.16\, (-2) $&$1.0\,$~~~~~~~~~~~~\\      
4........&$1.1~~~    $&$0.456~~ $&$17.5~  $&$19.27 $
&$~~47.12$&$\ldots~~$
   &$4.17\, (-2) $&$~1.0\,$~~~~~~~~~~~~\\
5........&$1.089  $&$0.539^{\rm \,c} $&$17.5~  $&$19.47 $
&$~~47.46$&$\ldots~~$
   &$4.09\, (-2) $&$~1.0\,$~~~~~~~~~~~~\\
6........&$1.089  $&$0.539^{\rm \,c} $&$18.86 $&$20.21 $
&$~~47.12$&$\ldots~~$
   &$4.09\, (-2) $&$~1.0\,$~~~~~~~~~~~~\\       
7........&$1.089  $&$~0.539^{\rm \,c} $&$19.56 $&$20.56 $
&$~~46.35$&$\ldots~~$
   &$4.09\, (-2) $&$~1.0\,$~~~~~~~~~~~~\\ 
$8^{\rm \,d}$......&$1.089  $&$1.435~ $&$18.86 $&$20.61 $
&$~~47.97$&$46.85$
   &$4.09\, (-2) $&$~1.0$ ~~[0.076]\\       
9........&$1.089  $&$1.838~ $&$18.86 $&$20.71 $
&$~~48.39$&$46.92$
   &$4.09\, (-2) $&$~1.0$ ~~[0.034]\\
10........&$1.089  $&$2.667~ $&$18.86 $&$20.85 $
&$~~48.66$&$47.06$
   &$4.09\, (-2) $&$~1.0$ ~~[0.025]\\ 
11........&$1.089  $&$4.543~ $&$18.86 $&$21.06 $
&$~~49.22$&$47.2~\,$
   &$4.09\, (-2) $&$~1.0~$ ~[0.01]\,~~\\
12........&$1.089  $&$6.433~$&$18.86 $&$21.21 $
&$~~49.54$&$47.3~\,$
   &$4.09\, (-2) $&$~1.0$ ~~[0.006]\\ 
13........&$1.089  $&$11.79\,~~  $&$18.86 $&$21.46 $
&$~~50.19$&$47.5~\,$
   &$4.09\, (-2) $&$~~~1.0$ ~~[0.002]\\          
\tableline
\end{tabular}
\tablenotetext{\rm a}{$\theta_1=89^\circ$, $\theta_2=91^\circ$ $\Rightarrow$
$\DD\theta=2^\circ$.}
\tablenotetext{\rm b}{Case numbers 1 through 7 are for PCS by
equatorially confined target
electrons.}
\tablenotetext{\rm c}{When the more exact value is used for $r=r_{\rm mb}
=1.091M$, $E_e\lra 0.512$~MeV $\simeq \mu_e$ (see Paper~1,
2002a; Bardeen et al. 1972),
as would be expected for equatorially
confined orbits.}
\tablenotetext{\rm d}{Case numbers 8 through 13 are for PCS by 
nonequatorially
confined target electrons.}
\end{center}
\end{table}
\clearpage
 
\begin{table}
\begin{center}
\caption{Model Parameters for 3C 279 (PPP)}
\begin{tabular}{crrrrrrrr}
\tableline\tableline
 $r=r_{\rm ph}$ &$(E_{\mp})_{\rm peak}$&$\log (\nu_{\rm peak})$
&$\log (L_{\rm peak})$
&$\log (L_{\rm obs})$
&$f_1=f_3^{\rm \,a}$
&$f_2=f_4~~$&$f_5~~~~~~$\\
Case no. &~(MeV)&~(Hz)
&$~~({\rm erg/ s})$&
$~~({\rm erg/ s})$&&&\\
\tableline
14$^{\rm b}$...... &$~~~1.289$ &$20.63 $&$46.04 $
&$\ldots~~~$&$~1.99\, (-2)$
&$~1.0\,~~~~~~~~~~~ $&$~1.0\,~~~~~~~~~~~$\\
15........&$174.6~~~ $&$22.59 $&$50.21 $
&$47.3~~\, $
&$1.99\, (-2) $&$~1.0$ ~~[0.1]~~~~&$~~1.0$ ~~[0.122]\\ 
16........&$306.7~~~ $&$22.84 $&$50.72 $
&$~~47.25~$
&$~1.99\, (-2) $&$~~~1.0$ ~~[0.06]\,~~&$~1.0$ ~~[0.095]\\ 
17........&$432.7~~~ $&$23.02 $&$51.07 $
&$47.2~~\,$
&$1.99\, (-2) $&$~1.0$ ~~[0.05]\,~~&$~1.0$ ~~[0.054]\\          
18........&$711.8~~~ $&$23.21 $&$51.50 $
&$~~47.18~$
&$~1.99\, (-2) $&$~~~1.0$ ~~[0.03]\,~~&$~1.0 $ ~~[0.053]\\ 
19........&$1068\,~~~~~ $&$23.40 $&$51.98 $
&$~~47.15~$
&$~1.99\, (-2) $&$~~~1.0$ ~~[0.02]\,~~&$~1.0$ ~~[0.037]\\ 
20........&$1673\,~~~~~  $&$23.60 $&$52.40 $
&$47.1~~\, $
&$1.99\, (-2) $&$~~~~1.0$ ~~[0.014]&$~1.0$ ~~[0.026]\\       
21........&$2469\,~~~~~  $&$23.77 $&$52.70 $
&$~~47.05~$
&$~1.99\, (-2) $&$~~~1.0$ ~~[0.01]\,~~&$~1.0$ ~~[0.022]\\                       
\tableline
\end{tabular}
\tablenotetext{\rm a}{$\theta_1=89.5^\circ$, 
$\theta_2=90.5^\circ$ $\Rightarrow$
$\DD\theta=1^\circ$.}
\tablenotetext{\rm b}{Case numbers 14 through 21 have infalling initial
(incident)  photon frequency, used in the ``secondary Penrose
Compton scattering'' (SPCS),  $\nu_{ph}\simeq 7.24\times
10^{18}$~Hz.}
\end{center}
\end{table}
 
\begin{table}
\begin{center}
\caption{Model Parameters for 3C 273 (PCS)}
\begin{tabular}{crrrrrrrr}
\tableline\tableline
 &$r$~~~&$E_e$~~~~&$~~\log (\nu_{\rm ph})$&
$\log (\nu_{\rm peak})$
&$\log (L_{\rm peak})$&$\log (L_{\rm obs})$
&$f_1^{\rm \,a}~~~~$
&$f_2$~~~~~~~ \\
 Case no. &$(M)$&(MeV)&(Hz)
&(Hz)&
$({\rm erg/ s})$&$({\rm erg/ s})$&&\\
\tableline
$1^{\rm \,b}$...... &$1.089  $&$~0.539^{\rm \,c} $&$16.23 $&$18.24 $
&$~~45.93$&$\ldots$~~~&$4.09\, (-2) $&$1.0\,$~~~~~~~~~~~~\\
2........&$1.099 $&$0.456~~ $&$17.09  $&$18.91 $
&$~~46.20$&$46.0~ $&$4.17\, (-2) $&$~~~1.0$ ~~[0.632]\\
3........&$1.089  $&$0.539^{\rm \,c} $&$17.09 $&$19.11 $
&$~~45.55$&$46.1~$
   &$4.09\, (-2) $&$~~~1.0$ ~~[0.359]\\    
4........&$1.089  $&$0.539^{\rm \,c} $&$17.38 $&$19.37 $
&$~~46.68$&$46.2~$
   &$4.09\, (-2) $&$~~~1.0$ ~~[0.330]\\          
5........&$1.089  $&$0.539^{\rm \,c} $&$17.86 $&$19.72 $
&$~~46.61$&$46.4~$
   &$4.09\, (-2) $&$~~~1.0$ ~~[0.611]\\
6........&$1.089  $&$0.539^{\rm \,c} $&$18.86 $&$20.21 $
&$~~46.27$&$\ldots$~~~&$4.09\, (-2) $&$~1.0\,$~~~~~~~~~~~~\\
7........&$1.089  $&$0.539^{\rm \,c} $&$19.56 $&$20.56 $
&$~~45.48$&$\ldots$~~~&$4.09\, (-2) $&$~1.0\,$~~~~~~~~~~~~\\
$8^{\rm \,d}$......&$1.089  $&$1.435~ $&$18.86 $&$20.61 $
&$~~47.12$&$46.2~$
   &$4.09\, (-2) $&$~~~1.0$ ~~[0.121]\\
9........&$1.089  $&$1.838~$&$18.86 $&$20.71 $
&$~~47.54$&$46.2~$
   &$4.09\, (-2) $&$~~~1.0$ ~~[0.046]\\
10........&$1.089  $&$2.667~ $&$18.86 $&$20.85 $
&$~~47.81$&$46.1~$
   &$4.09\, (-2) $&$~~~1.0$ ~~[0.020]\\         
11........&$1.089  $&$4.543~ $&$18.86 $&$21.06 $
&$~~48.37$&$46.08$
   &$4.09\, (-2) $&$~~~1.0$ ~~[0.005]\\
12........&$1.089  $&$6.433~$&$18.86 $&$21.21 $
&$~~48.69$&$~~46.05
  $ &$4.09\, (-2) $&$~1.0$ ~~[0.002]\cr        
13........&$1.089  $&$11.79\,~~ $&$18.86 $&$21.46 $
&$~~49.34$&$~~46.08
  $ &$4.09\, (-2) $&$~~~1.0$ ~~[0.001]\\
\tableline
\end{tabular}
\tablenotetext{\rm a}{$\theta_1=89^\circ$, $\theta_2=91^\circ$ $\Rightarrow$
$\DD\theta=2^\circ$.}
\tablenotetext{\rm b}{Case numbers 1 through 7 are for PCS by
equatorially confined target
electrons.}
\tablenotetext{\rm c}{When the more exact value is used for $r=r_{\rm mb}
=1.091M$, $E_e\lra 0.512$~MeV $\simeq \mu_e$ (see Paper~1;
Williams 2002a; Bardeen et al. 1972),
as would be expected for equatorially
confined orbits.}
\tablenotetext{\rm d}{Case numbers 8 through 13 are for PCS by 
nonequatorially
confined target electrons.}
\end{center}
\end{table}
\clearpage
 
\begin{table}
\begin{center}
\caption{Model Parameters for 3C 273 (PPP)}
\begin{tabular}{crrrrrrrr}
\tableline\tableline
 $r=r_{\rm ph}$ &$(E_{\mp})_{\rm peak}$&$\log (\nu_{\rm peak})$
&$\log (L_{\rm peak})$
&$\log (L_{\rm obs})$
&$f_1=f_3^{\rm \,a}$
&$f_2=f_4~~$&$f_5~~~~~~$\\
Case no. &~(MeV)&~(Hz)
&$~~({\rm erg/ s})$&
$~~({\rm erg/ s})$&&&\\
\tableline
14$^{\rm b}$...... &$~~~1.289$ &$20.63 $&$45.18 $
&$\ldots$~~~&$~1.99\, (-2)$
&$1.0\,~~~~~~~~~~~$&$1.0\,~~~~~~~~~~~$\\
15........&$~~~6.626 $&$21.21 $&$46.32 $
&$~~46.06~$
&$~1.99\, (-2) $&$1.0$ ~~[0.8]~~~~&$~~1.0$ ~~[0.859]\\
16........&$17.28~~  $&$21.62 $&$47.19 $
&$46.1~~\,  $
&$~1.99\, (-2) $&$~~1.0$ ~~[0.4]~~~~&$~~1.0$ ~~[0.508]\\      
17........&$73.37~~ $&$22.25 $&$48.58 $
&$45.7~~\,  $
&$~1.99\, (-2) $&$~~1.0$ ~~[0.1]~~~~&$~~1.0$ ~~[0.132]\\
18........&$106.1~~~   $&$22.37 $&$48.86 $
&$~~45.65~ $
&$~1.99\, (-2) $&$~~~1.0$ ~~[0.08]\,~~&$~1.0$ ~~[0.096]\\      
19........&$174.6~~~ $&$22.59 $&$49.36 $
&$45.6 ~~\,$
&$~1.99\, (-2) $&$~~~1.0$ ~~[0.05]\,~~&$~1.0$ ~~[0.069]\\
20........&$306.7~~~ $&$22.84 $&$49.86 $
&$45.5~~\, $
&$~1.99\, (-2) $&$~~~~1.0$ ~~[0.04]\,~~&$~1.0$ ~~[0.027]\\
21........&$432.7~~~ $&$23.02 $&$50.22 $
&$45.3~~\, $
&$~1.99\, (-2) $&$~~~~1.0$ ~~[0.03]\,~~&$~1.0$ ~~[0.014]\\     
22........&$711.8~~~$&$23.21 $&$50.65 $
&$~~45.25~$
&$~1.99\, (-2) $&$~~~~1.0$ ~~[0.022]&$~1.0 $ ~~[0.008]\\
23........&$1068\,~~~~~ $&$23.40 $&$51.13 $
&$45.2~~\,$
&$1.99\, (-2) $&$~~~~1.0$ ~~[0.015]&$~1.0$ ~~[0.005]\cr
24........&$1673\,~~~~~  $&$23.60 $&$51.55 $
&$45.0~~\, $
&$1.99\, (-2) $&$~~~~1.0$ ~~[0.009]&$~1.0$ ~~[0.004]\cr          
25........&$2469\,~~~~~$&$23.77 $&$51.85$
&$44.8~~\, $
&$~1.99\, (-2) $&$~~~~1.0$ ~~[0.008]&$~1.0$ ~~[0.001]\\
\tableline
\end{tabular}
\tablenotetext{\rm a}{$\theta_1=89.5^\circ$, 
$\theta_2=90.5^\circ$ $\Rightarrow$
$\DD\theta=1^\circ$.}
\tablenotetext{\rm b}{Case numbers 14 through 25 have infalling initial
(incident)  photon frequency, used in the ``secondary Penrose
Compton scattering'' (SPCS),  $\nu_{ph}\simeq 7.24\times
10^{18}$~Hz.}
\end{center}
\end{table}

\begin{figure} 
\epsscale{.65}
\plotone{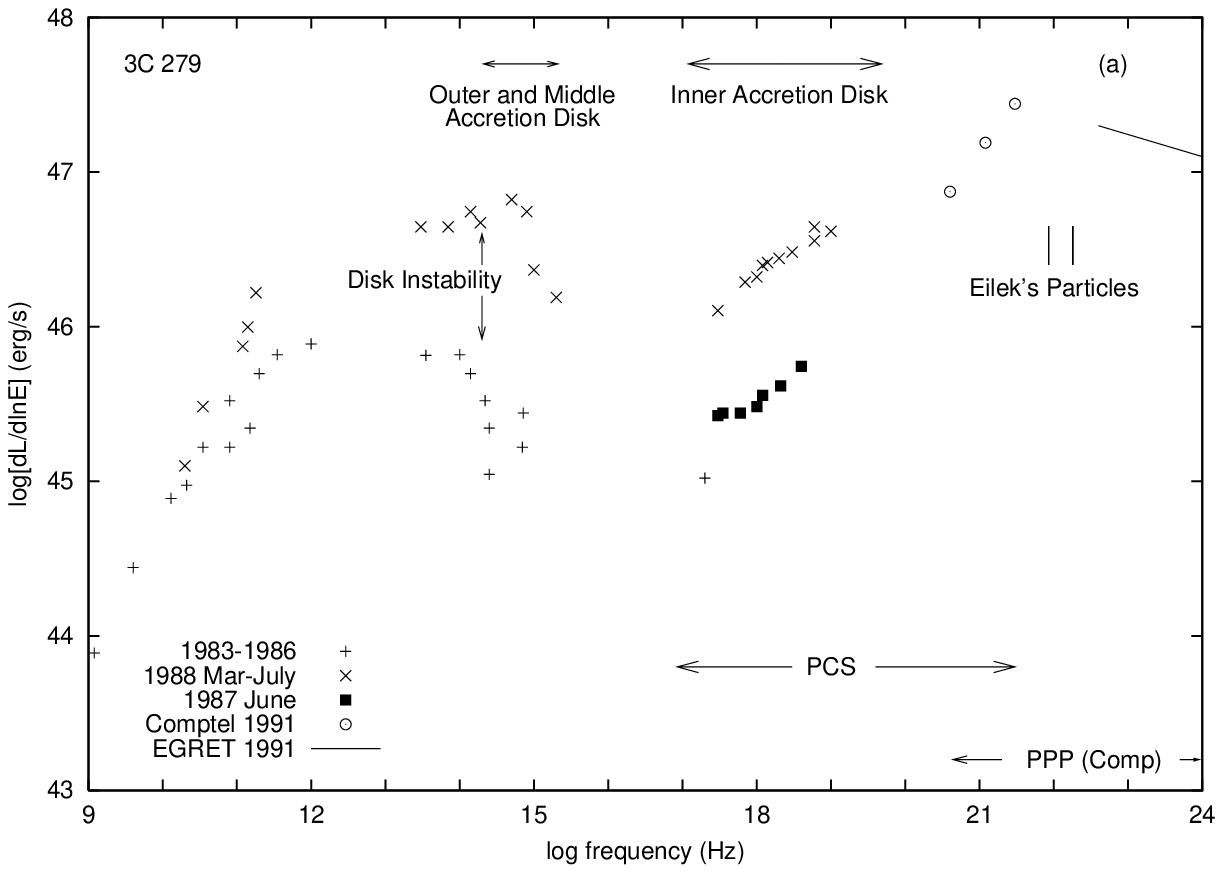}
\plotone{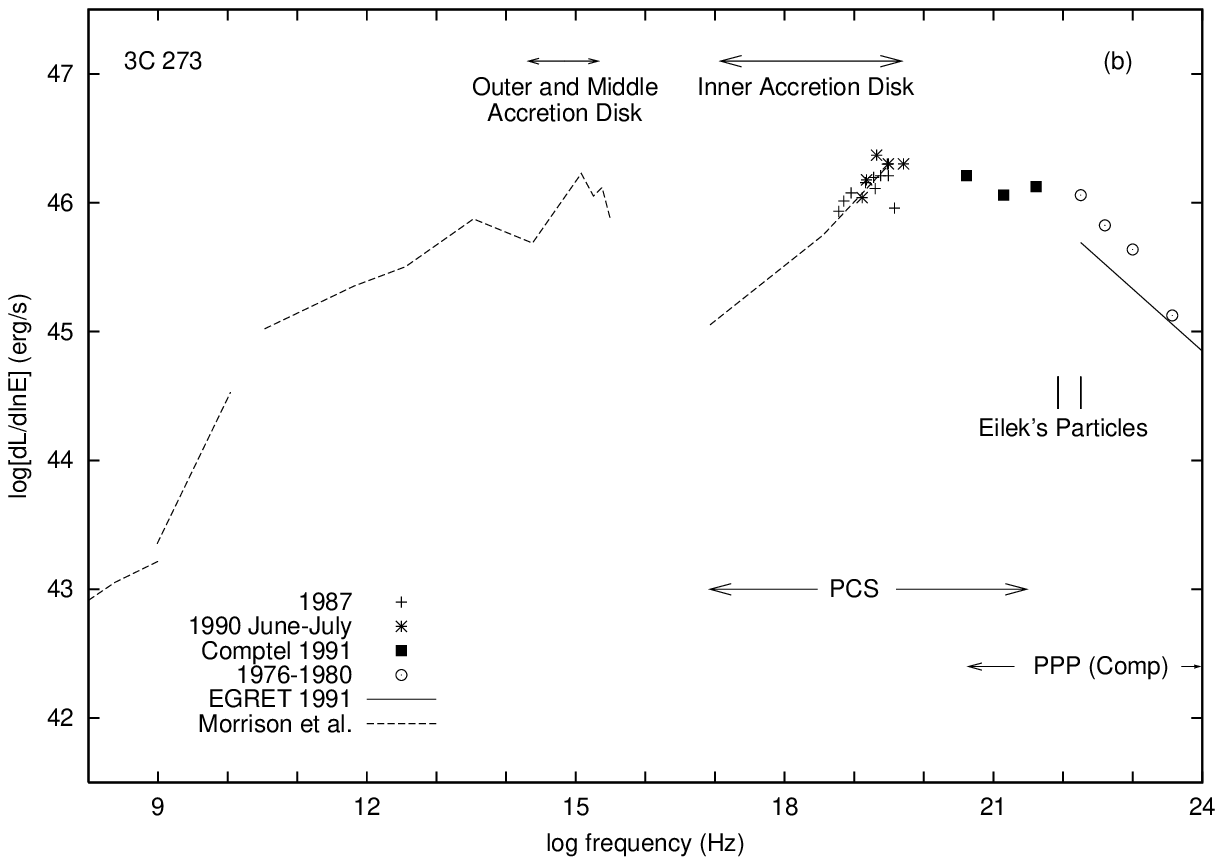}
\caption{ Schematic broad-band photon spectra  of
 quasars 3C 279 and 3C 273: comparing observations with theoretical
predictions.  The intervals indicated by horizontal arrows showing the
emission ranges due to Penrose processes [Penrose Compton scattering
(PCS) and Penrose pair production (PPP)] are
calculated based on a thin disk/ion corona
accretion model (see Table 1 and text).  General model proposed
intervals of accretion
disk emission are indicated for a $10^8 M_\odot$ Kerr black
hole; only
energy intervals of maximum brightness
are shown. Eilek's particle energy locations are
indicated by the parallel bars: the higher energy bar indicates
the $\gamma$-ray emission due to
the pion decays ($\pi^0\lra\gamma\gamma$), peaking at $\sim 75$~MeV;
and the lower energy bar indicates
the energy location of the $\, e^-e^+$
pairs produced in the subsequent  $\gamma\gamma$ pair production of the
$\pi^0$-induced photons, peaking at $\sim 35$~MeV; note, the                    ordinates and the
height of the bars have no meaning (in general $L_\gamma\sim 10^{43}~
{\rm erg \,s^{-1}}$; see text).
(a) 3C 279; references for spectral data are given in Hermsen et al.
(1993). (b) 3C 273; references for X-ray to $\gamma$-ray data are
given in Hermsen et al. (1993); spectral data for radio to X-ray
frequencies (dashed line) are taken from Morrison, Roberts,
and Alberto (1984).         \label{fig1}}
\end{figure}
\clearpage

\begin{figure} 
\epsscale{.55}
\plotone{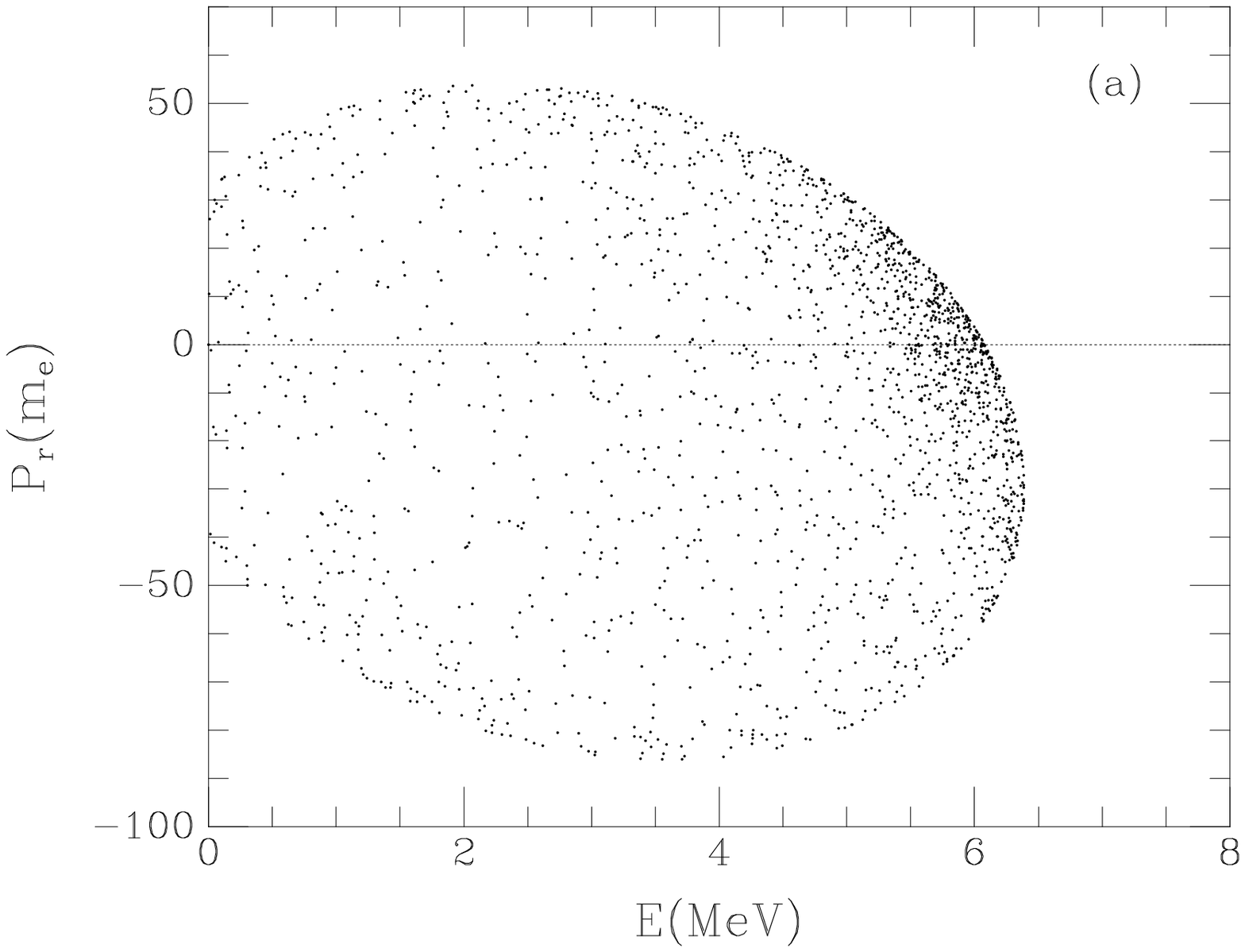}
\plotone{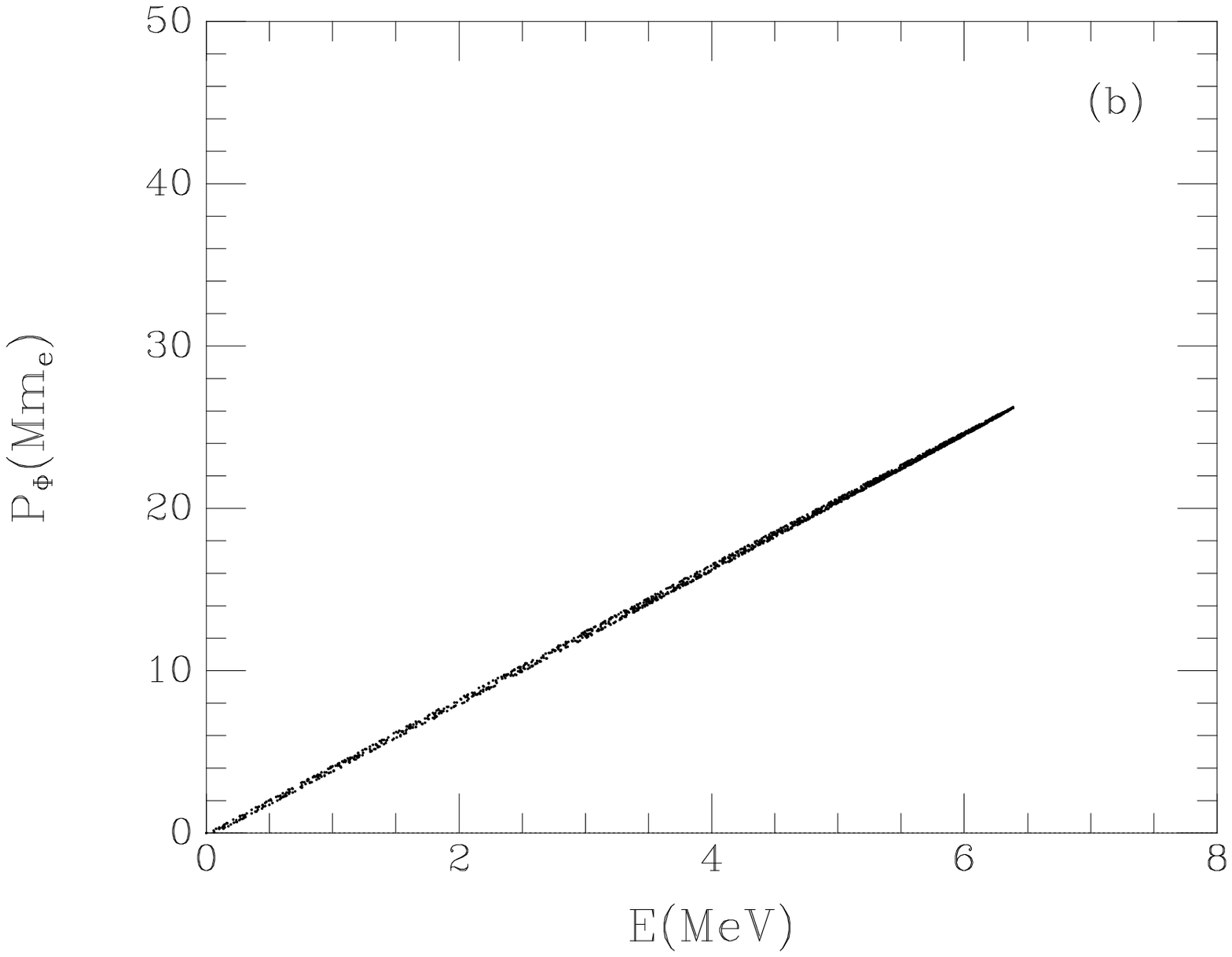}
\caption{ Compton
scattering at $r_{\rm mb}\sim 1.09M$: scatter
plots showing space momenta of
escaping photons scattered by orbiting nonequatorially
confined target electrons
(each point represents a particle from the 2000 scattering events). 
Initial
and final values are $E_{\rm ph}=0.03$~MeV (initial infalling photon),
$E_e\simeq 5.93$~MeV (nonequatorially confined target electron),
$L_e\simeq 24.1M m_e$, and $(E_{\rm ph}^\prime)_{\rm max}\simeq \break
6.25$~MeV (maximum energy of the escaping final photons).
(a) Radial momentum components:
$(P_{\rm ph}^\prime)_r$ vs. $E_{\rm ph}^\prime$.
(b) Azimuthal angular momentum components:
$L_{\rm ph}^\prime $ vs. $E_{\rm ph}^\prime$.
The units of $P_r$ and $L=P_\Phi$
are $m_e$ and $Mm_e$, respectively ($G=c=1$).  \label{fig2}} 
\end{figure}
\clearpage 

\begin{figure} 
\epsscale{.55}
\plotone{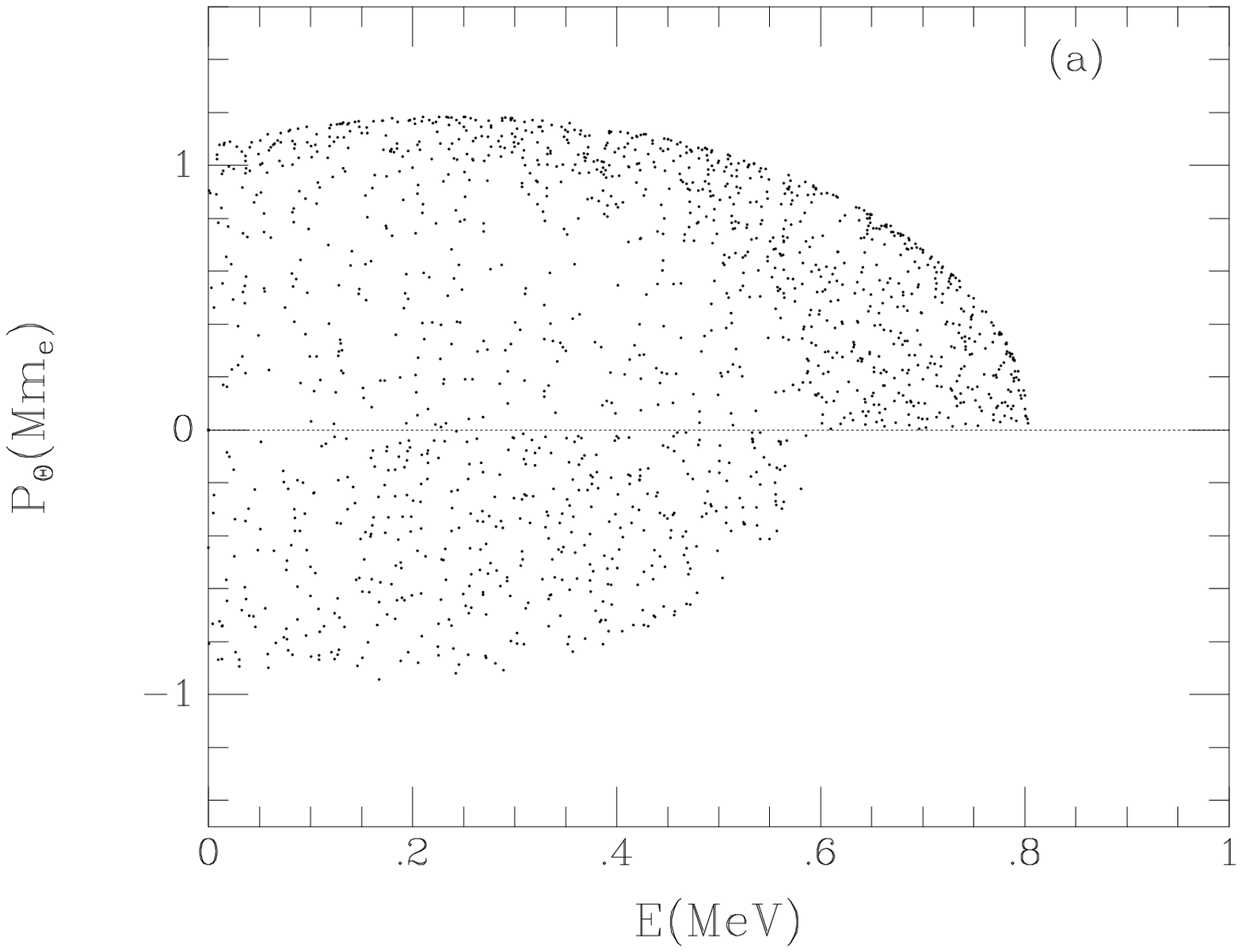}
\plotone{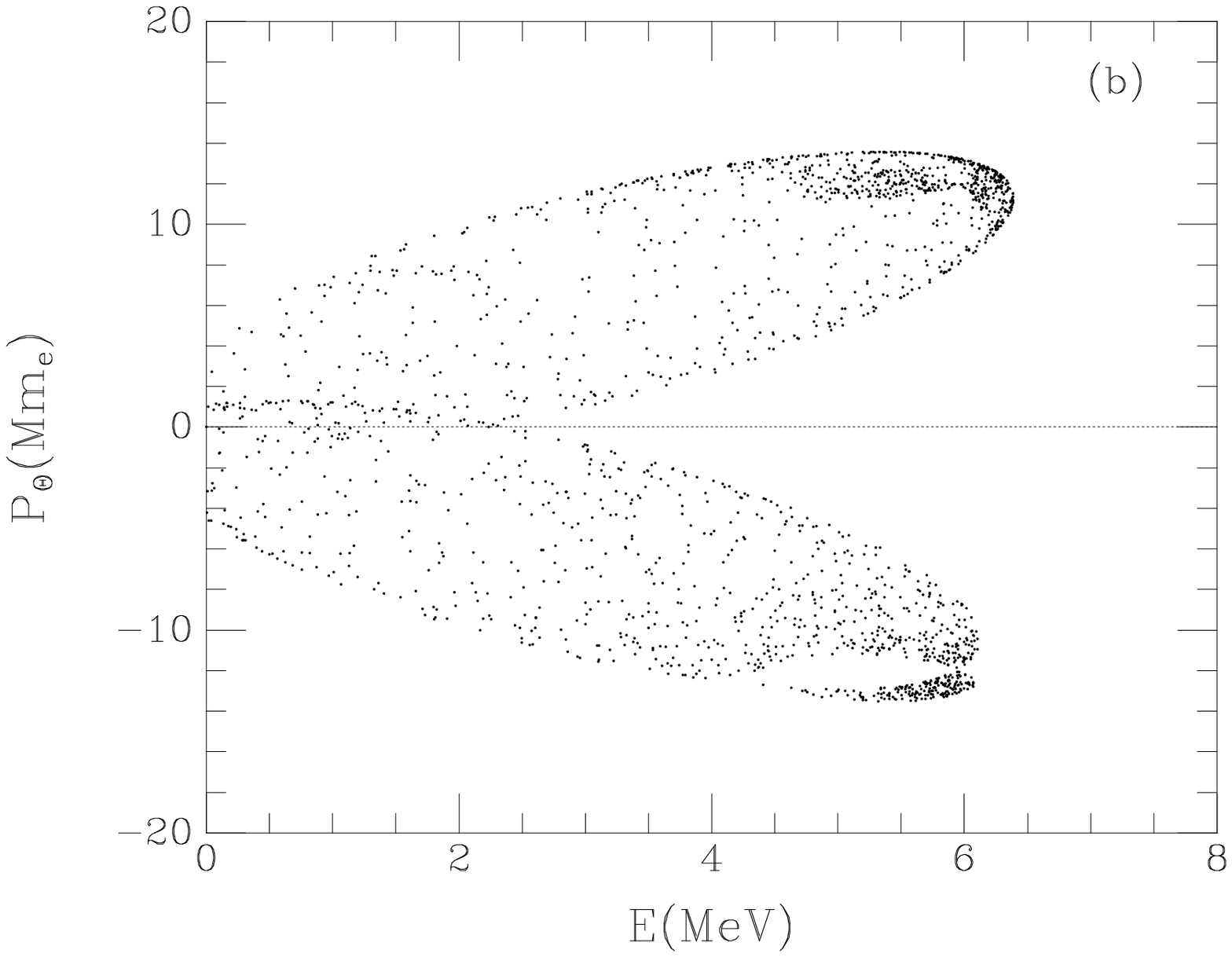}
\caption{ Compton scattering at
$r_{\rm mb}\sim 1.09M$: scatter plots
showing polar coordinate space momenta for escaping PCS photons,
$(P_{\rm ph}^\prime)_\Th ~(\equiv \sqrt{Q_{\rm ph}^\prime}\,)$
vs.~$E_{\rm ph}^\prime$. (a) $E_{\rm ph}=0.03$~MeV,
$E_e\simeq 0.539$~MeV (equatorial target electron), $(P_e)_\Theta=0$;
$(E_{\rm ph}^\prime)_{\rm max}\simeq 0.759$~MeV.
(b) Same as case presented in Fig.~2 for nonequatorially
confined targets, with $(P_e)_\Th~(\equiv \sqrt{Q_e}\,)=
\pm 12.43 Mm_e$ (half are give the positive value, the other half
the negative value); plotted here are the corresponding 
polar momenta. \label{fig3}}
\end{figure}
\clearpage  

\begin{figure} 
\epsscale{.55}
\plotone{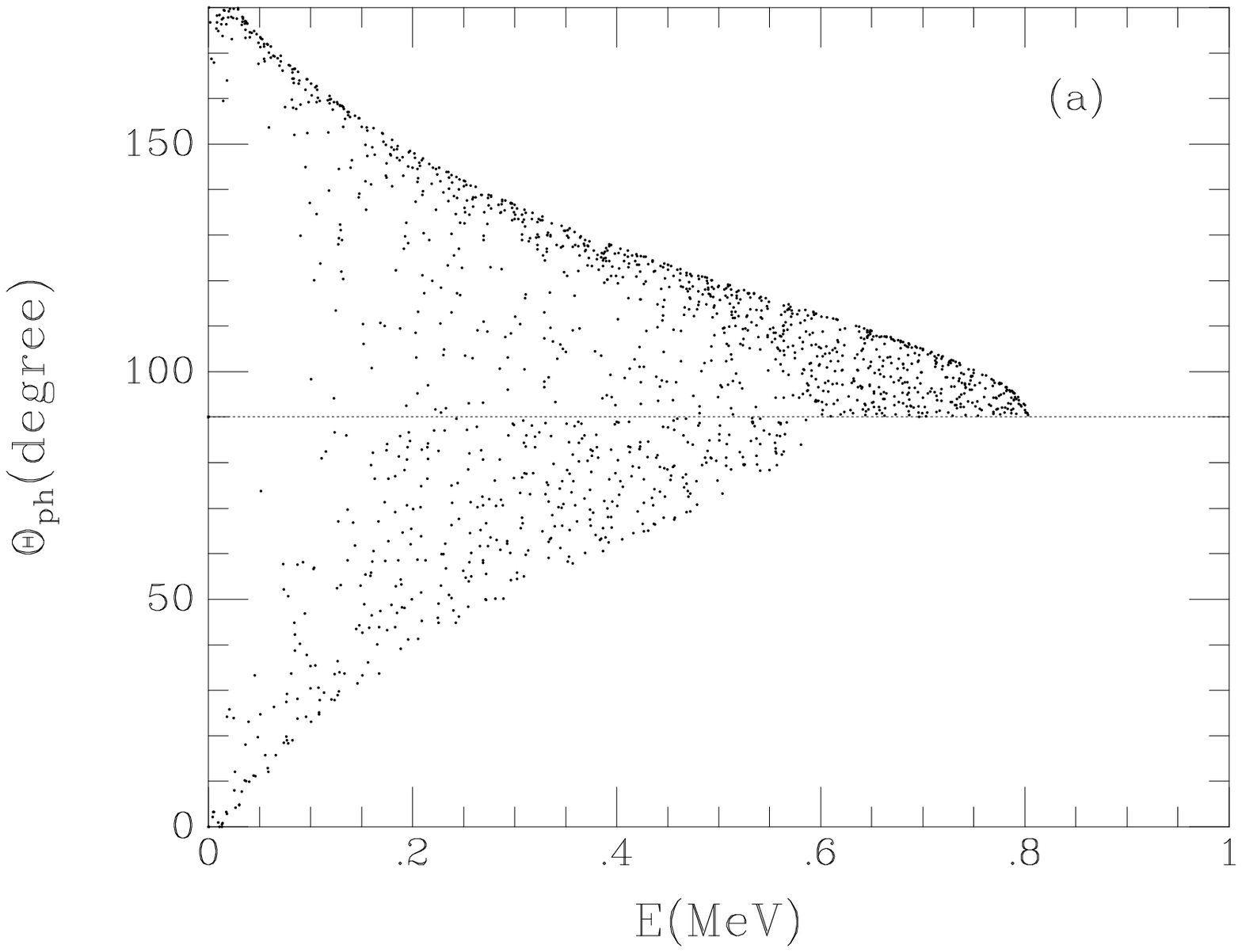}
\plotone{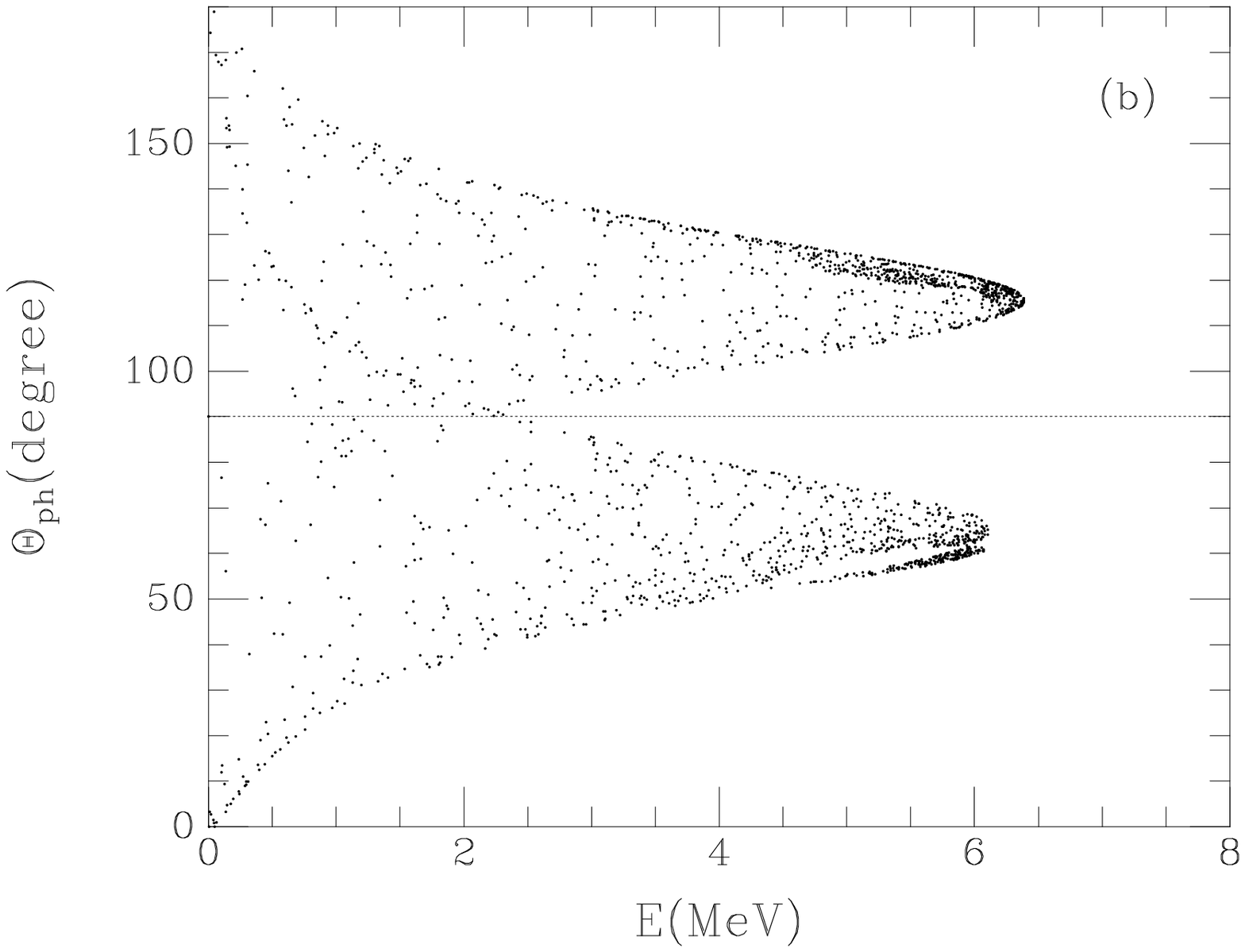}
\caption{Compton scattering: scatter plots
displaying polar angles above and below the equatorial plane
($\Th=90^\circ$) for escaping
photons, $\Th_{\rm ph}^\prime$~(degrees) vs.~$E_{\rm ph}^\prime$.
(a) and~(b)
are the same as cases presented in Figs.~3(a) and
3(b), respectively (compare for corresponding polar coordinate
momenta). \label{fig4}}
\end{figure}
\clearpage  

\begin{figure} 
\epsscale{.55}
\plotone{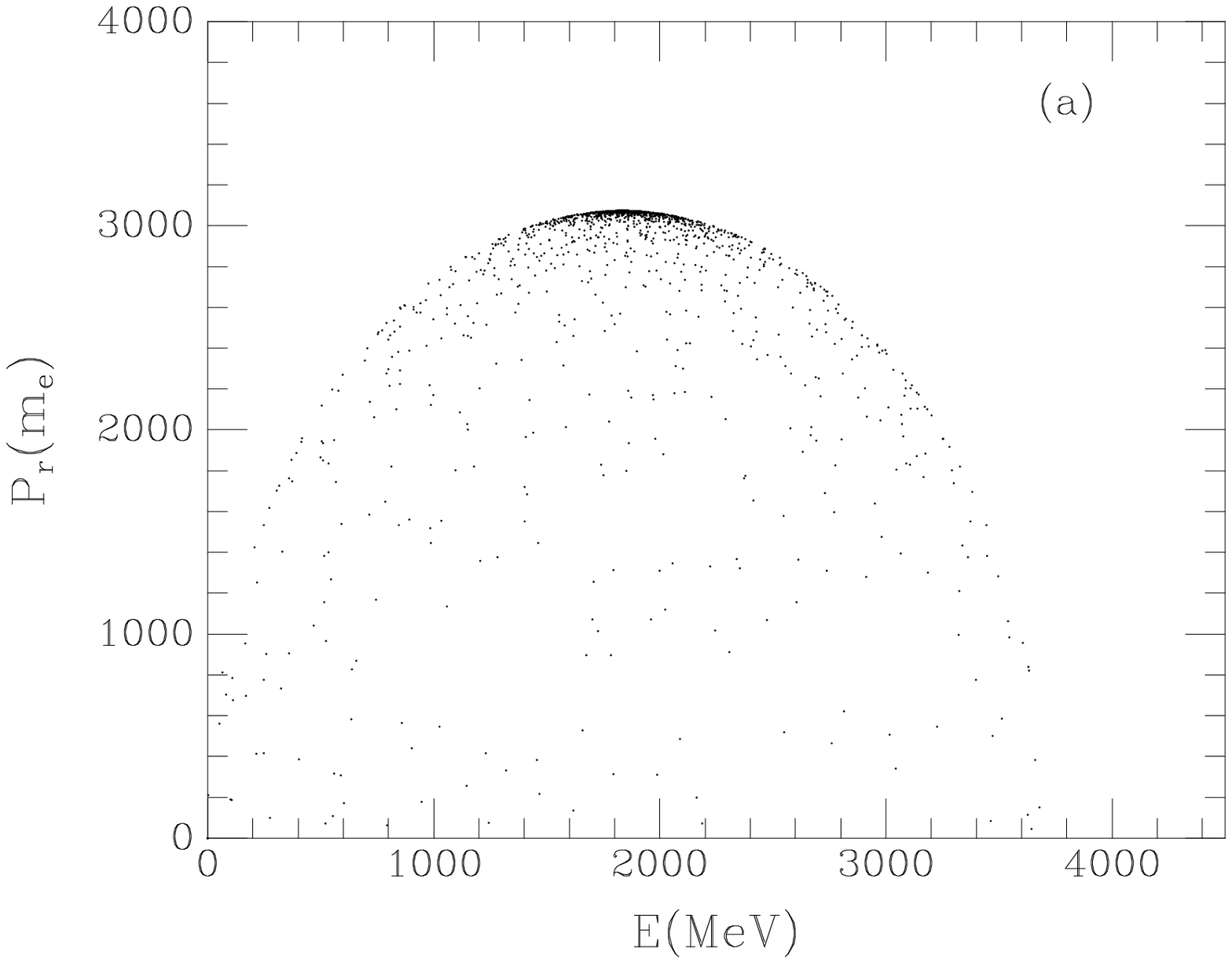}
\plotone{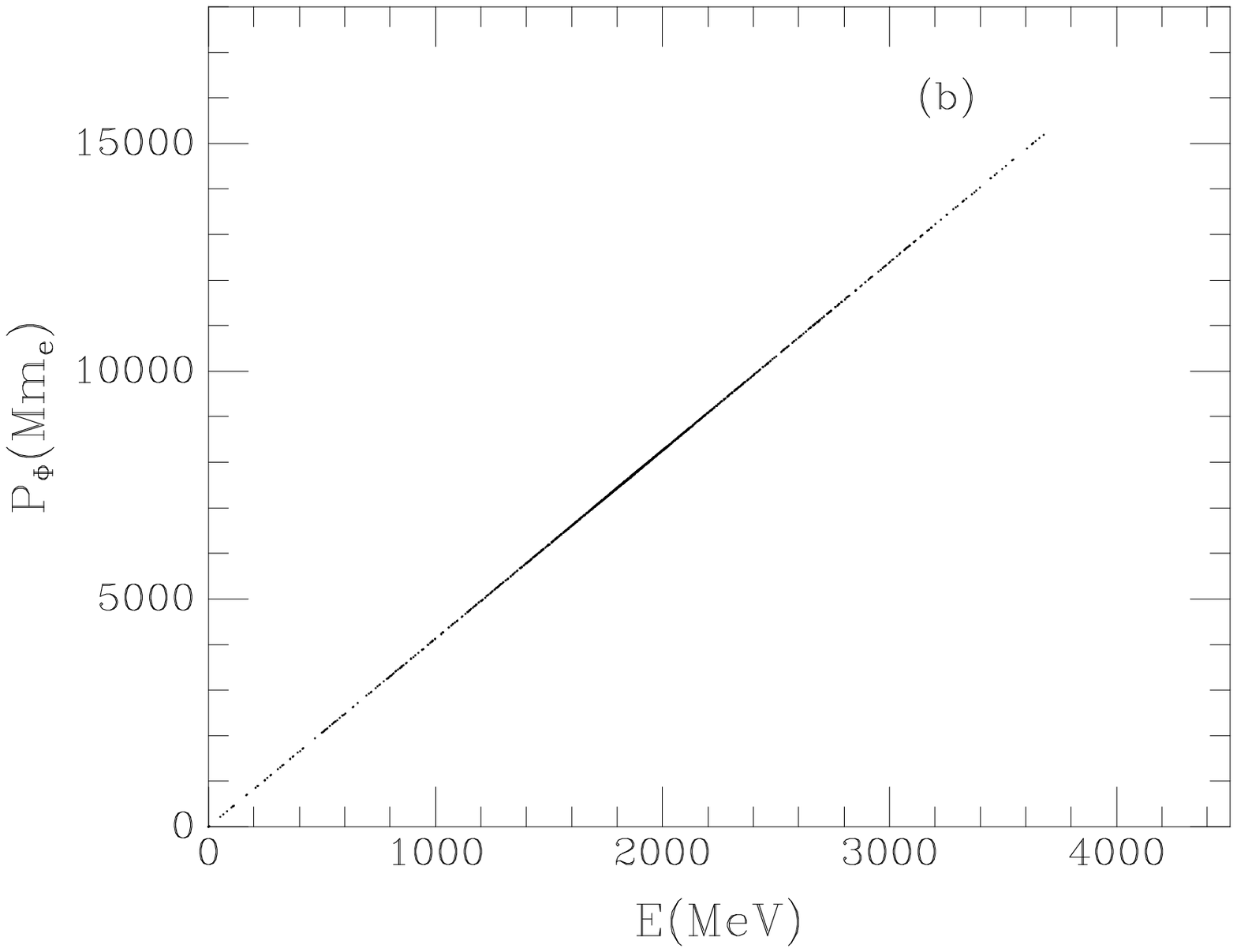}
\caption{Penrose pair production
($\gamma\gamma\lra e^-e^+$) at $r_{\rm ph}=1.074M$:
scatter plots showing
momentum components (each point represents a particle from 
the 2000 scattering events).
For the infalling photons:
$E_{\gamma 1}=0.03$~MeV, and for the target photons:
$E_{\gamma 2}\simeq 3.89$~GeV, 
$L_{\gamma 2}\simeq 1.6 \times 10^4Mm_e$, 
with $(P_{\gamma 2})_\Th~(\equiv\sqrt{Q_{\gamma 2}})=\pm 113 Mm_e$;
$(E_\mp)_{\rm max}\simeq 3.6$~GeV (maximum energy of the
escaping pairs).
(a) The radial momenta $(P_\mp)_r$
vs.~$E_\mp$. (b) The azimuthal
momenta $L_\mp$ vs.~$E_\mp$.
See Fig.~6(c) for the corresponding polar momenta.
The units of $P_r$ and $L$  are $m_e$ and $Mm_e$,
respectively ($G=c=1$).  \label{fig5}}
\end{figure}
\clearpage  

\begin{figure} 
\epsscale{.5}
\plotone{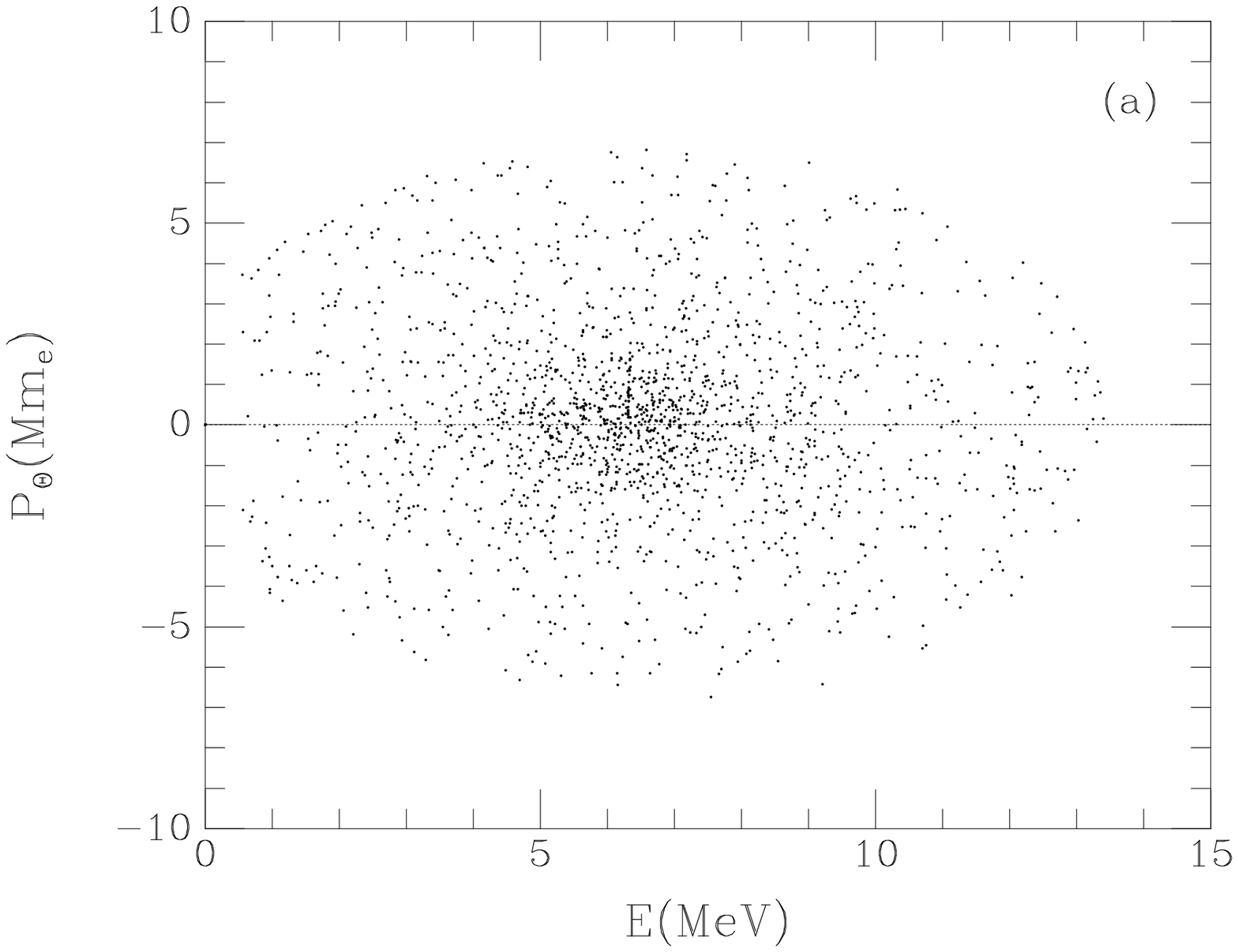}
\plotone{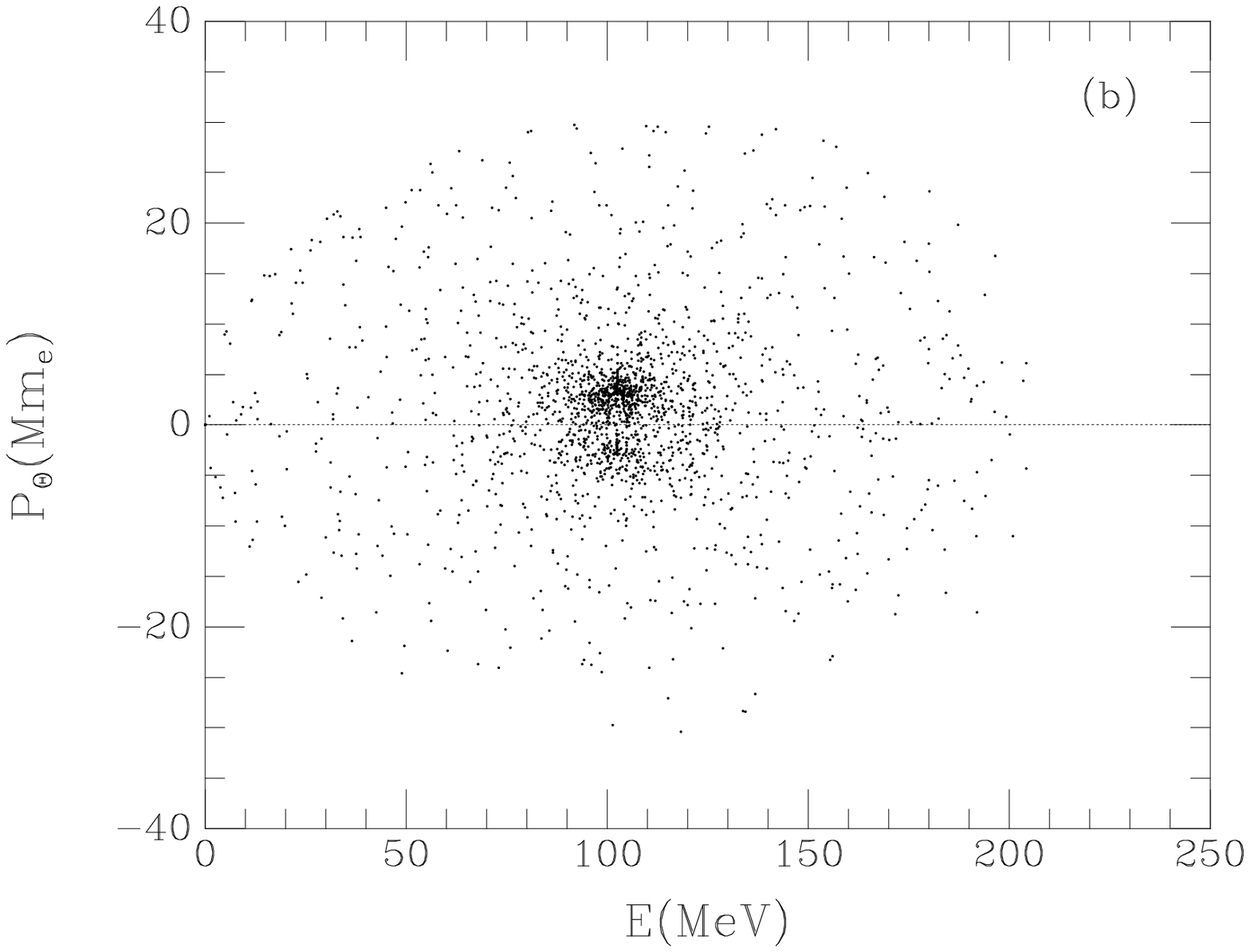}
\plotone{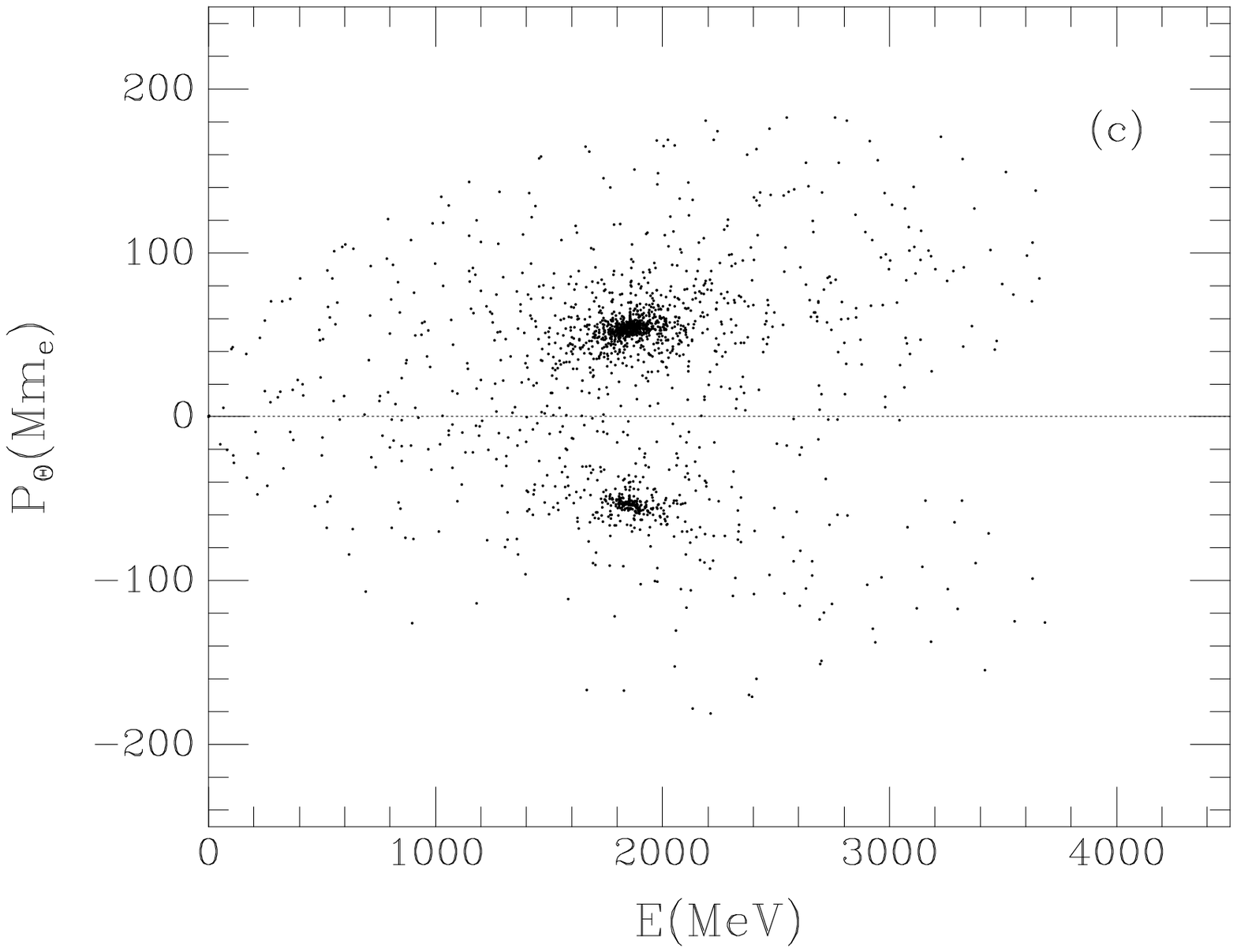}
\caption{Penrose pair production ($\gamma\gamma\lra e^-e^+$):
scatter plots showing polar coordinate space momenta for escaping
$e^-e^+$ pairs, $(P_\mp)_\Th ~(\equiv \sqrt{Q_\mp})$ vs.~$E_\mp $.
(a) $E_{\gamma 1}=0.03$~MeV, $E_{\gamma 2}\simeq 13.5$~MeV,
$L_{\gamma 2}\simeq 55.6 Mm_e$, with 
$(P_{\gamma 2})_\Th=\pm 0.393Mm_e$;
$(E_\mp)_{\rm max}\simeq 13.34$~MeV. (b) $E_{\gamma 1}=0.03$~MeV,
$E_{\gamma 2}\simeq 207$~MeV, $L_{\gamma 2}\simeq 849 Mm_e$, with
$(P_{\gamma 2})_\Th=\pm 6Mm_e$; $(E_\mp)_{\rm max}\simeq 200$~MeV.
(c) $E_{\gamma 1}=0.03$~MeV, $E_{\gamma 2}\simeq 3.89$~GeV;
$(E_\mp)_{\rm max}\simeq 3.6$~GeV;
these are the corresponding polar momenta for Fig.~5. 
\label{fig6}}
\end{figure}
\clearpage  

\begin{figure} 
\epsscale{.5}
\plotone{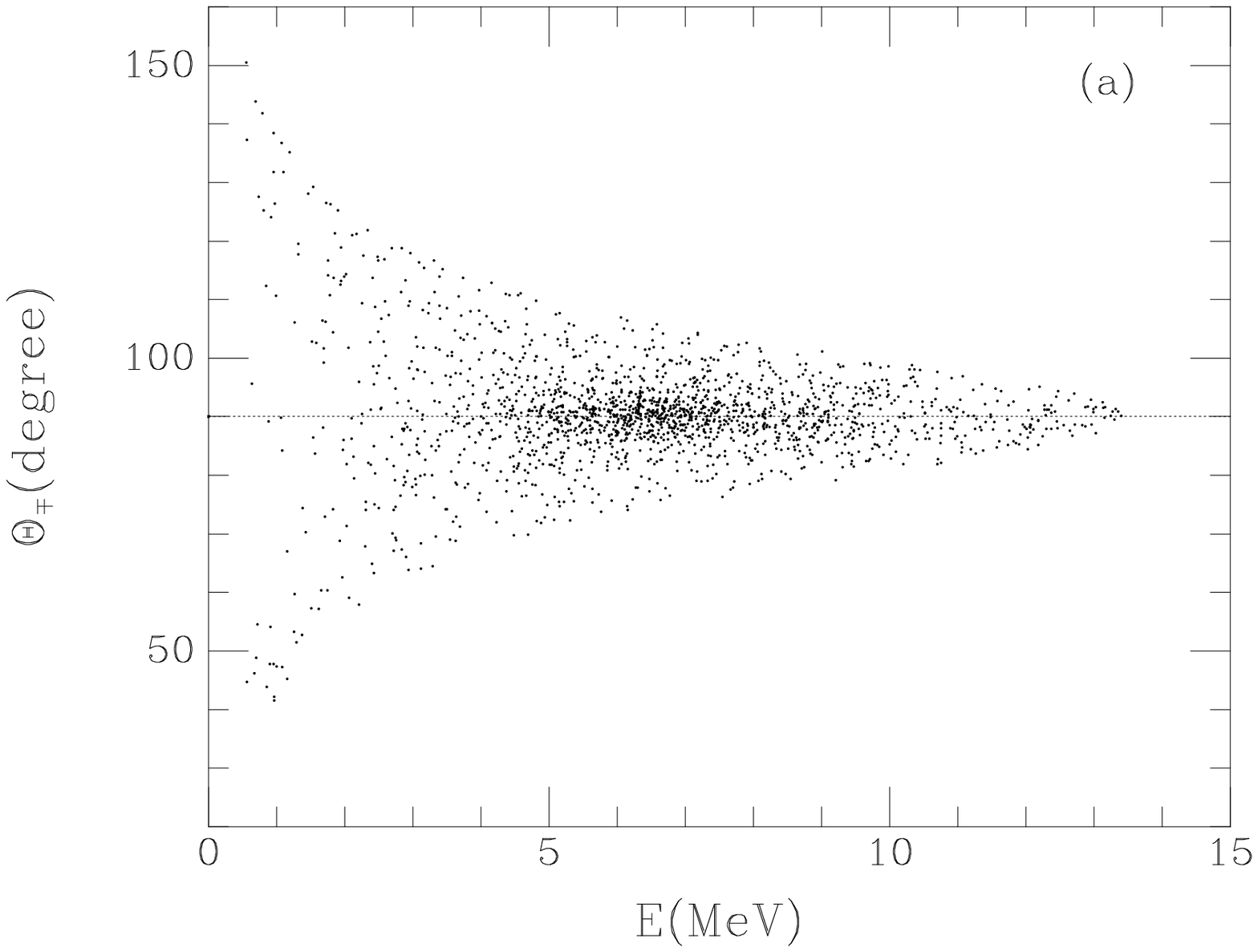}
\plotone{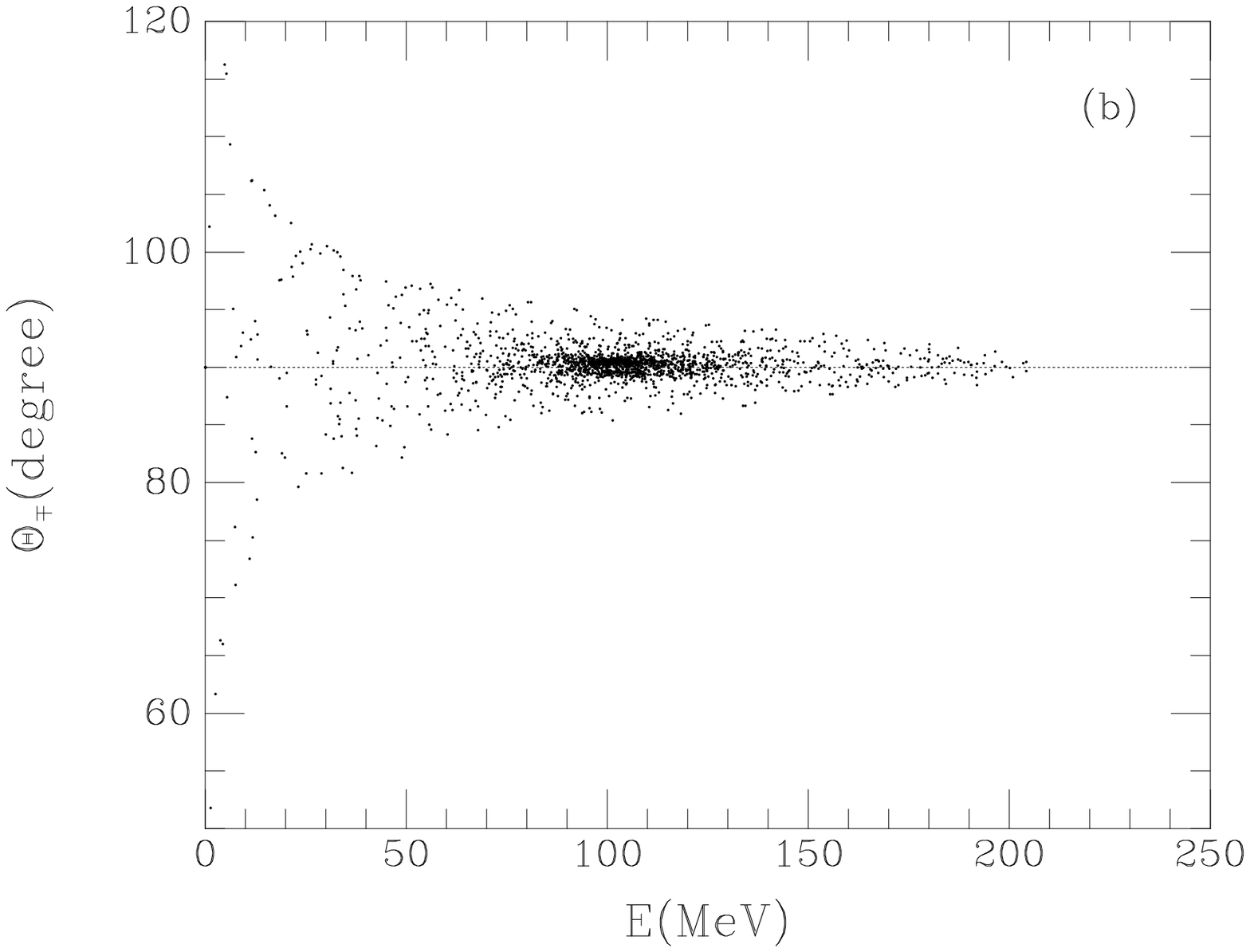}
\plotone{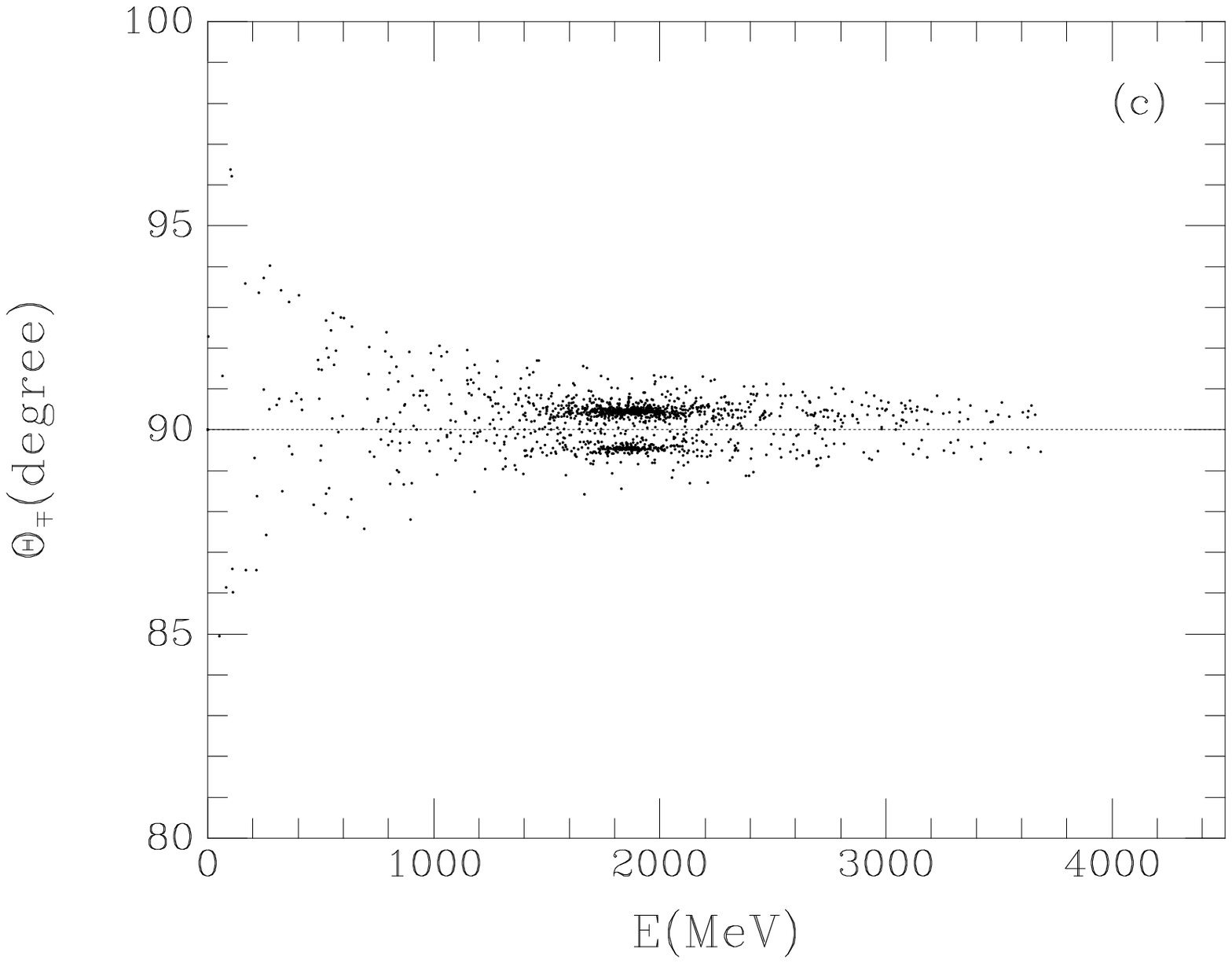}        
\caption{Penrose pair production ($\gamma\gamma\lra e^-e^+$):
scatter plots displaying polar
angles above and below the equatorial
plane ($\Th=90^\circ$) for escaping
$e^-e^+$ pairs,
$\Th_\mp$~(degrees) vs.~$E_\mp$.
(a), (b), and~(c)
are the same as cases presented in Figs.~6(a), 6(b), and~6(c),
respectively (compare for
the corresponding polar coordinate momenta). \label{fig7}}
\end{figure}
\clearpage  

\begin{figure} 
\epsscale{.7}
\plotone{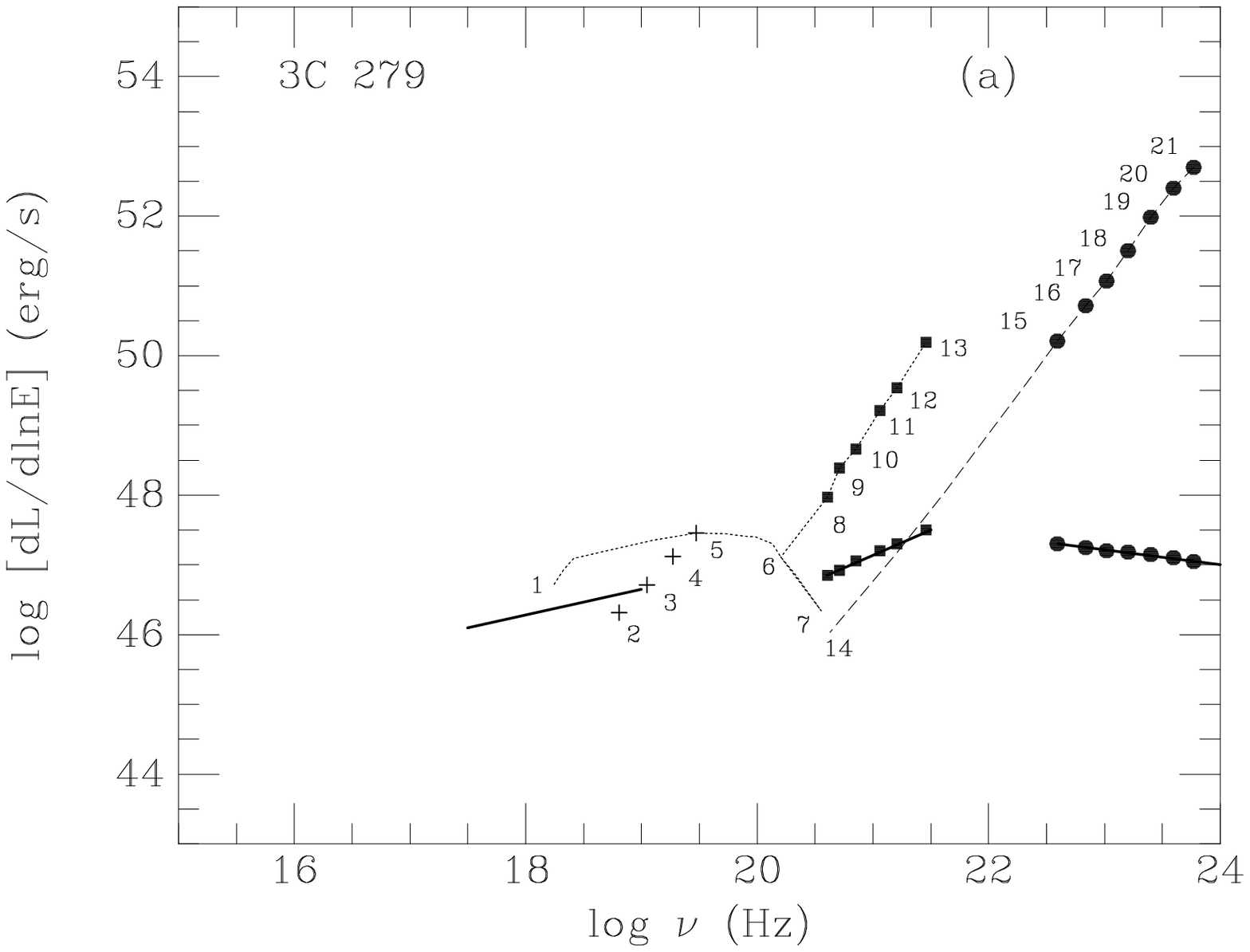}
\plotone{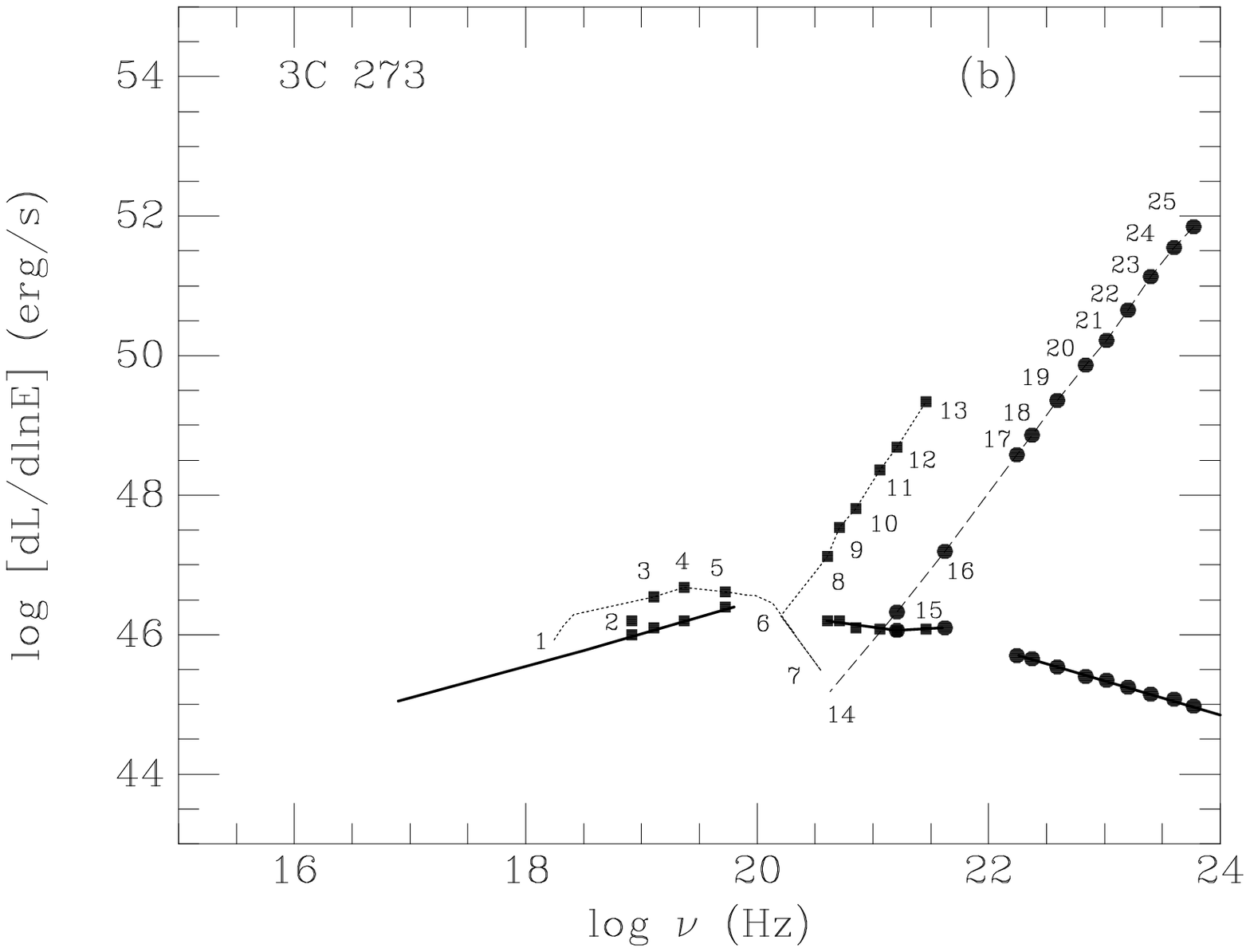}
\caption{Comparing the theoretical spectra
with observations for 3C 279 and 3C 273. The calculated
PCS and PPP (\gggg) luminosity spectra are represented by the solid
squares and large solid dots, respectively.
The observed spectra is indicated
by the solid line.  The upper curves with the solid squares and
solid dots superimposed on the dotted line and the
dashed line, respectively,
for PCS and PPP, are the spectra calculated from this model.
Superimposed on the lower solid line of the observations are
solid squares and solid dots from the model calculations 
that have been adjusted to agree with
observations.   These adjustments depend on the fraction of
particles that actually undergoes scattering events.  This fraction
is defined by the $f_n$ parameters. The numbers on the figures
correspond to the case numbers on Tables~2 through~5
(see text for further details).  (a)~Spectra
for quasar 3C 279. (b)~Spectra for quasar 3C 273. \label{fig8}}
\end{figure}
\clearpage  
 
\end{document}